\documentclass[aps,prd,longbibliography,nofootinbib]{revtex4-2}

\usepackage{graphicx}
\usepackage{amsmath,amssymb,bm}
\usepackage{siunitx}
\usepackage{booktabs}
\usepackage{microtype}
\usepackage{natbib}
\usepackage{hyperref}
\usepackage{comment}
\hypersetup{colorlinks=true,allcolors=blue}

\newcommand{\dd}{\mathrm{d}}

\newcommand{\cH}{\mathcal{H}}
\newcommand{\cL}{\mathcal{L}}

\newcommand{\cO}{\mathcal{O}}

\newcommand{\cS}{\mathcal{S}}

\newcommand{\LCDM}{\Lambda{\rm CDM}}
\newcommand{\wCDM}{w{\rm CDM}}
\newcommand{\wowaCDM}{w_0w_a{\rm CDM}}
\newcommand{\Mpl}{M_{\rm Pl}}
\newcommand{\Omm}{\Omega_{\rm m}}
\newcommand{\OmDE}{\Omega_{\rm DE}}
\newcommand{\Omk}{\Omega_k}
\newcommand{\fsig}{f\sigma_8}
\newcommand{\rd}{r_{\rm d}}
\newcommand{\rs}{r_{\rm s}}
\newcommand{\zeff}{z_{\rm eff}}
\newcommand{\DL}{D_{\rm L}}
\newcommand{\DM}{D_{\rm M}}
\newcommand{\DA}{D_{\rm A}}

\newcommand{\muMG}{\mu_{\rm MG}} 
\newcommand{\Rb}{R_b}           
\newcommand{\Rshift}{\mathcal{R}} 

\begin{document}

\title{Dark Energy After DESI DR2: \\ Observational Status, Reconstructions, and Physical Models}

\author{Slava G. Turyshev} 
\affiliation{ 
Jet Propulsion Laboratory, California Institute of Technology,\\
4800 Oak Grove Drive, Pasadena, CA 91109-0899, USA
}

\date{\today}

\begin{abstract}
We review late-time cosmic acceleration after DESI Data Release 2 (DR2), emphasizing the interplay between Type Ia supernovae (SNe Ia), anisotropic BAO, CMB calibration, and perturbation-sensitive probes (RSD and weak lensing). DESI DR2 delivers percent-level BAO distance ratios over $0\lesssim z\lesssim2.5$, including a Ly$\alpha$-forest anchor at $z_{\rm eff}=2.33$. In CMB-calibrated combinations, flat $\Lambda$CDM exhibits a mild parameter mismatch, while allowing evolving dark energy (e.g.\ CPL $w_0$--$w_a$) can improve the fit; the preference is dataset-dependent and is particularly sensitive to redshift-dependent SN calibration/selection residuals at the few$\times10^{-2}$\,mag level.  To sharpen likelihood-level interpretation, we provide two diagnostics: (i) an $r_d$-independent BAO-shape observable, $F_{\rm AP}(z)\equiv D_{\rm M}(z)/D_{\rm H}(z)$, constructed directly from published $(D_{\rm M}/r_d,\,D_{\rm H}/r_d)$ with covariance propagation; and (ii) a linear-response map from SN Hubble-diagram systematics $\delta\mu(z)$ to induced biases in $(w_0,w_a)$, yielding an explicit calibration requirement for DESI-era claims of evolving $w(z)$. We synthesize parametric and non-parametric reconstructions of $w(z)$ and $\rho_{\rm DE}(z)$ and map the resulting phenomenology to microphysical dark-energy and modified-gravity models subject to perturbation stability and gravitational-wave propagation constraints.

\end{abstract}

\maketitle


\section{Introduction}
\label{sec:intro}

The late-time accelerated expansion of the Universe is empirically established by Type~Ia supernovae (SNe~Ia) distances \cite{Riess1998,Perlmutter1999} and is now constrained by a mature, internally redundant set of probes: (i) SNe~Ia (relative luminosity distances), (ii) baryon acoustic oscillations (BAO) (standard-ruler ratios), (iii) the cosmic microwave background (CMB) (early-time calibration and the acoustic scale), and (iv) growth/lensing observables (redshift-space distortions, cosmic shear, and galaxy--galaxy lensing). In an FLRW spacetime within general relativity (GR), the background acceleration equation\footnote{\textit{Conventions.}
We use GR for general relativity, FLRW for Friedmann--Lema\^{\i}tre--Robertson--Walker, CMB for cosmic microwave background, BAO for baryon acoustic oscillations, SN/SNe~Ia for Type~Ia supernova(e), RSD for redshift-space distortions, and AP for Alcock--Paczynski. CPL denotes the Chevallier--Polarski--Linder form $w(a)=w_0+w_a(1-a)$. EFT denotes effective field theory. } is
\begin{equation}
\frac{\ddot a}{a}=-\frac{4\pi G}{3}\sum_i\left(\rho_i+3p_i\right),
\label{eq:accel}
\end{equation}
so acceleration requires a dominant component with effective equation of state $w\equiv p/\rho<-1/3$.
The concordance model $\LCDM$ realizes this with a cosmological constant $\Lambda$ ($w=-1$), but interpreting $\rho_\Lambda=\Lambda/(8\pi G)$ as vacuum energy is radiatively unstable and technically unnatural
\cite{Weinberg1989CC,Carroll2001CCReview,PeeblesRatra2003DE,Frieman2008DEReview}.

The core \emph{inference problem} is not merely whether $w(z)$ deviates from $-1$, but whether any preference for
evolving $w(z)$ is robust to (a) the BAO ruler calibration (via $\rd$), (b) SN cross-calibration and selection,
and (c) perturbation-level closure using growth and lensing. At likelihood level, BAO constrain $\DM(z)/\rd$ and $D_{\rm H}(z)/\rd$, while SNe constrain the \emph{shape} of $\DL(z)$ up to an overall absolute-magnitude calibration; consequently, late-time dark-energy parameters are entangled with $\rd$ (early-time physics) and with sub-percent photometric/selection systematics in SN compilations. This degeneracy structure makes multi-probe consistency checks essential.

A key recent input is the BAO program of the Dark Energy Spectroscopic Instrument (DESI), in particular DESI Data Release~2 (DR2) \cite{DESIDR2Key,DESIDR2LyA,DESIDR2BAOValidation}. DESI DR2 provides percent-level BAO distance ratios over $0\lesssim z\lesssim 2.5$ and, when combined with CMB data, exhibits a mild ($\simeq 2.3\sigma$) discrepancy between the BAO-preferred and CMB-preferred parameter regions in flat $\LCDM$. Allowing dynamical dark energy in $\wowaCDM$ improves the joint fit and is reported to be preferred over $\LCDM$ at dataset-dependent significance.\footnote{CPL/$\wowaCDM$  is used here as a phenomenological compression of the late-time background expansion, not as a physical theory of dark energy; physical inference requires a fully specified perturbation sector and full likelihood analysis.} In CPL and related phenomenological parameterizations, many DESI-era combined fits fall in an effective region with $w_0>-1$ and $w_a<0$ \cite{DESIDR2Key,Lodha2025ExtendedDE,ParkdeCruzPerezRatra2024NonDESI}. However, the mapping from this effective CPL quadrant to a physical dark-energy model is not one-to-one: recent analyses of physically specified quintessence/$\phi$CDM models and model-independent reconstructions show that canonical or nonphantom scalar-field dynamics can remain viable for current data in appropriately defined model spaces \cite{TadaTerada2024QuintessentialDESI,ParkRatra2025PhiCDM,WangEtAl2026QuintessenceReconstruction}. Moreover, BAO+CMB combinations can already exhibit a preference for evolving-$w$ parameterizations, while the inclusion and calibration of SNe materially changes the inferred significance, preferred direction in parameter space, and whether the anomaly localizes to very low redshift \cite{ParkdeCruzPerezRatra2024NonDESI,ShlivkoPoulin2026OmTugOfWar,Efstathiou2025Systematics,Vincenzi2025CompareSN,PopovicEtAl2025DESDovekie,Dhawan2025Axis}.

Independent DESI+SN+CMB re-analyses indicate that any preference for departures from $w=-1$ is driven predominantly by the lowest-redshift SN anchor and is confined (if present) to very low redshifts, $z\lesssim\mathcal{O}(0.1)$, raising the possibility of residual local-universe systematics (or a local-structure effect) rather than a smooth, global evolution of $w(z)$ over the full DESI lever arm \cite{Gialamas2025DESI2024BAO_local}.
Complementary low-redshift closure tests from weak lensing (WL)+ clustering (e.g.\ DES Year~6 $3\times2$\,pt) probe both geometry and growth
and therefore directly test whether a background-level anomaly is consistent with perturbations and gravitational dynamics \cite{DESY6_3x2pt_2026}.
Modern ground-based CMB likelihoods (e.g.\ ACT DR6 and SPT-3G) provide additional early-time calibration context in such combined analyses \cite{ACTDR6Power2025,SPT3GDR1_2025}.
Independent cosmic-shear programs (e.g.\ KiDS-Legacy and HSC Y3) provide further low-$z$ cross-checks on growth and geometry \cite{KiDSLegacy2025,HSCY3ClusteringZ2025}.
Finally, next-generation space surveys (e.g.\ Euclid) motivate near-term improvements in joint-likelihood frameworks and systematic-error propagation \cite{Euclid2025Overview}.

It is also important to place the DESI-era discussion in broader context. Comparable hints of dynamical or non-constant dark energy appeared already in several pre-DESI or DESI-independent analyses using combinations of SNe, BAO, Alcock--Paczynski-shape information, and other low-redshift probes \cite{CaoRatra2023LowerZExpansion,DongEtAl2023TomographicAP,VanRaamsdonkWaddell2023DecreasingDE,RubinEtAl2023UnionThroughUNITY,deCruzPerezParkRatra2024XCDM,ParkdeCruzPerezRatra2024NonDESI}. DESI DR1/DR2 did not create this issue \emph{ex nihilo}; rather, DESI sharpened it by providing a cleaner and more precise BAO lever arm with which to test whether such hints survive increasingly stringent cross-probe consistency checks.

From a theory standpoint, the phenomenology is sharply constraining: (i) canonical single-field quintessence cannot realize $w<-1$ or a phantom crossing without additional degrees of freedom or non-canonical structure, and (ii) modified gravity explanations must satisfy
stringent stability requirements and gravitational-wave propagation constraints \cite{Turyshev_FunPhys_2025}.
Therefore, the correct mapping from $(H(z),\DM(z),\fsig(z),\text{lensing})$ to model space must be done consistently at the perturbation level,
not solely through background fits.

Here we focus on late-time acceleration under the minimal assumptions of an FLRW background and (unless stated otherwise) metric gravity,
and we separate the discussion into two logically distinct layers:
(i) \emph{likelihood-level observables} (SN, BAO, CMB compressed information, and growth/lensing) together with their dominant parameter
degeneracies and calibration anchors, and (ii) \emph{phenomenological-to-physical mappings} that translate reconstructed $H(z)$, $\rho_{\rm DE}(z)$,
and $w(z)$ into microphysical models subject to perturbation stability and gravitational-wave propagation bounds. This separation is essential because DESI-era claims of evolving $w(z)$ can be driven by calibration/selection systematics at the few$\times 10^{-2}$ mag level in the SN sector and by early-time assumptions that fix $\rd$.

In addition to synthesizing the current observational status and the model landscape, we include two compact analyses that make the discussion more falsifiable at the percent level. First, from anisotropic BAO measurements we form the Alcock--Paczynski (AP) combination $F_{\rm AP}(z)=D_{\rm M}(z)H(z)/c$ \cite{AlcockPaczynski1979} in which the sound horizon $\rd$ cancels, providing a clean discriminator between (a) late-time changes in $E(z)$ and (b) early-time physics that shifts $\rd$. Second, we quantify how small redshift-dependent SN distance-modulus systematics propagate into biases in $(w_0,w_a)$ under linear response, giving an explicit error-budget target for calibration/selection modeling when interpreting joint BAO+SN+CMB preferences for evolving $w(z)$.

Beyond synthesizing post--DESI-DR2 constraints and model space, we provide three reusable,
likelihood-level ingredients:
(i) an $r_d$-independent anisotropic-BAO ``shape'' data vector based on $F_{\rm AP}(z)=D_{\rm M}(z)/D_{\rm H}(z)$ with covariance propagation from published $(D_{\rm M}/r_d,\,D_{\rm H}/r_d)$ products\footnote{Notations: $r_d$ ($r_s$) is  the drag-epoch (recombination) sound horizon; $D_{\rm H}$ is the  Hubble distance $c/H(z)$; $D_{\rm M}$ is the transverse comoving distance; $D_A$ ($D_{\rm L}$) is the angular-diameter (luminosity) distance; 
$z_{\rm eff}$ is the effective redshift of a measurement bin; 
$f\sigma_8$ is the RSD growth observable;
$S_8$ is the $\sigma_8\sqrt{\Omega_m/0.3}$;
AIC/BIC is the Akaike/Bayesian information criterion;
GW/EM denotes the gravitational-wave (GW) / electromagnetic (EM)};
(ii) a nuisance-marginalized linear-response map that converts redshift-dependent SN
distance-modulus systematics into parameter biases in $(w_0,w_a)$; and
(iii) an explicit illustrative late-time interacting-thawer construction that realizes
CPL-like trends while leaving early-time ruler physics essentially unchanged. These additions are designed to separate \emph{data-combination sensitivity} from \emph{model-space interpretation} in a way that is lightweight (analytic) but technically faithful to DESI-era systematics requirements.

This paper is organized as follows.
Section~\ref{sec:background} establishes conventions for background evolution, distances, curvature, and null tests, and makes explicit where differentiation enters $w(z)$ reconstruction.
Section~\ref{sec:likelihoods} summarizes the likelihood-level observables for SNe, BAO, CMB, and growth/lensing, emphasizing the $(H_0,\rd)$ and calibration/selection degeneracies that control joint inferences.
Section~\ref{sec:status} synthesizes contemporary constraints with emphasis on DESI DR2 and perturbation-sensitive closure tests (DES Y6 and other shear programs), and identifies which combinations drive the reported preference for evolving $w(z)$. 
Section~\ref{sec:recon} reviews parametrized reconstructions (e.g.\ the Chevallier--Polarski--Linder (CPL) $w_0$--$w_a$ form) and non-parametric reconstructions of $w(z)$ and/or $\rho_{\rm DE}(z)$, including prior sensitivity and the interpretation of phantom crossing.
Section~\ref{sec:models} maps reconstruction-level phenomenology to physically motivated models (scalar-field dark energy, coupled dark sectors, early-time modifications shifting $\rd$, and modified gravity),
including stability conditions and gravitational-wave consistency requirements.
Section~\ref{sec:stats} summarizes robust model-comparison practice (likelihood improvements, information criteria/evidence, and posterior predictive checks).
Finally, Section~\ref{sec:concl} provides 
conclusions and a quantitative outlook for discriminating scenarios with upcoming data.

\section{Background dynamics, distances, and null tests}
\label{sec:background}

\subsection{Friedmann equation and dark-energy evolution}

Assuming an FLRW spacetime with curvature $k$ and scale factor $a(t)$, define $1+z\equiv a^{-1}$ and $H\equiv \dot a/a$.
The expansion rate can be written
\begin{align}
H^2(z) &= H_0^2\Big[\Omm(1+z)^3 + \Omega_r(1+z)^4 + \Omk(1+z)^2 + \OmDE\,f_{\rm DE}(z)\Big],
\label{eq:friedmann}
\\
f_{\rm DE}(z) &\equiv \frac{\rho_{\rm DE}(z)}{\rho_{\rm DE}(0)}
= \exp\!\Big(3\int_0^z \frac{1+w(z')}{1+z'}\,\dd z'\Big).
\label{eq:fde}
\end{align}
The deceleration parameter is
\begin{equation}
q(z)\equiv -\frac{\ddot a}{aH^2}= -1 + (1+z)\frac{1}{H(z)}\frac{\dd H}{\dd z}.
\label{eq:qdef}
\end{equation}

The present dark-energy density is
\begin{equation}
\rho_{\rm DE,0}=\OmDE\,\rho_{\rm crit,0},\qquad \rho_{\rm crit,0}\equiv \frac{3H_0^2}{8\pi G}.
\end{equation}
For Planck 2018 baseline $\LCDM$ values \cite{Planck2018VI}, $\rho_{\rm DE,0}$ corresponds to an energy scale
$(\rho_{\rm DE,0})^{1/4}\sim\mathrm{meV}$.

\subsection{Distances and curvature}

Define the comoving radial distance
\begin{equation}
\chi(z)=\int_0^z \frac{c\,\dd z'}{H(z')}.
\end{equation}
The transverse comoving distance is
\begin{equation}
\DM(z)=
\begin{cases}
\frac{c}{H_0}\frac{1}{\sqrt{\Omk}}\,\sinh\!\left(\sqrt{\Omk}\,\frac{H_0}{c}\chi\right), & \Omk>0,\\
\chi, & \Omk=0,\\
\frac{c}{H_0}\frac{1}{\sqrt{|\Omk|}}\,\sin\!\left(\sqrt{|\Omk|}\,\frac{H_0}{c}\chi\right), & \Omk<0,
\end{cases}
\end{equation}
with $\DA(z)=\DM(z)/(1+z)$ and $\DL(z)=(1+z)\DM(z)$.

\subsection{The ${\rm Om}(z)$ diagnostic and notation}

A widely used null test for (spatially flat) $\LCDM$ is the Sahni--Shafieloo--Starobinsky diagnostic \cite{Sahni2008Om}
\begin{equation}
{\rm Om}(z)\equiv \frac{E^2(z)-1}{(1+z)^3-1},\qquad E(z)\equiv \frac{H(z)}{H_0}.
\label{eq:Omz}
\end{equation}
We reserve the symbol $h$ for the dimensionless Hubble constant $h\equiv H_0/(100~{\rm km\,s^{-1}\,Mpc^{-1}})$, and use $E(z)$ for the dimensionless expansion history.\footnote{For flat $\LCDM$ (neglecting radiation at low $z$), ${\rm Om}(z)\equiv\Omm$ is constant. Roman typography is deliberate: ${\rm Om}$ names a diagnostic and should not be confused with $\Omm$.}

A reconstruction identity follows from defining an effective dark-energy density
\begin{equation}
\rho_{\rm DE}(z)=3\Mpl^2\Big[H^2(z)-H_0^2\Omm(1+z)^3-H_0^2\Omk(1+z)^2-H_0^2\Omega_r(1+z)^4\Big],
\end{equation}
so that
\begin{equation}
w(z)=-1+\frac{1}{3}(1+z)\frac{\dd\ln\rho_{\rm DE}}{\dd z}.
\label{eq:wFromRho}
\end{equation}
This makes explicit that reconstructing $w(z)$ from background data requires differentiation and is therefore prior- and noise-sensitive.

\section{Observables and likelihood structure}
\label{sec:likelihoods}

\subsection{SNe Ia: Hubble diagram likelihood}

SNe constrain the luminosity distance through the distance modulus
\begin{equation}
\mu(z)=m_{\rm B} - M_{\rm B} = 5\log_{10}\!\Big(\frac{\DL(z)}{10~{\rm pc}}\Big),
\qquad
\DL(z)=(1+z)\,\DM(z).
\end{equation}
Modern analyses apply light-curve standardization and propagate a structured covariance:
\begin{equation}
m_{\rm B}^{\rm corr}=m_{\rm B}^{*}+\alpha x_1-\beta c+\Delta_M+\Delta_{\rm bias}+\cdots.
\end{equation}
Given residual vector $\Delta\bm{\mu}$ and total covariance $\bm{C}_{\rm SN}$, the Gaussian likelihood is
\begin{equation}
-2\ln\cL_{\rm SN} = \Delta\bm{\mu}^{\,T}\bm{C}_{\rm SN}^{-1}\Delta\bm{\mu} + \ln\det \bm{C}_{\rm SN} + \mathrm{const}.
\end{equation}
Because $M_{\rm B}$ is degenerate with $H_0$ in SN-only analyses, SNe primarily constrain the \emph{shape} of $\DL(z)$. Key contemporary SN resources include Pantheon+ \cite{Brout2022PantheonPlus,Scolnic2022PantheonPlusData}, Union3/UNITY \cite{RubinEtAl2023UnionThroughUNITY,Rubin2025Union3}, and the DES 5-year sample. For the DES sample specifically, it is useful to distinguish the original DES-SN5YR cosmology release from the later DES recalibration/reanalysis (``DES-Dovekie''), which updates the cross-calibration and weakens the nominal evidence for evolving dark energy relative to the original DES-SN5YR processing \cite{Abbott2024DESY5SN,PopovicEtAl2025DESDovekie}.

A useful scale translation is that a coherent offset $\Delta\mu$ in distance modulus corresponds to a fractional luminosity-distance shift
\begin{equation}
\frac{\Delta \DL}{\DL}
=
\frac{\ln 10}{5}\,\Delta\mu
\simeq 0.4605\,\Delta\mu,
\end{equation}
so that $\Delta\mu=0.02$ mag implies $\Delta \DL/\DL\simeq 0.92\%$ and $\Delta\mu=0.04$ mag implies $\simeq 1.84\%$. At sufficiently low redshift where $\DL\simeq cz/H_0$, the same offset maps to $\Delta H_0/H_0\simeq -\Delta\DL/\DL$. Therefore, few$\times 10^{-2}$ mag cross-calibration and selection residuals are \emph{directly} relevant for percent-level inferences of late-time deviations in $E(z)$ when SNe are combined with BAO and CMB.

\subsubsection{Analytic marginalization of cross-survey SN calibration/bias offsets.}
To propagate cross-survey calibration and/or bias-correction uncertainties coherently, it is useful to include
explicit additive offsets for each SN subset (survey, calibration pipeline, or bias-correction family).
Let the SN data vector be $\bm{\mu}^{\rm obs}$ and define
\begin{equation}
\bm{\mu}^{\rm obs} = \bm{\mu}^{\rm th}(\bm{\theta}) + \bm{A}\,\bm{\Delta} + \bm{\epsilon},
\label{eq:sn_offset_model}
\end{equation}
where $\bm{\theta}$ are cosmological parameters, $\bm{\Delta}$ is an $N_s$-vector of nuisance offsets,
and $\bm{A}$ is the $(N_{\rm SN}\times N_s)$ ``assignment matrix'' with $A_{is}=1$ if SN $i$ belongs to subset $s$ and $A_{is}=0$ otherwise. Assume a Gaussian prior $\bm{\Delta}\sim{\cal N}(\bm{0},\bm{\Pi})$, where $\bm{\Pi}$ encodes prior variances and correlations
(e.g.\ shared calibration modes across subsets).

Because Eq.~\eqref{eq:sn_offset_model} is linear in $\bm{\Delta}$, the nuisance parameters can be marginalized analytically:
the net effect is a low-rank covariance update,
\begin{equation}
\bm{C}_{\rm SN}^{\rm marg} = \bm{C}_{\rm SN} + \bm{A}\,\bm{\Pi}\,\bm{A}^{T},
\label{eq:sn_cov_marg}
\end{equation}
and the SN likelihood is obtained by replacing $\bm{C}_{\rm SN}\rightarrow \bm{C}_{\rm SN}^{\rm marg}$ in the Gaussian expression. Eq.~\eqref{eq:sn_cov_marg} provides a practical mechanism to test, in a controlled way, whether few$\times10^{-2}$ mag-level cross-calibration/bias differences are sufficient to account for the dataset dependence of evolving-$(w_0,w_a)$ preferences in DESI+SN+CMB combinations.

\subsubsection{Linear-response mapping from redshift-dependent SN systematics to $(w_0,w_a)$ biases.}

Because SN-only fits marginalize the absolute magnitude $M_{\rm B}$, a \emph{constant} offset in $\mu(z)$ is largely absorbed.
Cosmological parameters are biased by the \emph{redshift dependence} of residual systematics, which we denote $\delta\mu_{\rm sys}(z)$.

To first order, a small systematic shift biases parameters $p_i\in\{w_0,w_a,\ldots\}$ by
\begin{equation}
\delta p_i = \left(F^{-1}\right)_{ij}\,
\sum_{\alpha,\beta}
\frac{\partial \mu(z_\alpha)}{\partial p_j}\,
\left(C_{\rm SN}^{-1}\right)_{\alpha\beta}\,
\delta\mu_{\rm sys}(z_\beta),
\qquad
F_{ij}=
\sum_{\alpha,\beta}
\frac{\partial \mu(z_\alpha)}{\partial p_i}\,
\left(C_{\rm SN}^{-1}\right)_{\alpha\beta}\,
\frac{\partial \mu(z_\beta)}{\partial p_j},
\label{eq:sn_param_bias}
\end{equation}
where $\alpha,\beta$ index SNe or redshift bins.
This expression makes explicit that even few$\times10^{-2}$\,mag coherent residuals can bias $(w_0,w_a)$ at the level relevant for DESI-era
percent-scale distance constraints.

For intuition, Table~\ref{tab:sn_w0wa_sensitivity} reports illustrative derivatives of \emph{relative} moduli
$\Delta\mu(z)\equiv \mu(z)-\mu(z_{\rm ref})$ (so that a constant offset cancels) with $z_{\rm ref}=0.1$ approximating the low-$z$ anchor. The numbers are evaluated for a flat $\Lambda$CDM (i.e., cosmological constant + cold dark matter) fiducial ($\Omega_m=0.315$, $h=0.674$, $w_0=-1$, $w_a=0$) and are intended as order-of-magnitude calibration targets rather than a substitute for a full end-to-end SN covariance analysis.

\begin{table}[t]
\caption{\label{tab:sn_w0wa_sensitivity}
Illustrative sensitivity of SN relative moduli $\Delta\mu(z)\equiv\mu(z)-\mu(z_{\rm ref})$ to $(w_0,w_a)$ in flat $w_0w_a$CDM, evaluated at a $\Lambda$CDM fiducial with $z_{\rm ref}=0.1$. The last two columns give the single-parameter shift that reproduces a redshift-dependent offset $\Delta\mu_{\rm sys}=+0.02$\,mag between $z$ and $z_{\rm ref}$.}
\begin{tabular}{lcccc}
\hline
$z$ & $\partial \Delta\mu/\partial w_0$ & $\partial \Delta\mu/\partial w_a$ &
$\delta w_0$ (if $w_a$ fixed) & $\delta w_a$ (if $w_0$ fixed) \\
\hline\hline
0.3 & $-0.143$ & $-0.016$ & $-0.14$ & $-1.2$ \\
0.5 & $-0.230$ & $-0.036$ & $-0.087$ & $-0.56$ \\
1.0 & $-0.310$ & $-0.071$ & $-0.065$ & $-0.28$ \\
2.0 & $-0.298$ & $-0.089$ & $-0.067$ & $-0.22$ \\
\hline
\end{tabular}
\end{table}

\subsubsection{Covariance-projected parameter bias with $M_{\rm B}$ and calibration offsets marginalized.}
\label{sec:sn_bias_covproj}

Eq.~(\ref{eq:sn_param_bias}) is exact at linear order, but in practice the relevant inverse covariance is the one
\emph{after} marginalizing over SN nuisance directions that absorb calibration-like modes, including the absolute magnitude $M_{\rm B}$ and (optionally) survey/pipeline offsets $\bm{\Delta}$ as in Eq.~(\ref{eq:sn_offset_model}).

Let the SN residual vector be
$\Delta\bm{\mu}\equiv \bm{\mu}^{\rm obs}-\bm{\mu}^{\rm th}(\bm{\theta})-\bm{B}\bm{\xi}$,
where $\bm{\xi}$ collects nuisance parameters (e.g.\ $M_{\rm B}$ and subset offsets) and $\bm{B}$ is the corresponding design/assignment matrix (one column for $M_{\rm B}$, plus one column per subset if used). For Gaussian SN errors with covariance $\bm{C}_{\rm SN}$, analytic marginalization over $\bm{\xi}$ yields the
projection matrix
\begin{equation}
\bm{P}
\equiv
\bm{C}_{\rm SN}^{-1}
-
\bm{C}_{\rm SN}^{-1}\bm{B}
\big(\bm{B}^T\bm{C}_{\rm SN}^{-1}\bm{B}\big)^{-1}
\bm{B}^T\bm{C}_{\rm SN}^{-1},
\label{eq:SN_projection}
\end{equation}
which removes all modes in the column space of $\bm{B}$. The Fisher matrix for cosmological parameters $p_i$ built from SN data alone is then
\begin{equation}
F_{ij}^{\rm SN}=
\Big(\frac{\partial \bm{\mu}}{\partial p_i}\Big)^T
\bm{P}
\Big(\frac{\partial \bm{\mu}}{\partial p_j}\Big),
\label{eq:SN_Fisher_proj}
\end{equation}
and the nuisance-marginalized linear bias induced by a systematic Hubble-diagram residual $\delta\bm{\mu}_{\rm sys}$
is
\begin{equation}
\delta p_i =
\left(F^{\rm SN}\right)^{-1}_{ij}
\Big(\frac{\partial \bm{\mu}}{\partial p_j}\Big)^T
\bm{P}\,\delta\bm{\mu}_{\rm sys}.
\label{eq:SN_bias_proj}
\end{equation}
Eqs.~(\ref{eq:SN_projection})--(\ref{eq:SN_bias_proj}) provide a likelihood-faithful way to propagate realistic,
correlated SN systematics (Pantheon+/DES-SN5YR/Union3 covariances) into biases in $(w_0,w_a)$ after marginalizing over calibration-like nuisance directions.

\subsubsection{Calibration requirement as a bound on systematic eigenmodes.}
\label{sec:sn_requirement_eigenmodes}

To turn Eq.~(\ref{eq:SN_bias_proj}) into an explicit \emph{requirement}, we expand plausible redshift-dependent systematics in a basis of modes that match the structured SN covariance. A natural choice is to use eigenmodes of the \emph{systematic} covariance (or of calibration/selection sub-blocks)
provided by contemporary SN analyses:
\begin{equation}
\delta\bm{\mu}_{\rm sys} = \sum_{n=1}^{N_{\rm mode}} a_n\,\bm{e}_n,
\qquad
\bm{C}_{\rm sys}\bm{e}_n=\lambda_n\bm{e}_n,
\qquad
\bm{e}_n^T\bm{e}_m=\delta_{nm}.
\label{eq:sys_eigen}
\end{equation}
For each mode, the induced parameter shift is linear:
\begin{equation}
\delta p_i = \sum_n a_n\,b_{i}^{(n)},
\qquad
b_{i}^{(n)}\equiv
\left(F^{\rm SN}\right)^{-1}_{ij}
\Big(\frac{\partial \bm{\mu}}{\partial p_j}\Big)^T
\bm{P}\,\bm{e}_n .
\label{eq:mode_to_bias}
\end{equation}
Demanding that biases remain below a chosen tolerance,
$|\delta w_0|<\Delta w_0^{\rm tol}$ and $|\delta w_a|<\Delta w_a^{\rm tol}$,
implies per-mode amplitude bounds, e.g.
\begin{equation}
|a_n| \lesssim
\min\!\bigg(
\frac{\Delta w_0^{\rm tol}}{|b_{w_0}^{(n)}|},
\frac{\Delta w_a^{\rm tol}}{|b_{w_a}^{(n)}|}
\bigg).
\label{eq:mode_bound}
\end{equation}
This procedure addresses three methodological requirements simultaneously: (i) it uses the \emph{actual} SN covariance and redshift distribution, (ii) it marginalizes nuisance directions consistent with SN analyses, and (iii) it tests whether the dominant calibration/selection eigenmodes are aligned with those that bias $(w_0,w_a)$.

\subsubsection{From response coefficients to an explicit calibration requirement.}

\begin{figure}[t]
  \centering
    \includegraphics[width=0.60\linewidth]{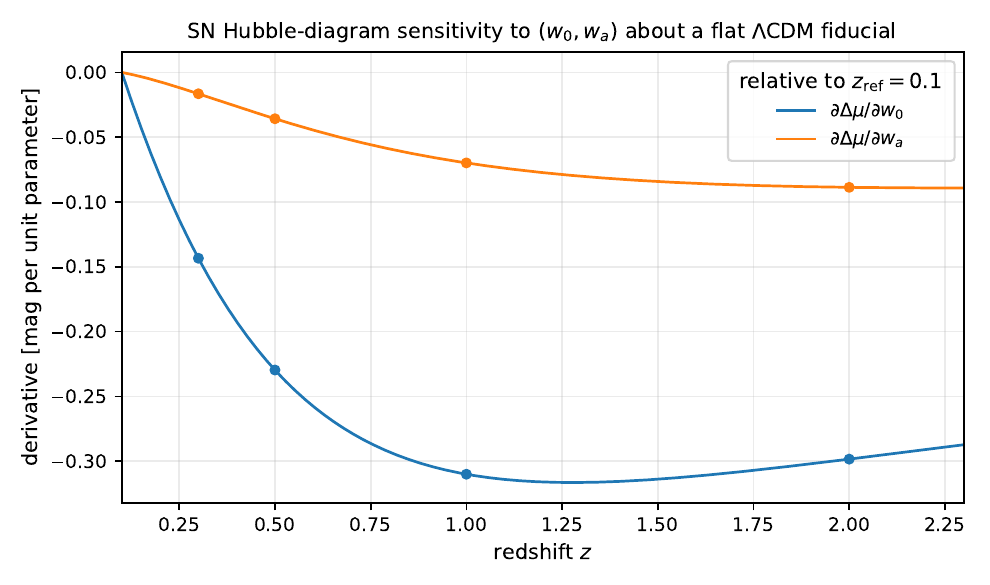}
    \vskip -10pt
  \caption{Response of relative SN distance moduli $\Delta\mu(z)\equiv \mu(z)-\mu(z_{\rm ref})$ (with $z_{\rm ref}=0.1$) to CPL parameters $(w_0,w_a)$ around a flat $\Lambda$CDM fiducial. These derivatives enter the linear-response bias mapping in Eq.~(\ref{eq:sn_param_bias}) and quantify the redshift dependence required for SN calibration/selection systematics to bias $(w_0,w_a)$. Markers indicate the redshifts used in Table~\ref{tab:sn_w0wa_sensitivity}.}
  \label{fig:SN_response}
\end{figure}

To leading order, a coherent redshift-dependent systematic in relative moduli can be projected onto
dark-energy parameters via
\[
\Delta\mu_{\rm sys}(z)\approx
\frac{\partial \Delta\mu}{\partial w_0}\,\delta w_0+
\frac{\partial \Delta\mu}{\partial w_a}\,\delta w_a,
\]
with local response coefficients illustrated in Table~\ref{tab:sn_w0wa_sensitivity}
(evaluated around a flat $\Lambda$CDM fiducial and referenced to $z_{\rm ref}=0.1$).
At $z\simeq 1$ the table gives
$\partial\Delta\mu/\partial w_0\simeq -0.31$ and $\partial\Delta\mu/\partial w_a\simeq -0.071$.
Thus, requiring parameter biases below target levels $(|\delta w_0|,|\delta w_a|)$ implies the approximate local bound
\[
|\Delta\mu_{\rm sys}(z\simeq 1)| \lesssim
\min\!\big(0.31\,|\delta w_0|,\;0.071\,|\delta w_a|\big)\,{\rm mag},
\]
with the important caveat that a full requirement should be derived from the full SN covariance and the actual
systematic mode shape $\delta\mu_{\rm sys}(z)$ (cf.\ Eq.~(\ref{eq:sn_param_bias})).

For example, using the local response coefficients in Table~\ref{tab:sn_w0wa_sensitivity} at $z\simeq 1$,
a coherent relative-modulus residual of $+0.02$ mag corresponds to a single-parameter shift
$\delta w_0\simeq -0.065$ (holding $w_a$ fixed) or $\delta w_a\simeq -0.28$ (holding $w_0$ fixed),
highlighting the need to control redshift-dependent SN systematics at the $(1$--$2)\times10^{-2}$ mag level.

\subsection{BAO: distance ratios to $\rd$}

BAO analyses are typically performed relative to a fiducial cosmology and report anisotropic dilation parameters
\begin{equation}
\alpha_\perp(z)\equiv
\frac{\DM(z)/\rd}{\DM^{\rm fid}(z)/\rd^{\rm fid}},\qquad
\alpha_\parallel(z)\equiv 
\frac{D_{\rm H}(z)/\rd}{D_{\rm H}^{\rm fid}(z)/\rd^{\rm fid}},
\qquad
D_{\rm H}(z)\equiv \frac{c}{H(z)}.
\label{eq:aaD}
\end{equation}
Some analyses also quote an isotropic combination using the volume-averaged distance
\begin{equation}
D_{\rm V}(z)\equiv \left[z\,\DM^2(z)\,D_{\rm H}(z)\right]^{1/3},
\qquad
\frac{D_{\rm V}(z)}{\rd}\ \ \ \text{(isotropic BAO)}.
\end{equation}
These definitions make explicit that BAO constrain a \emph{dimensionless} distance ladder relative to $\rd$, with early-time physics entering through $\rd$.

It is useful to factor out the overall $H_0$ scaling by writing $H(z)=H_0 E(z)$ with dimensionless $E(z)$. Then
\begin{equation}
\frac{\DM(z)}{\rd}
=
\frac{c}{H_0\,\rd}\,
\cS_k\!\Big(\int_0^z \frac{\dd z'}{E(z')}\Big),
\qquad
\frac{D_{\rm H}(z)}{\rd}=\frac{c}{H_0\,\rd}\frac{1}{E(z)},
\end{equation}
where $\cS_k(x)=x$ for $\Omk=0$ and is the usual $\sin/\sinh$ curvature mapping otherwise. Thus, \emph{BAO-only} constraints largely determine the product $H_0\,\rd$ (and the shape of $E(z)$), while disentangling $H_0$ from $\rd$ requires early-time calibration (CMB) and/or additional absolute-scale information.

The early-time calibration enters through the drag-epoch sound horizon,
\begin{equation}
\rd=\int_{z_d}^{\infty}\frac{c_s(z)}{H(z)}\,\dd z,\qquad
c_s(z)=\frac{c}{\sqrt{3\big(1+\Rb(z)\big)}},\qquad \Rb(z)\equiv \frac{3\rho_b}{4\rho_\gamma}.
\end{equation}
Thus BAO are geometric but inherit sensitivity to early-time physics through $\rd$.
DESI DR2 reports BAO from $>14$ million galaxies and quasars and provides dedicated validation \cite{DESIDR2Key,DESIDR2BAOValidation}.
The DR2 Ly$\alpha$ analysis reports at $\zeff=2.33$ \cite{DESIDR2LyA}
\begin{equation}
\frac{D_{\rm H}(\zeff)}{\rd}= 8.632 \pm 0.098 \pm 0.026,\qquad
\frac{\DM(\zeff)}{\rd}= 38.99 \pm 0.52 \pm 0.12,
\end{equation}
with statistical and theoretical-systematic components quoted separately.

\subsubsection{$\rd$-independent anisotropy: the Alcock--Paczynski parameter.}

From anisotropic BAO one can form the dimensionless Alcock--Paczynski (AP) combination \cite{AlcockPaczynski1979,Ballinger1996}
\begin{equation}
F_{\rm AP}(z)\equiv \frac{(1+z)D_{\rm A}(z)H(z)}{c}
= \frac{D_{\rm M}(z)}{D_{\rm H}(z)}
= \frac{D_{\rm M}(z)/\rd}{D_{\rm H}(z)/\rd},
\label{eq:FAP_def}
\end{equation}
in which both $H_0$ and the early-time ruler $\rd$ cancel. Consequently, $F_{\rm AP}(z)$ primarily probes the \emph{shape} of the late-time expansion history $E(z)$ (and curvature, if allowed), providing a useful discriminator between genuine late-time dynamics and early-time modifications that shift $\rd$.

As $F_{\rm AP}(z)=D_{\rm M}(z)/D_{\rm H}(z)$ is independent of both $H_0$ and $r_d$, it isolates the
\emph{late-time} expansion-shape information contained in anisotropic BAO. Any modification that primarily rescales the BAO ruler $r_d$ (through early-time physics) shifts both $D_{\rm M}/r_d$ and $D_{\rm H}/r_d$ in the same direction, while leaving their ratio approximately unchanged. Therefore, statistically significant deviations in $F_{\rm AP}(z)$ point to changes in the late-time expansion history (and/or curvature if allowed), strengthening the interpretation in terms of late-time dynamics rather than a pure ruler rescaling.

\begin{figure}[t]
  \centering
  \includegraphics[width=0.60\linewidth]{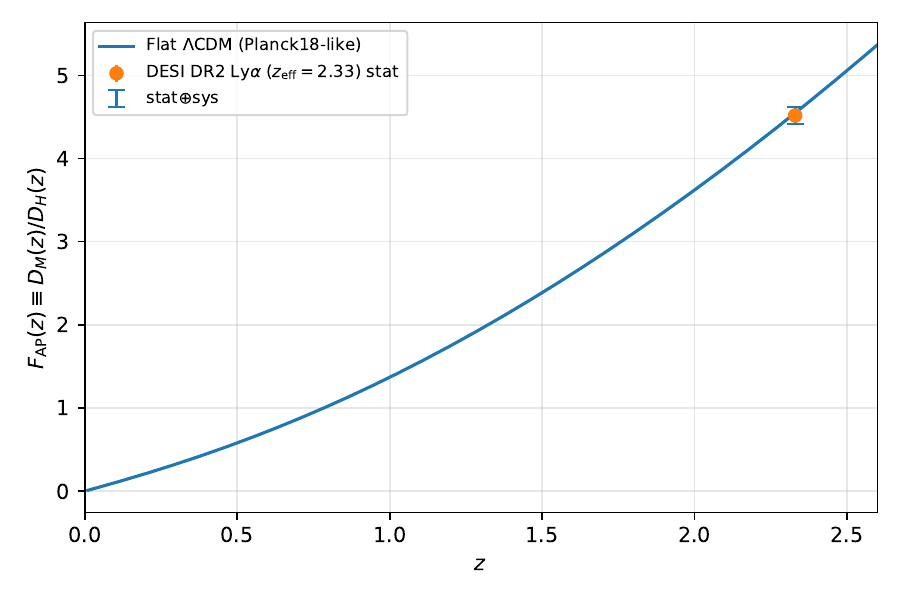}
  \vspace{-1.2ex}
\caption{$r_d$-independent anisotropic-BAO ``shape'' observable
  $F_{\rm AP}(z)\equiv D_{\rm M}(z)/D_{\rm H}(z)$.
The DESI DR2 Ly$\alpha$ point at $z_{\rm eff}=2.33$ is computed directly from the published
anisotropic ratios $(D_{\rm M}/r_d,\,D_{\rm H}/r_d)$   values (with covariance propagated as in Eq.~(\ref{eq:FAP_fracvar}) and statistical+theory-systematic terms added in quadrature for the plotted error bar), giving $F_{\rm AP}(2.33)=4.518\pm0.095_{\rm stat}\pm0.019_{\rm sys}$
(Table~\ref{tab:FAPvector}).  The curve shows the flat $\Lambda$CDM prediction for a Planck-like $\Omega_m$ (background shape only; independent of $H_0$ and $r_d$).}
  \label{fig:FAP_DR2}
\end{figure}

Using the DR2 Ly$\alpha$ distance ratios at $\zeff=2.33$ \cite{DESIDR2LyA},  statistical errors
$\sigma(D_{\rm H}/r_d)=0.098$ and $\sigma(D_{\rm M}/r_d)=0.52$ and the published statistical correlation
$\rho_{\rm MH}=-0.457$,
we obtain
\begin{equation}
F_{\rm AP}(\zeff)
=\frac{D_{\rm M}(\zeff)}{D_{\rm H}(\zeff)}
=\frac{(D_{\rm M}/r_d)}{(D_{\rm H}/r_d)}
=\frac{38.99}{8.632}
=4.518,
\qquad
\sigma_{\rm stat}\!\left[F_{\rm AP}(\zeff)\right]=0.095,
\qquad
\sigma_{\rm sys}\!\left[F_{\rm AP}(\zeff)\right]\simeq 0.019,
\label{eq:FAP_LyA_covcorr}
\end{equation}
If the anisotropic-BAO uncertainties were uncorrelated, independent-error propagation would give
$ \sigma_{\rm stat}^{(\rho=0)}\!\left[F_{\rm AP}(\zeff)\right]\simeq 0.079.$
However, the published statistical correlation $\rho_{\rm MH}=-0.457$ increases the uncertainty.
Using Eq.~(\ref{eq:FAP_fracvar}) yields
$ \sigma_{\rm stat}\!\left[F_{\rm AP}(\zeff)\right]\simeq 0.095.$ Propagating the quoted theoretical-systematic components in quadrature gives
$\sigma_{\rm sys}\!\left[F_{\rm AP}(\zeff)\right]\simeq 0.019,$ noting that a likelihood-level treatment should propagate the full $(D_{\rm M}/r_d,\,D_{\rm H}/r_d)$ covariance. A likelihood-level evaluation should use the full $(D_{\rm M}/r_d,\,D_{\rm H}/r_d)$ covariance; Eq.~(\ref{eq:FAP_def}) is intended as a compact, $r_d$-independent summary statistic.
Figure~\ref{fig:FAP_DR2} shows the DR2 Ly$\alpha$ anchor as an illustrative example;
the same construction applies bin-by-bin whenever $(D_{\rm M}/r_d,\;D_{\rm H}/r_d)$ and their covariance
are published (Appendix~\ref{app:FAPvector}).

DESI provides $f_{\rm AP}$ (or equivalent) for specific cases, but the consistently defined,
$r_d$-independent \emph{shape} data vector can be constructed for \emph{any} anisotropic BAO distance product whenever $(D_{\rm M}/r_d,\,D_{\rm H}/r_d)$ and their covariance are reported. Our contribution is to (i) provide a covariance-preserving mapping from published anisotropic-BAO distance products to an $r_d$-independent ``shape'' statistic $F_{\rm AP}(z)$ and its covariance, and (ii) illustrate the construction explicitly for the DR2 Ly$\alpha$ bin at $z_{\rm eff}=2.33$ (Appendix~\ref{app:FAPvector}). Extending the same construction to the full DR2 redshift-bin set
is straightforward whenever the full anisotropic-BAO covariance matrices are available.

\subsubsection{Baseline $\Lambda$CDM check.}

Because $F_{\rm AP}(z)=D_{\rm M}(z)/D_{\rm H}(z)$ is independent of both $H_0$ and the early-time ruler $r_d$,
it provides a particularly clean consistency check of the {\em late-time} shape $E(z)$. For a spatially flat model ($\Omega_k=0$) one may write the identity $F_{\rm AP}(z)=E(z)\int_0^z \dd z'/E(z')$, showing explicitly that $F_{\rm AP}$ depends only on the dimensionless expansion history.

As a concrete benchmark, adopting the Planck 2018 baseline flat $\Lambda$CDM parameters
(e.g.\ $\Omega_m\simeq0.315$, $h\simeq0.674$), one finds
\[
F_{\rm AP}^{\Lambda{\rm CDM}}(z=2.33)\simeq 4.55.
\]
Comparing with the DR2 Ly$\alpha$ estimate above,
$F_{\rm AP}^{\rm obs}(2.33)=4.518\pm0.095_{\rm stat}\pm0.019_{\rm sys}$,
the difference is
\[
\Delta F_{\rm AP}\equiv F_{\rm AP}^{\rm obs}-F_{\rm AP}^{\Lambda{\rm CDM}}
\simeq -0.03 \pm 0.10,
\]
consistent at $\lesssim 0.3\sigma$ (stat$\oplus$sys).
This illustrates both the utility and the present limitation of $F_{\rm AP}$: it is an $r_d$-independent discriminator of late-time dynamics versus early-time shifts in $r_d$,
but a single high-$z$ point is not yet strongly discriminating among smooth late-time models.

\subsubsection{Tomographic $\Omega_{m0}$ from $F_{\rm AP}(z)$ in flat $\Lambda$CDM}
\label{sec:Om_from_FAP}

A complementary use of the $r_d$-independent AP ratio is that, \emph{within spatially flat $\Lambda$CDM},
$F_{\rm AP}(z)$ provides a one-parameter constraint on the present matter density $\Omega_{m0}$.
Neglecting radiation at $z\lesssim 3$,
\begin{equation}
E(z)=\sqrt{\Omega_{m0}(1+z)^3+\left(1-\Omega_{m0}\right)},
\qquad
F_{\rm AP}^{\Lambda{\rm CDM}}(z;\Omega_{m0})
=E(z)\int_0^z \frac{\dd z'}{E(z')}.
\label{eq:FAP_LCDM}
\end{equation}
For fixed $z$, $F_{\rm AP}^{\Lambda{\rm CDM}}(z;\Omega_{m0})$ is monotonic in $\Omega_{m0}$ over the
cosmologically relevant range, so each measurement of $F_{\rm AP}(z_i)$ can be inverted to define
a tomographic estimator $\widehat{\Omega}_{m0}^{{\rm(AP)}}(z_i)$ via
\begin{equation}
F_{\rm AP}^{\Lambda{\rm CDM}}(z_i;\widehat{\Omega}_{m0}^{{\rm(AP)}})=F_{\rm AP}^{\rm obs}(z_i).
\label{eq:Om_from_FAP_def}
\end{equation}
In $\Lambda$CDM one should recover a redshift-independent $\widehat{\Omega}_{m0}^{{\rm(AP)}}(z)$ across bins; a drift with $z$ provides a null test that does not introduce the additional parameter degeneracies inherent to CPL fits. This one-parameter ``AP inversion'' tomographic null test has been advocated and applied in
\cite{OColgain:2024DESIconfirmLCDM,OColgain:2025HowMuchEvolvedDR1}. Table~\ref{tab:Om_from_FAP_example} 
illustrates inversion of the $r_d$-independent AP ratio into an effective flat-$\Lambda$CDM matter density from Eq.~(\ref{eq:Om_from_FAP_def}). For internal consistency, Table~\ref{tab:Om_from_FAP_example} uses the covariance-aware propagated uncertainty $\sigma_{\rm stat}[F_{\rm AP}(2.33)]\simeq 0.095$ from Eq.~(\ref{eq:FAP_fracvar}), rather than the uncorrelated estimate $\sigma_{\rm stat}^{(\rho=0)}[F_{\rm AP}(2.33)]\simeq 0.079$.

\begin{table}[t]
\caption{\label{tab:Om_from_FAP_example}
Illustrative inversion of the $r_d$-independent AP ratio into an effective flat-$\Lambda$CDM matter density,
$\widehat{\Omega}_{m0}^{\,{\rm(AP)}}$, defined by Eq.~(\ref{eq:Om_from_FAP_def}). Values shown use the Ly$\alpha$ DR2 bin and are intended as a compact consistency check (a decisive test requires the full multi-bin $F_{\rm AP}(z)$ vector and its covariance).}
\begin{tabular}{cccc}
\hline
$z_{\rm eff}$ & $F_{\rm AP}(z)$ & $\widehat{\Omega}_{m0}^{\,{\rm(AP)}}$ & Assumptions \\ \hline\hline
2.33 & ~~$4.518\pm0.095~({\rm stat})\pm0.019~({\rm sys})$ &
~~$0.304^{+0.037}_{-0.033}~({\rm stat})\pm0.007~({\rm sys})$ &
~~ flat $\Lambda$CDM; radiation neglected \\
\hline
\end{tabular}
\end{table}

For a single redshift bin with Gaussian uncertainty $\sigma_{F_{\rm AP}}$,
one may define the 1D likelihood
\begin{equation}
-2\ln\mathcal{L}(\Omega_{m0})
\frac{\left[F_{\rm AP}^{\rm obs}(z)-F_{\rm AP}^{\Lambda{\rm CDM}}(z;\Omega_{m0})\right]^2}{\sigma_{F_{\rm AP}}^2},
\label{eq:FAP_Om_like_1d}
\end{equation}
or, for a multi-bin vector $\bm{y}={\ln F_{\rm AP}(z_i)}$ with covariance $\bm{C}y$ [Eqs.~(\ref{eq:ylnFAP})--(\ref{eq:Cy_from_Cd}) below],
\begin{equation}
-2\ln\mathcal{L}(\Omega_{m0})
\left(\bm{y}-\bm{y}_{\rm th}(\Omega_{m0})\right)^T
\bm{C}y^{-1}
\left(\bm{y}-\bm{y}_{\rm th}(\Omega_{m0})\right),
\label{eq:FAP_Om_like_vec}
\end{equation}
which yields an $r_d$- and $H_0$-independent constraint on $\Omega_{m0}$ under the flat-$\Lambda$CDM assumption.

\subsubsection{Covariance-aware propagation for $F_{\rm AP}$.}

The uncertainty on $F_{\rm AP}(z)=D_{\rm M}(z)/D_{\rm H}(z)$ should be propagated using the published
anisotropic-BAO covariance between $(D_{\rm M}/r_d)$ and $(D_{\rm H}/r_d)$. A convenient expression in terms of reported fractional uncertainties and correlation
coefficient is Eq.~(\ref{eq:FAP_fracvar}). Define
\begin{equation}
x \equiv \ln\!\Big(\frac{D_{\rm M}}{r_d}\Big), \qquad
y \equiv \ln\!\Big(\frac{D_{\rm H}}{r_d}\Big),
\qquad\Rightarrow\qquad
\ln F_{\rm AP} = x-y ,
\label{eq:FAP_logdefs}
\end{equation}
so that, for a single redshift bin,
\begin{equation}
{\rm Var}\!\left(\ln F_{\rm AP}\right)
=
{\rm Var}(x)+{\rm Var}(y)-2\,{\rm Cov}(x,y).
\label{eq:FAP_logvar}
\end{equation}
Equivalently, in terms of fractional uncertainties on the reported anisotropic BAO observables,
\begin{equation}
\Big(\frac{\sigma(F_{\rm AP})}{F_{\rm AP}}\Big)^2
\simeq
\Big(\frac{\sigma(D_{\rm M}/r_d)}{D_{\rm M}/r_d}\Big)^2
+
\Big(\frac{\sigma(D_{\rm H}/r_d)}{D_{\rm H}/r_d}\Big)^2
-2\,\rho_{\rm MH}\,
\Big(\frac{\sigma(D_{\rm M}/r_d)}{D_{\rm M}/r_d}\Big)
\Big(\frac{\sigma(D_{\rm H}/r_d)}{D_{\rm H}/r_d}\Big),
\label{eq:FAP_fracvar}
\end{equation}
where $\rho_{\rm MH}$ is the correlation coefficient between $(D_{\rm M}/r_d)$ and $(D_{\rm H}/r_d)$ in that bin
(read directly from the published anisotropic BAO covariance matrix).

\subsubsection{An $r_d$-independent BAO likelihood block from anisotropic BAO}

Because $F_{\rm AP}(z)=D_{\rm M}(z)/D_{\rm H}(z)$ is independent of both $H_0$ and $r_d$, anisotropic BAO measurements can be split into (i) a ``shape'' block that is $r_d$-free and (ii) a ``scale'' block that retains the $(H_0 r_d)^{-1}$ dependence.

Let the anisotropic BAO data vector in $N$ redshift bins be
\begin{equation}
\bm{d}\equiv
\bigg(
\ln\!\frac{D_{\rm M}(z_1)}{r_d},\ldots,\ln\!\frac{D_{\rm M}(z_N)}{r_d},
\ln\!\frac{D_{\rm H}(z_1)}{r_d},\ldots,\ln\!\frac{D_{\rm H}(z_N)}{r_d}
\bigg)^T,
\end{equation}
with covariance matrix $\bm{C}_d$ (as provided by the BAO analysis).
Define the derived $r_d$-independent vector
\begin{equation}
\label{eq:ylnFAP}
\bm{y}\equiv
\left(
\ln F_{\rm AP}(z_1),\ldots,\ln F_{\rm AP}(z_N)
\right)^T,
\qquad
\ln F_{\rm AP}(z_i)=\ln\!\frac{D_{\rm M}(z_i)}{r_d}-\ln\!\frac{D_{\rm H}(z_i)}{r_d}.
\end{equation}
In matrix form, $\bm{y}=\bm{A}\bm{d}$ with $\bm{A}=(\bm{I}_N,\,-\bm{I}_N)$, so that the covariance is
\begin{equation}
\bm{C}_y=\bm{A}\bm{C}_d\bm{A}^T.
\label{eq:Cy_from_Cd}
\end{equation}
A Gaussian $r_d$-independent BAO likelihood block is then
\begin{equation}
-2\ln\mathcal{L}_{\rm AP}
=
\left(\bm{y}-\bm{y}_{\rm th}\right)^T
\bm{C}_y^{-1}
\left(\bm{y}-\bm{y}_{\rm th}\right),
\label{eq:Lap}
\end{equation}
computed from the cosmological model prediction $F_{\rm AP}(z)$. This block isolates late-time expansion-shape information and is therefore particularly useful for
disentangling genuine late-time dynamics from early-time physics that shifts $r_d$.

\subsection{CMB: early-time calibration and compressed parameters}

The CMB tightly constrains the acoustic angular scale
\begin{equation}
\theta_* \equiv \frac{\rs(z_*)}{\DM(z_*)},
\end{equation}
and combinations close to the CMB ``shift parameter''
$\Rshift \equiv \sqrt{\Omega_m H_0^2}\,D_{\rm M}(z_*)/c$. For late-time dark-energy inference, the CMB primarily (i) calibrates $\rs$ and $\rd$ and (ii) constrains $\Omm h^2$ and $\Omega_b h^2$. Planck 2018 remains the baseline reference \cite{Planck2018VI}, with recent ground-based CMB datasets used in joint low-$z$ comparisons (see \cite{DESY6_3x2pt_2026} and references therein).

Recent high-resolution ground-based CMB measurements provide complementary late-time leverage through improved small-scale temperature/polarization, foreground separation, and CMB lensing reconstruction. In particular, ACT DR6 and SPT-3G DR1 deliver independent constraints that are routinely combined with Planck in contemporary joint analyses, including the DES Y6 combined-CMB baseline used for low-$z$ comparisons \cite{ACTDR6Power2025,SPT3GDR1_2025,DESY6_3x2pt_2026}. For evolving dark-energy and modified-gravity tests, the primary role remains early-time calibration (fixing $\rs$ and $\rd$), but consistency across CMB datasets is increasingly important when interpreting percent-level shifts in late-time parameters.

For clarity, the compressed-CMB discussion in this review is intended mainly as a likelihood-level summary for near-$\Lambda$CDM late-time phenomenology. Compressed distance priors and shift-parameter combinations are not guaranteed to remain accurate in arbitrary physical scalar-field, interacting-dark-sector, or modified-gravity models, especially when perturbations, early-time physics, or lensing are modified. Accordingly, recent physical-model studies that test dynamical dark energy and lensing-systematics dependence have typically relied on full CMB likelihoods rather than only on compressed late-time summaries \cite{ParkdeCruzPerezRatra2024AL,RoyChoudhuryOkumura2024ExtendedParams,RoyChoudhury2025DESIDR2Extended,RoyChoudhuryOkumuraUmetsu2025NPDDE}. Precision constraints on such physical models should therefore use the full CMB likelihood together with a full Einstein--Boltzmann treatment.

A further caution is the long-discussed CMB lensing-smoothing anomaly. In extended fits to Planck data, the phenomenological lensing consistency parameter $A_L$ is often preferred above unity at roughly the $\sim 2\sigma$ level, and several recent analyses show that allowing $A_L$ to vary can weaken the apparent significance of dynamical dark energy in $\wowaCDM$ and related parameterizations \cite{ParkdeCruzPerezRatra2024AL,ParkRatra2025ExcessSmoothing,RoyChoudhuryOkumura2024ExtendedParams,RoyChoudhury2025DESIDR2Extended}. Conversely, in some physically restricted nonphantom/quintessence-like analyses, accommodating the data can require both viable dynamical dark energy and a nontrivial lensing-anomaly budget \cite{ParkRatra2025PhiCDM,RoyChoudhuryOkumuraUmetsu2025NPDDE}. We therefore treat the $A_L$ issue as part of the CMB-side robustness budget when interpreting DESI-era evidence for evolving dark energy.

\subsection{Growth, lensing, and modified gravity}

Linear matter perturbations evolve (sub-horizon) as
\begin{equation}
\delta_m''(a) + \Big[\frac{3}{a} + \frac{\dd \ln H}{\dd a}\Big]\delta_m'(a)
- \frac{3}{2}\frac{\Omm H_0^2}{a^5 H^2(a)}\,\muMG(a,k)\,\delta_m(a)=0,
\end{equation}
where primes are $\dd/\dd a$ and $\muMG(a,k)=1$ in GR with minimally coupled matter.

In redshift space, linear peculiar velocities introduce anisotropy in the observed galaxy power spectrum.
At leading order (Kaiser limit), one may write
\begin{equation}
P_s(k,\mu,z)=\Big(b_1(z)+f(z)\mu^2\Big)^2 P_m(k,z),
\label{eq:kaiser}
\end{equation}
where $\mu\equiv \hat{\bm{k}}\!\cdot\!\hat{\bm{n}}$ is the cosine to the line of sight, $b_1$ is the linear bias, and $f\equiv \dd\ln D/\dd\ln a$.
The multipoles measured in practice are
\begin{equation}
P_\ell(k,z)=\frac{2\ell+1}{2}\int_{-1}^{1}\dd\mu\ P_s(k,\mu,z)\,{\cal L}_\ell(\mu),
\end{equation}
with ${\cal L}_\ell$ Legendre polynomials.
Beyond linear order, realistic full-shape analyses incorporate nonlinear matter clustering, scale-dependent bias, and velocity-dispersion (Finger-of-God; FoG) effects, and are typically validated on simulation-based mocks. These observables break degeneracies that are invisible to background-only distance probes and provide a direct closure test for clustering dark energy and modified gravity \cite{Kaiser1987,Hamilton1998}.

Weak-lensing cosmic shear and galaxy--galaxy lensing are commonly analyzed via tomographic two-point functions.
In harmonic space (Limber approximation with $\ell\rightarrow \ell+1/2$),
the shear power spectrum for source bins $(i,j)$ is
\begin{equation}
C_\ell^{\gamma\gamma,ij}
\simeq
\int_0^{\chi_H}\frac{\dd\chi}{\chi^2}\,
W_i(\chi)\,W_j(\chi)\,
P_{\delta\delta}\!\Big(k=\frac{\ell+1/2}{\chi},z(\chi)\Big),
\label{eq:cl_shear}
\end{equation}
where $W_i(\chi)$ is the lensing efficiency kernel.
The widely quoted late-time clustering combination is
\begin{equation}
S_8 \equiv \sigma_8\sqrt{\frac{\Omm}{0.3}},
\end{equation}
which partially captures the $\sigma_8$--$\Omm$ degeneracy direction in shear-only constraints.
A full 3$\times$2pt analysis combines shear--shear, galaxy--galaxy lensing, and galaxy clustering, enabling internal consistency tests for bias and systematics and strengthening discrimination among background-only extensions and perturbation-level new physics \cite{Limber1953,LoVerdeAfshordi2008,DESY6_3x2pt_2026}.

Redshift-space distortions (RSD) constrain $\fsig(z)$; weak lensing probes combinations of metric potentials. Phenomenological modified gravity descriptions often use $\muMG(a,k)$  and the slip parameter $\eta(a,k)\equiv\Phi/\Psi$.
Effective field theory (EFT) and review treatments are given in \cite{Gubitosi2013EFT,BelliniSawicki2014Alpha,Clifton2012MG,Joyce2015MGReview}. Multi-messenger bounds from GW170817 constrain gravitational-wave propagation in the late Universe \cite{Abbott2017GW170817,Baker2017GWConstraints}.

\paragraph{Growth index as a one-parameter closure test.}
A compact phenomenological summary of late-time growth is the growth index $\gamma$, defined by
\begin{equation}
f(z)\equiv \frac{\dd\ln D}{\dd\ln a} \simeq \Omega_m(z)^{\gamma},
\qquad
\Omega_m(z)=\frac{\Omm(1+z)^3}{E^2(z)}.
\label{eq:gamma_def}
\end{equation}
For GR+$\LCDM$ with smooth dark energy and scale-independent linear growth at late times,  one expects $\gamma\simeq 0.55$ (with weak model dependence) \cite{WangSteinhardt1998,Linder2005Growth}. Joint fits that prefer evolving $w(z)$ but remain consistent with GR should yield $\gamma$ consistent with this expectation,
whereas many modified-gravity scenarios map into an effective $\Delta\gamma\neq 0$.

\paragraph{An $E_{\rm G}$-type statistic linking lensing to RSD.}
Combinations of galaxy--galaxy lensing (sensitive to $\Phi+\Psi$) and redshift-space distortions (sensitive to $f$) enable bias-reduced tests of gravity on linear scales.
A commonly used observable is an $E_{\rm G}$-type combination, schematically of the form
\begin{equation}
E_{\rm G} \ \propto\ \frac{C_\ell^{\kappa g}}{\beta\,C_\ell^{gg}},
\qquad
\beta(z)\equiv \frac{f(z)}{b_1(z)},
\label{eq:EG_obs}
\end{equation}
where $C_\ell^{\kappa g}$ is a convergence--galaxy cross-spectrum and $C_\ell^{gg}$ is a galaxy auto-spectrum. In GR with negligible anisotropic stress and on sufficiently linear scales, $E_{\rm G}$ has a nearly scale-independent prediction, while departures in $\mu(a,k)$ and/or $\eta(a,k)$ generically induce scale/redshift dependence.

\subsection{Standard sirens: absolute distances independent of $\rd$ and SN calibration}
\label{sec:standardsirens}

Compact-binary gravitational-wave (GW) signals provide a direct luminosity distance $\DL^{\rm GW}$ from the waveform amplitude (a ``standard siren''), independent of the SN distance ladder and independent of the early-time ruler $\rd$ \cite{Schutz1986}. With an electromagnetic counterpart (or statistical host association), one obtains a redshift and can constrain $H_0$ and, in principle, late-time expans ion. The first joint GW+electromagnetic (EM) standard-siren $H_0$ measurement was demonstrated with GW170817 \cite{Abbott2017H0Siren}. In the longer term, a population of standard sirens can provide a powerful cross-check on whether apparent DESI+SN+CMB preferences for evolving $w(z)$ are driven by distance-ladder systematics or reflect genuine late-time dynamics.

\section{Contemporary constraints: DESI DR2 and complementary probes}
\label{sec:status}

Table~\ref{tab:datasets_summary} provides a compact dataset/constraint summary. For clarity, the datasets used in the present discussion do not all play the same inferential role. We treat
Pantheon+ and Union/UNITY-style SN compilations as primary or robustness-level distance-ladder anchors because they explicitly propagate selection and calibration structure through the likelihood, while the original DES-SN5YR release is retained mainly as a \emph{historical comparison} because part of the recent literature debate concerns the impact of its calibration pipeline relative to the later DES recalibration/reanalysis (DES-Dovekie). Likewise, DES Y6 $3\times2$pt is treated here as a primary perturbation-sensitive closure test, whereas KiDS-Legacy and HSC Y3 provide independent weak-lensing cross-checks, and peculiar-velocity $f\sigma_8$ measurements together with cosmic-chronometer $H(z)$ determinations supply additional low-$z$ growth and geometry consistency information.

\begin{table*}[t]
\caption{\label{tab:datasets_summary}
Compact dataset/constraint summary used in DESI-era dark-energy inference. Section~A lists distance-ladder and early-time calibration anchors; Section~B lists perturbation-sensitive closure tests (growth, lensing, and cross-calibration systematics). Here $D_{\rm H}(z)\equiv c/H(z)$, $D_{\rm M}(z)\equiv(1+z)D_{\rm A}(z)$, and $\rd$ is the drag-epoch sound horizon. We define $S_8\equiv\sigma_8\sqrt{\Omega_m/0.3}$.
Quoted values are as reported by the cited analyses for their stated model/dataset definitions (details and comparisons are likelihood-dependent).}
\setlength{\tabcolsep}{4pt}
\renewcommand{\arraystretch}{1.12}
\small
\begin{tabular}{lll}
\hline
Probe / dataset & Key measurement(s) & Selected quantitative statements \\
\hline\hline

\multicolumn{3}{@{}l@{}}{\textit{A. Distance-ladder and early-time calibration anchors}}\\
\hline
Planck 2018 ($\Lambda$CDM) &
CMB anisotropies &
Baseline $\Lambda$CDM parameter constraints (e.g.\ $H_0\simeq 67.4\pm$
\\

& &
$0.5~{\rm km\,s^{-1}\,Mpc^{-1}}$,
$\Omega_m\simeq 0.315\pm0.007$) \cite{Planck2018VI}
\\

DESI DR2 BAO (key) &
BAO distances, $0\lesssim z\lesssim 2.5$ &
Mild $\simeq 2.3\sigma$ discrepancy with CMB-preferred parameters in 
\\
&& flat $\Lambda$CDM; in $w_0w_a$CDM,
preferred over $\Lambda$CDM at $3.1\sigma$ for 
\\
&& BAO+CMB and $2.8$--$4.2\sigma$ with SNe (SN-sample dependent) \cite{DESIDR2Key}
\\

DESI DR2 BAO &
Pipeline/systematics validation &
Dedicated validation suite for galaxy+QSO BAO measurements
\\
(validation) & &and nuisance modeling \cite{DESIDR2BAOValidation}
\\

DESI DR2 Ly$\alpha$ BAO &
$\DM(\zeff)/r_d,\ D_{\rm H}(\zeff)/r_d$ at  &
$D_{\rm H}/r_d = 8.632\pm0.098\pm0.026$,
$\DM/r_d = 38.99\pm0.52\pm0.12$; 
\\
& $\zeff=2.33$ & sub-percent isotropic precision \cite{DESIDR2LyA}
\\

Pantheon+ (SNe) &
SN Hubble diagram to $z\simeq 2.26$ &
SN-only: $\Omega_m=0.334\pm0.018$ (flat $\Lambda$CDM);
SN-only: $w_0=$
\\
&&
$-0.90\pm0.14$ (flat $w$CDM [constant-$w$ dark energy extension]) \cite{Brout2022PantheonPlus}
\\

DES-SN5YR (SNe) &
1635 DES SNe + low-$z$ anchor &
SN-only: $\Omega_m=0.352\pm0.017$ (flat $\Lambda$CDM);
extended-model fits 
\\
 &&
 exhibit strong $(\Omega_m,w)$ and $(w_0,w_a)$ degeneracies \cite{Abbott2024DESY5SN}
\\

Union3/UNITY (SNe) &
Hierarchical SN population+ &
Public distances and joint inference framework; emphasizes end
\\
&
selection model &
to-end selection and calibration propagation \cite{Rubin2025Union3}
\\

\hline
\multicolumn{3}{@{}l@{}}{\textit{B. Perturbation-sensitive closure tests (growth, lensing, systematics)}}\\
\hline
DES Y6 (3$\times$2pt) &
Shear+clustering+galaxy-- &
In $\Lambda$CDM: $S_8=0.789\pm0.012$,
$\Omega_m=0.333^{+0.023}_{-0.028}$; in $w$CDM: \\
&
galaxy lensing &
$w=-1.12^{+0.26}_{-0.20}$.
Also reports combined-probe fits including DESI 
\\
& & DR2 BAO and modern CMB \cite{DESY6_3x2pt_2026}
\\

KiDS-Legacy (cosmic &
Shear 2-pt, tomographic &
Reports $S_8=0.815^{+0.016}_{-0.021}$ (cosmic shear only; definition $S_8\equiv $
\\

shear) &&
$ \sigma_8\sqrt{\Omega_m/0.3}$) \cite{KiDSLegacy2025}
\\

HSC Y3 cosmic shear  &
Shear 2-pt with clustering-$z$  &
Reanalysis with DESI-based clustering-$z$ calibration reports $S_8=$
\\
(updated calibration) &calibration &
$0.805\pm0.018$ and discusses multi-survey combinations \cite{HSCY3ClusteringZ2025}
\\

SN systematics  &
Cross-calibration and selection &
Evidence for evolving DE depends on SN compilation and bias
\\

sensitivity & & corrections; few$\times10^{-2}$ mag-level effects can shift
$(w_0,w_a)$ posteri- 
\\
 & & ors \cite{Efstathiou2025Systematics,Vincenzi2025CompareSN,Popovic2025DESDovekie}
\\

\hline
\end{tabular}
\end{table*}

\subsection{DESI DR2: background-only implications and reconstruction-level phenomenology}

At the background level, DESI BAO constrain $\DM(z)/\rd$ and  $D_{\rm H}(z)/\rd$, so late-time inferences are entangled with the early-time calibration of $\rd$ by the CMB. In $\LCDM$, DESI DR2 BAO are well-fit but exhibit a mild discrepancy with CMB-preferred parameters; allowing $\wowaCDM$ improves the joint fit and favors $w_0>-1$, $w_a<0$ \cite{DESIDR2Key}. Extended reconstructions (including ${\rm Om}(z)$-based diagnostics) report qualitatively consistent low-$z$ deviations across multiple priors and methods \cite{Lodha2025ExtendedDE}. A key methodological requirement is therefore to demonstrate that the preference is not driven by calibration/selection residuals in the SN component of the combined likelihood.

If the preference for evolving $w(z)$ in joint BAO+CMB(+SN) fits is primarily absorbing a mismatch in the early-time ruler $\rd$, then $\rd$-independent BAO combinations such as $F_{\rm AP}(z)=D_{\rm M}(z)/D_{\rm H}(z)$ should remain close to their $\Lambda$CDM predictions at the same redshifts (modulo curvature). Conversely, a statistically significant shift in $F_{\rm AP}(z)$ points to a genuine change in the late-time expansion shape $E(z)$ and/or curvature, and therefore strengthens the case that the preference is not solely an early-time $\rd$ effect. This motivates routinely reporting and comparing both $(D_{\rm M}/\rd, D_{\rm H}/\rd)$ and $F_{\rm AP}(z)$ when interpreting DESI-era tensions.

\subsection{DESI DR2 analysis closure: validation, nuisance structure, and robustness}
\label{sec:desi_validation}

Because DR2-level BAO errors are at (sub-)percent precision over much of $0\lesssim z\lesssim 2.5$, late-time inference is increasingly limited by
(i) modeling of broadband/nuisance terms in the BAO template fits, (ii) redshift-dependent selection and completeness, and (iii) covariance validation.
DESI DR2 therefore provides a dedicated validation analysis that stress-tests the BAO pipeline against suite-level systematics, including mock-based closure, nuisance marginalization stability, and consistency across tracer samples (galaxy and QSO) and redshift bins \cite{DESIDR2BAOValidation}. In the context of evolving-$w(z)$ claims from combined BAO+CMB(+SN), this validation layer is essential: it reduces the plausibility that the inferred preference is driven by BAO pipeline pathologies, shifting attention to cross-calibration and selection modeling
in the SN component and to early-time calibration assumptions (via $\rd$).

\subsection{Complementary low-$z$ closure tests: DES Y6 and combined-probe fits}

DES Year 6 $3\times2$\,pt provides a growth+geometry cross-check that is sensitive to dark-energy clustering and modified gravity, not just distances.\footnote{Here ``$3\times2$\,pt'' denotes the joint analysis of three two-point functions: cosmic shear, galaxy--galaxy lensing, and galaxy clustering.} It reports $S_8$ and $\Omm$ constraints in $\LCDM$ and a broad but informative constraint on constant-$w$ in $\wCDM$, and quantifies parameter differences
with respect to a combined-CMB baseline (Planck+ACT DR6+SPT-3G DR1) \cite{DESY6_3x2pt_2026}.
Importantly, DES Y6 also reports joint fits combining $3\times2$\,pt with DESI DR2 BAO and other low-$z$ probes, producing some of the tightest
$\LCDM$ constraints to date and a $w$ constraint consistent with $-1$ in constant-$w$ extensions \cite{DESY6_3x2pt_2026}.
This creates a concrete, quantitative framework for testing whether DESI DR2’s preference for evolving $w(z)$ persists once perturbation-sensitive information is included.

These perturbation-sensitive data help discriminate smooth evolving dark energy from clustering dark energy ($c_s^2\ll 1$) and from modified gravity where $(\mu,\eta)\neq(1,1)$ can mimic effective $w(z)$ evolution at the background level.

While cosmic shear and galaxy--galaxy lensing provide uniquely valuable low-redshift geometry-plus-growth closure tests, they should not be treated in isolation as the only perturbation-sensitive probes. Independent growth information also comes from peculiar-velocity measurements and compilations of $\fsig$, while direct expansion-rate information from cosmic-chronometer $H(z)$ measurements remains an important low-redshift cross-check \cite{Turner2024PeculiarVelocity,SaidEtAl2024DESIPVFundamentalPlane,Moresco2024CosmicChronometers}. At the same time, recent KiDS-Legacy consistency analyses indicate that current WL catalogs are becoming increasingly internally consistent and more compatible with external probes, even though the WL systematic floor remains an active area of development \cite{Stolzner2025KiDSLegacyConsistency,deCruzPerezParkRatra2022Lensing}.

\section{Parametrizations and reconstructions}
\label{sec:recon}

Throughout this paper we use CPL/$w_0$--$w_a$ fits as a low-dimensional phenomenological compression of the late-time background expansion history, not as a claim that the dark-energy microphysics is literally described by a perfect fluid with arbitrarily prescribed perturbation properties. This distinction matters because the perturbation closure adopted for phenomenological $w(z)$ fits need not coincide with that of any explicit scalar-field or modified-gravity model. For that reason, we separate likelihood-level phenomenology from model-level interpretation, and we regard physical conclusions as secure only when they survive an explicit perturbation-level analysis in a fully specified model \cite{TadaTerada2024QuintessentialDESI,ParkRatra2025PhiCDM,RoyChoudhuryOkumuraUmetsu2025NPDDE}.

\subsection{CPL and phantom crossing redshift}

The Chevallier--Polarski--Linder (CPL) parametrization \cite{Chevallier2001,Linder2003}
(often denoted $w_0w_a$CDM) is
\begin{equation}
w(a)=w_0+w_a(1-a)=w_0+w_a\frac{z}{1+z}.
\end{equation}
It implies
\begin{equation}
f_{\rm DE}(a)=a^{-3(1+w_0+w_a)}\exp\!\Big[3w_a(a-1)\Big].
\end{equation}
The effective phantom crossing $w(z_\times)=-1$ occurs at
\begin{equation}
z_\times=\frac{-1-w_0}{w_a+1+w_0},
\end{equation}
when the RHS is positive. Because single-field canonical quintessence cannot cross $w=-1$, evidence for crossing motivates either extended field content, non-canonical terms, interactions, or modified gravity.

For CPL, it is often useful to define the \emph{pivot} equation of state $w_p\equiv w(a_p)$ at the scale factor $a_p$ where $w_p$ is best constrained.
Writing $w(a)=w_0+w_a(1-a)$, one finds
\begin{equation}
a_p = 1 + \frac{{\rm Cov}(w_0,w_a)}{{\rm Var}(w_a)},\qquad
w_p = w_0 + w_a(1-a_p),
\end{equation}
so that ${\rm Var}(w_p)$ is minimized by construction. A compact survey-performance metric is the Dark Energy Task Force (DETF) figure of merit
\begin{equation}
{\rm FoM}_{\rm DETF}\equiv \frac{1}{\sigma(w_p)\,\sigma(w_a)},
\end{equation}
which is proportional to the inverse area of the $(w_0,w_a)$ error ellipse under Gaussian assumptions \cite{Albrecht2006DETF}.
In a DESI+SN+CMB context, reporting $(w_p,a_p)$ (or equivalently $z_p$) alongside $(w_0,w_a)$ improves interpretability and comparison across datasets and priors.

\subsection{Non-parametric strategies and prior sensitivity}

Binned $w(z)$ or $\rho_{\rm DE}(z)$, principal components, and Gaussian processes reduce functional bias, but Eq.~\eqref{eq:wFromRho} shows that
$w(z)$ reconstruction is differentiation-sensitive.
Robust inference is therefore typically performed at the level of $H(z)$, distances, or $\rho_{\rm DE}(z)$ with explicit priors rather than by differentiating noisy distance fits. DESI DR2 extended analyses provide a useful benchmark comparison of multiple reconstruction approaches \cite{Lodha2025ExtendedDE}.

\section{Physical models and perturbation-level consistency}
\label{sec:models}

In this section we will discuss model classes and map DESI-era phenomenology into concrete microphysical requirements and falsifiers. At the background level, distance data constrain the {\em shape} of the expansion history $E(z)$ and (through BAO) ratios to the sound horizon $r_d$; at the perturbation level, growth and lensing test whether the same model can reproduce $\fsig(z)$ and lensing potentials without instabilities or conflicts with multi-messenger bounds.

A useful way to organize the model space is by the answer to three questions:

\begin{enumerate}
\item \textit{Late-time shape vs early-time ruler.}
Is the apparent preference for evolving $w(z)$ driven by a genuine modification of late-time $E(z)$, or is it largely absorbing an early-time shift in the BAO ruler $r_d$? $r_d$-independent anisotropic-BAO diagnostics (e.g.\ $F_{\rm AP}$) directly probe the late-time shape and therefore help separate these possibilities.

\item \textit{Can the model realize (or mimic) phantom crossing?}
If future combined analyses robustly prefer an effective crossing of $w=-1$, then canonical single-field quintessence is excluded. One must invoke either
(i) multiple fields / non-canonical structure, (ii) interacting dark sectors (where $w_{\rm eff}$ can cross $-1$ without a fundamental phantom), or
(iii) modified gravity where the inferred effective $w(z)$ is not a microphysical
equation of state.

\item \textit{What happens to perturbations?}
Many background-level explanations are only viable if they pass perturbation-level closure: consistent linear growth, stable sound speeds/no ghosts, and compatibility
with gravitational-wave propagation constraints. In practice, this means that full-shape clustering, RSD, and $3\times 2$pt measurements and are essential components of a perturbation-level model validation test.
\end{enumerate}

The rest of this section follows this logic: for each class we summarize (a) the minimal ingredients needed to reproduce the relevant background behavior,
(b) the characteristic perturbation signatures, and (c) the most direct observational discriminants in near-term data.

\subsection{If evolving $w(z)$ persists: what microphysics is minimally required?}

Several DESI-era combined-probe fits expressed in CPL-like parameterizations prefer an effective region with $w_0>-1$ and $w_a<0$, which in that phenomenological language often corresponds to an apparent crossing of $w=-1$ at finite redshift. However, this statement is parameterization-dependent and should not be over-interpreted as a direct microphysical detection. Recent analyses of physically specified scalar-field models and quintessence reconstructions show that canonical or nonphantom dynamics can remain compatible with current data in appropriately defined model spaces \cite{TadaTerada2024QuintessentialDESI,ParkRatra2025PhiCDM,WangEtAl2026QuintessenceReconstruction,RoyChoudhuryOkumuraUmetsu2025NPDDE}. A robust detection of persistent phantom behavior would still exclude canonical single-field quintessence, but current DESI-era inferences have not yet reached that model-independent level of certainty.

\subsection{Cosmological constant and vacuum energy naturalness}

A cosmological constant contributes
\begin{equation}
T^{(\Lambda)}_{\mu\nu}=-\rho_\Lambda g_{\mu\nu},\qquad w=-1,\qquad \rho_\Lambda=\frac{\Lambda}{8\pi G},
\end{equation}
but the vacuum-energy interpretation is radiatively unstable, motivating dynamical or gravitational alternatives \cite{Weinberg1989CC,Carroll2001CCReview}. Table~\ref{tab:model_discriminants} summarizes model classes and leading discriminants beyond background distances.

\subsection{Early-time physics that shifts $\rd$ and its impact on late-time inference}
\label{sec:early_rd}

Because BAO constrain $\DM(z)/\rd$ and $D_{\rm H}(z)/\rd$, any modification of the pre-recombination expansion rate $H(z)$ can shift $\rd$ and thereby propagate into inferred late-time parameters even if $E(z)$ at $z\lesssim 2$ is unchanged.
A broad class of scenarios can be parameterized by an additional early component with fractional contribution
\begin{equation}
f_{\rm new}(z)\equiv \frac{\rho_{\rm new}(z)}{\rho_{\rm tot}(z)}\qquad (z\gtrsim 10^3),
\end{equation}
which increases $H(z)$ and typically reduces $\rd=\int c_s/H\,\dd z$. Early-dark-energy (EDE) models implement this idea with a component that is non-negligible around matter--radiation equality and then rapidly dilutes, aiming to reconcile early- and late-Universe distance ladders by decreasing $\rd$ while permitting a larger inferred $H_0$ in CMB+BAO fits \cite{Poulin2019EDE}. However, joint fits including CMB lensing and late-time large-scale-structure data often restrict the allowed EDE fraction and can reintroduce tension through shifts in $\sigma_8$-related parameters. A key driver is that many pre-recombination modifications that reduce $r_d$ enhance the CMB early integrated Sachs--Wolfe (eISW) contribution; maintaining an acceptable CMB fit then tends to push the inferred physical CDM density $\omega_c$ upward, which in turn raises the predicted clustering amplitude (e.g.\ $\sigma_8$ and $S_8$) and tightens LSS consistency constraints \cite{Hill2020EDE,Kamionkowski2023EDEReview,Vagnozzi2021eISW,PedrottiVagnozzi2024HubbleMultiD}. For a DESI DR2-driven dynamical-$w(z)$ interpretation, explicitly testing whether the preference can be absorbed by a modest $\rd$ shift (i.e.\ early-time physics) versus requiring genuine late-time dynamics is thus a necessary model-discrimination step.

\begin{table*}[t]
\caption{\label{tab:model_discriminants}
Model classes and leading discriminants beyond background distances.
The intent is to map DESI-era background phenomenology into \emph{perturbation-level} requirements and falsifiers.}
\setlength{\tabcolsep}{3pt}
\begin{tabular*}{\textwidth}{llll}
\hline
Class & Background signature & Perturbations / gravity signature & Sharp cross-checks \\ \hline\hline
$\LCDM$ &
$w=-1$, $\rho_{\rm DE}=\mathrm{const.}$ &
$\mu=\eta=1$; GR growth &
Consistency of geometry+growth; \\

& & &sirens give $\DL^{\rm GW}=\DL^{\rm EM}$.\\[2pt]

Smooth quintessence &
$w(z)>-1$; no phantom  &
$c_s^2\simeq 1$; weak clustering; GR  &
No persistent $w<-1$; growth index \\
& crossing & growth &
 near GR; stability without tuning.\\[2pt]

Clustering DE &
$w(z)>-1$ possible; effec-  &
$c_s^2\ll 1$; modifies lensing/growth  &
3$\times$2pt vs distances; scale dependence \\

(k-essence) &
tive features in $E(z)$ & at fixed $H(z)$ &
in growth/lensing.\\[2pt]

Interacting dark  &
$w_{\rm eff}(z)$ can mimic crossing &
Modified growth + momentum &
RSD+$3\times2$pt closure; consistency of \\
sector & &transfer constraints & bias and velocity fields.\\[2pt]

Early-time $r_d$ shift   &
Background fits partly  &
Shifts early-time structure, often  &
CMB lensing + LSS + BAO consis-\\

 (EDE-like) & absorbed by $\rd$ &
$\sigma_8$-linked &
tency; direct tests of $\rd$ priors.\\[2pt]

Modified gravity &
Effective $w(z)$ from  &
$\mu(a,k)\neq1$, $\eta(a,k)\neq1$; &
Sirens: $\DL^{\rm GW}\neq\DL^{\rm EM}$; growth/lensing \\ 

(scalar-tensor) &
 distances may cross $-1$ &
GW friction possible & scale tests; EFT stability priors.\\ 

\hline
\end{tabular*}
\end{table*}

\subsection{Canonical quintessence: when does the CPL quadrant make sense?}
\label{sec:quintessence_expanded}

Canonical scalar-field dark energy (quintessence) in its modern cosmological form was introduced in  \cite{PeeblesRatra1988TimeVariableCC,RatraPeebles1988RollingScalar}. The CPL sign pattern $w_0>-1$, $w_a<0$ can arise naturally for thawing trajectories, but this correspondence is only approximate: it is a projection of a physical history onto a two-parameter phenomenological basis, not a one-to-one model identifier \cite{TadaTerada2024QuintessentialDESI,ParkRatra2025PhiCDM}.

For a minimally coupled canonical scalar $\phi$,
\begin{equation}
S=\int \dd^4x\sqrt{-g}\Big[\frac{\Mpl^2}{2}R-\frac{1}{2}(\nabla\phi)^2-V(\phi)\Big]+S_m[g_{\mu\nu},\psi_m],
\end{equation}
with
\begin{equation}
\rho_\phi=\frac{1}{2}\dot\phi^2+V,\qquad p_\phi=\frac{1}{2}\dot\phi^2-V,\qquad w_\phi\in[-1,1].
\end{equation}
Given inferred $w(a)$ and $\rho_\phi(a)$,
\begin{equation}
V(a)=\frac{1-w(a)}{2}\rho_\phi(a),\qquad
\Big(\frac{\dd\phi}{\dd\ln a}\Big)^2 = 3\Mpl^2\,\Omega_\phi(a)\big[1+w(a)\big].
\end{equation}
Canonical quintessence cannot cross $w=-1$ without additional degrees of freedom or non-canonical structure \cite{Copeland2006QuintessenceReview,Zlatev1999Tracker}.

Canonical quintessence is the minimal dynamical alternative to a cosmological constant: a minimally coupled scalar with a standard kinetic term.
Its defining microphysical property is
\begin{equation}
w_\phi \equiv \frac{p_\phi}{\rho_\phi} \ge -1,
\end{equation}
so any {\em robust} evidence for persistent $w<-1$ or a stable crossing immediately excludes this class.

\paragraph{Thawing vs.\ freezing (and why $w_0>-1$, $w_a<0$ can be natural).}
A helpful phenomenological classification is:

\begin{itemize}
\item \textit{Thawing:} $w_\phi(z)\simeq -1$ at high redshift, with departures from
$-1$ developing only at late times when Hubble friction weakens. When projected
onto CPL, thawing trajectories often appear in the quadrant $w_0>-1$ and $w_a<0$
because the field is closer to $\Lambda$ in the past and less so today.

\item \textit{Freezing / tracker:} $w_\phi$ can be less negative at intermediate
redshift and evolve toward $-1$ at late times, often producing different
effective CPL trends. Tracker behavior can reduce sensitivity to initial
conditions but tends to be more constrained by early dark-energy fractions.
\end{itemize}

This classification is useful because it connects a {\em sign pattern} in $(w_0,w_a)$ to dynamics without over-interpreting a two-parameter fit.

\paragraph{Representative potentials and what they predict.}
Well-studied examples include (i) inverse power-law (tracker) potentials,
(ii) exponentials (scaling solutions), and (iii) hilltop / PNGB-like potentials (thawing-like behavior). For canonical quintessence the sound speed is
$c_s^2\simeq 1$, so dark-energy clustering is weak on sub-horizon scales; as a result, background fits typically imply only modest changes to growth relative to GR unless the expansion history itself changes appreciably.

\paragraph{Near-term discriminants.}
If a thawing-like interpretation remains viable, then:
(i) $F_{\rm AP}(z)$-type BAO shape diagnostics should show coherent late-time shape deviations rather than pure $r_d$ rescaling;
(ii) growth constraints should remain close to GR expectations (e.g.\ growth-index consistency) because there is no additional force or slip beyond GR; and
(iii) any apparent phantom crossing in reconstructions would point away from this class and toward interactions/multi-field structure/modified gravity.

\subsection{K-essence, clustering, and stability}

For $P(\phi,X)$ with $X\equiv -\tfrac12(\nabla\phi)^2$,
\begin{equation}
c_s^2=\frac{P_{,X}}{P_{,X}+2XP_{,XX}},
\end{equation}
and stability requires $P_{,X}>0$ (no ghosts) and $c_s^2>0$ (no gradient instabilities).
If $c_s^2\ll1$, dark energy can cluster on sub-horizon scales, modifying growth and lensing at fixed background expansion.

\subsection{No-go theorems for single-field phantom crossing and minimal extensions}

A recurring phenomenological feature in several DESI DR2-driven reconstructions is an \emph{effective} phantom crossing $w(z_\times)=-1$.
It is therefore important to state explicitly what classes of microphysics can and cannot realize such behavior.
For a broad class of single-field models with Lagrangian $P(\phi,X)$ (including k-essence), smooth evolution across $w=-1$ is obstructed under standard stability assumptions; see e.g.\ the ``no-go'' result of Vikman \cite{Vikman2005NoGo}. Consequently, persistent evidence for a crossing motivates at least one of:
(i) multiple fields (``quintom''-type sectors), (ii) non-minimal couplings and/or higher-derivative interactions arranged to remain ghost-free, or (iii) modified gravity in which the \emph{effective} $w(z)$ inferred from distances is not a microphysical equation of state.

A minimal multi-field illustration is a two-field sector with one canonical and one phantom-like mode (or an effective phantom mode emerging from higher-derivative structure). At the level of background evolution, such sectors can realize $w(z)$ crossing while maintaining a healthy EFT below a cutoff if the would-be ghost is avoided or sequestered. Early ``quintom'' phenomenology was explored in \cite{Feng2005Quintom}.

If future combined analyses continue to indicate an effective crossing of $w=-1$, then at least one of the following must be true:

\begin{enumerate}
\item \textit{Additional degrees of freedom:} multi-field ``quintom''-type sectors
(or effective multi-field behavior) allow stable crossings at the expense of
additional parameters and stability bookkeeping.

\item \textit{Interactions in the dark sector:} the microphysical $w$ can remain
non-phantom while the inferred $w_{\rm eff}$ crosses $-1$ due to $Q\neq 0$.
In this case, growth/lensing become part of the model definition.

\item \textit{Modified gravity / higher-derivative EFT:} the object inferred as $w(z)$ from background distances is not a microphysical equation of state. Then one expects correlated signatures such as gravitational slip, modified
growth, and potentially a modified gravitational-wave luminosity distance.
\end{enumerate}

This classification is useful because it transforms the qualitative phrase ``phantom crossing'' into a falsifiable set of model requirements.

\subsection{Interacting dark sectors: apparent phantom behavior without a phantom}
\label{sec:interacting_expanded}

A physically distinct mechanism for apparent $w(z)$ evolution is energy--momentum exchange within the dark sector. At background level one may write
\begin{equation}
\label{eq:darksector_cont_c}
\dot\rho_c + 3H\rho_c = Q,\qquad
\dot\rho_{\rm DE} + 3H\big(1+w\big)\rho_{\rm DE} = -Q,
\end{equation}
where $Q$ is an interaction term (often parameterized as $Q=\beta H\rho_c$, $Q=\beta H\rho_{\rm DE}$, or $Q=\beta H(\rho_c+\rho_{\rm DE})$).
The interaction modifies both the expansion history and the mapping between background fits and an \emph{effective} equation of state:
\begin{equation}
\label{eq:weff_general}
w_{\rm eff}(z)=w(z)-\frac{Q}{3H\rho_{\rm DE}},
\end{equation}
so that a fit with $w_{\rm eff}(z)$ may mimic phantom-like behavior even when the underlying $w(z)\ge -1$.
This degeneracy is broken by perturbations: the same interaction generically modifies growth and lensing,
and therefore must be confronted with $\fsig(z)$ and $3\times2$\,pt constraints. A canonical reference for coupled quintessence is \cite{Amendola2000CoupledDE}.

Interacting dark energy provides a qualitatively distinct route to effective $w(z)$ evolution: the dark-energy equation of state can remain non-phantom at the microphysical level while the {\em inferred} effective equation of state crosses $-1$ because energy--momentum is exchanged with dark matter. Vacuum-decay (time-dependent $\Lambda$) models can be cast in an equivalent exchange form with $w=-1$ and $Q\equiv -\dot\rho_\Lambda$ (energy transfer between a vacuum-like component and matter); early examples include the Bose-condensate evaporation picture \cite{DymnikovaKhlopov2000VacuumEvap}.

\paragraph{Covariant definition.}
A robust definition starts from covariant non-conservation in each dark sector:
\begin{equation}
\label{eq:Qcov}
\nabla_\mu T^{\mu\nu}_{(c)} = Q^\nu,\qquad
\nabla_\mu T^{\mu\nu}_{({\rm DE})} = -Q^\nu,
\end{equation}
with baryons and radiation conserved separately. For an FLRW background and an interaction parallel to the CDM 4-velocity,
$Q^\nu = Q\,u^\nu_{(c)}$, Eq.~(\ref{eq:Qcov}) reduces directly to the continuity equations Eqs.~(\ref{eq:darksector_cont_c}). The effective equation of state inferred from a background-only fit is then Eq.~(\ref{eq:weff_general}).

\paragraph{Why perturbations are essential.}
The same interaction modifies growth (and, depending on the covariant structure, can introduce a fifth force and/or momentum exchange). Some popular choices of
$Q$ that appear innocuous at background level can generate large-scale perturbation instabilities or unphysical behavior if implemented inconsistently. Therefore, viability must be assessed in a perturbation-complete framework.

\paragraph{Observational discriminants.}
Interacting models generically predict correlated signatures:
\begin{itemize}
\item modified $\fsig(z)$ and changes to the clustering amplitude relative to a background-only fit;
\item potential scale dependence in growth/lensing depending on the interaction structure and DE sound speed;
\item shifts in the inferred $\Omega_m(z)$ history that can be tested by joint distance+growth consistency (e.g.\ DES-like $3\times2$pt closure tests).
\end{itemize}

In the context of DESI-era evolving-$w$ claims, interacting dark sectors are therefore attractive because they can mimic an effective CPL trend while making sharp predictions for growth and lensing that are testable with the same data ecosystem.

\paragraph{Effective phantom behavior}

A fundamental phantom scalar with negative kinetic term is ghost-unstable \cite{Caldwell2002PhantomMenace}.
Effective $w<-1$ or phantom crossing may instead arise from multi-field sectors, interactions, or modified gravity where the inferred effective $w(z)$ differs from microphysics. Perturbation-level viability requires ghost/gradient stability and consistency with gravitational-wave constraints \cite{Abbott2017GW170817,Baker2017GWConstraints,Joyce2015MGReview}.

\subsection{Modified gravity and EFT constraints}

Scalar-tensor theories admit an EFT organization \cite{Gubitosi2013EFT,BelliniSawicki2014Alpha} (see reviews in \cite{Clifton2012MG,Joyce2015MGReview}.)
Representative viable subclasses include screened $f(R)$ models \cite{HuSawicki2007fR} and infrared modifications such as the Dvali--Gabadadze--Porrati (DGP) braneworld model \cite{Dvali2000DGP}, subject to strong late-time constraints from multi-messenger observations and large-scale structure.

In addition to constraints on the tensor speed (e.g.\ from GW170817), modified gravity generically predicts a modified \emph{friction term} in the gravitational-wave propagation equation on cosmological backgrounds. A common parametrization writes (schematically, in conformal time)
\begin{equation}
h'' + \big[2+\nu(a)\big]\cH\,h' + c_T^2 k^2 h = 0,
\end{equation}
where $\nu(a)=0$ and $c_T=1$ in GR.
When $\nu(a)\neq 0$, the gravitational-wave luminosity distance differs from the electromagnetic one:
\begin{equation}
\DL^{\rm GW}(z)=\DL^{\rm EM}(z)\,
\exp\!\Big[\frac{1}{2}\int_0^z \frac{\nu(z')}{1+z'}\,\dd z'\Big].
\end{equation}
Thus, standard-siren measurements provide a direct consistency test that is complementary to background probes and to growth/lensing, and can decisively discriminate modified-gravity explanations of evolving-$w(z)$ phenomenology \cite{Belgacem2018GWLumDist}.

\subsection{Reproducible model-comparison block for DESI-era combinations}
\label{sec:modelcomp_repro}

To make model-comparison statements transparent and reproducible, one should report for each dataset combination:
(i) the best-fit $\chi^2_{\min}$ in $\Lambda$CDM and in the extension,
(ii) $\Delta\chi^2 \equiv \chi^2_{\min}(\Lambda{\rm CDM})-\chi^2_{\min}({\rm ext})$,
(iii) a probe-by-probe decomposition $\Delta\chi^2=\sum_X \Delta\chi^2_X$,by probe $X\in\{$BAO, SN, CMB, growth/lensing$\}$, where applicable,
and (iv) information criteria (AIC/BIC) and/or Bayesian evidence ratios. In this work we focus on likelihood-level diagnostics and systematics projections; when quoting numerical model-comparison metrics we defer to the corresponding survey likelihood analyses and references.
For $k$ additional parameters and $N_{\rm data}$ total data points,
\begin{equation}
{\rm AIC}=\chi^2_{\min}+2k_{\rm tot},\qquad
{\rm BIC}=\chi^2_{\min}+k_{\rm tot}\ln N_{\rm data},
\label{eq:AICBIC}
\end{equation}
where $k_{\rm tot}$ is the total number of parameters in the model (including nuisance parameters that are fit). For nested models, we quote the Gaussian-equivalent significance  computed from the $\chi^2_k$ distribution [Eq.~\eqref{eq:Nsigma_mapping}]; we do \emph{not} use the shorthand $N_\sigma\simeq\sqrt{\Delta\chi^2}$ when $k\neq 1$. When Bayesian evidence ratios are available (from nested sampling), we report $\Delta\ln Z\equiv \ln Z_{\rm ext}-\ln Z_{\LCDM}$ for the stated prior ranges.

With
\(
\Delta\chi^2\equiv \chi^2_{\min}(\Lambda{\rm CDM})-\chi^2_{\min}({\rm ext})
\)
and
\(
\Delta k \equiv k_{\rm ext}-k_{\Lambda{\rm CDM}},
\)
the adopted sign convention implies
\begin{equation}
\Delta{\rm AIC}
\equiv {\rm AIC}_{\rm ext}-{\rm AIC}_{\Lambda{\rm CDM}}
= -\Delta\chi^2 + 2\Delta k.
\label{eq:deltaAIC_from_deltachi2}
\end{equation}
For the CPL extension relative to $\Lambda$CDM, $\Delta k=2$, so
\begin{equation}
\Delta{\rm AIC}=4-\Delta\chi^2.
\label{eq:deltaAIC_CPL}
\end{equation}
Thus $\Delta{\rm AIC}<0$ corresponds to AIC preference for the extension, equivalently
$\Delta\chi^2>4$ for the two-parameter $(w_0,w_a)$ case.

\subsection{A minimal late-transition interacting thawer (LTIT) model}
\label{sec:LTIT_model}

Motivated by the repeated appearance of CPL-like trends with $w_0>-1$ and $w_a<0$ in DESI-era combined analyses, we consider a model that:
(i) leaves early-time physics (and thus $r_d$) essentially unchanged,
(ii) produces late-time apparent evolution in an effective $w_{\rm eff}(z)$, and
(iii) makes perturbation-level predictions testable with RSD and lensing.

Consider a canonical scalar $\phi$ with potential $V(\phi)$ and a coupling only to cold dark matter (baryons uncoupled), implemented via a conformal factor in the DM sector:
\begin{equation}
\label{eq:LTIT_action}
S = \int \dd^4x\sqrt{-g}\Big[\frac{\Mpl^2}{2}R - \frac12(\nabla\phi)^2 - V(\phi)\Big]
+ S_b[g_{\mu\nu},\psi_b] + S_c[\tilde g_{\mu\nu},\psi_c],
\qquad
\tilde g_{\mu\nu}= C^2(\phi)\,g_{\mu\nu}.
\end{equation}
Define the (field-dependent) coupling
\begin{equation}
\beta(\phi) \equiv \Mpl \frac{\dd \ln C(\phi)}{\dd \phi}.
\end{equation}
A late-time ``trigger'' is encoded by choosing
\begin{equation}
\beta(\phi)=\frac{\beta_0}{2}\Big[1+\tanh\!\Big(\frac{\phi-\phi_t}{\Delta\phi}\Big)\Big],
\label{eq:beta_tanh}
\end{equation}
so the coupling is negligible at early times and turns on only when $\phi$ approaches $\phi_t$ (typically at $z\lesssim \cO(1)$ for thawing-like evolution). We call this the late-transition interacting thawer (LTIT) model.

On an FLRW background, the interacting-sector continuity equations are given by Eqs.~(\ref{eq:darksector_cont_c})
with the replacements $(\rho_{\rm DE},w)\rightarrow(\rho_\phi,w_\phi)$.
For the LTIT coupling in Eq.~(\ref{eq:beta_tanh}), the interaction source term is
\begin{equation}
Q_{\rm LTIT} = \beta(\phi)\frac{\rho_c}{\Mpl}\,\dot\phi,
\label{eq:Q_LTIT}
\end{equation}
so that the inferred effective equation of state becomes
\begin{equation}
w_{\rm eff}^{\rm LTIT}(z)=w_\phi(z)-\frac{Q_{\rm LTIT}}{3H\rho_\phi}.
\label{eq:weff_LTIT}
\end{equation}

If a CPL fit prefers $w_0>-1$ but $w_a<0$, a thawing-like scalar naturally keeps $w_\phi\simeq -1$ at higher $z$ while deviating at low $z$. The LTIT coupling in Eqs.~(\ref{eq:beta_tanh})--(\ref{eq:Q_LTIT}) then introduces additional late-time redshift dependence (only after the trigger), allowing $w_{\rm eff}(z)$ to mimic a stronger CPL-like evolution and, if required, an apparent crossing.

The LTIT model is not a purely background-level modification and makes correlated predictions:
\begin{enumerate}
\item It predicts correlated changes in growth via modified dark-matter dilution and, depending on the covariant completion, an additional effective force in the dark sector. Therefore, it is directly constrained by $\fsig(z)$ and $3\times2$\,pt measurements.
\item Because the coupling is designed to be late-time, it does not resolve any discrepancy through an early-time shift of $r_d$. Accordingly, $r_d$-independent BAO-shape diagnostics (e.g.\ $F_{\rm AP}$) should track the implied change in $E(z)$.
\item A complete analysis must verify perturbation stability (no ghosts or gradient instabilities) and check that the chosen covariant interaction avoids large-scale instabilities.
\end{enumerate}

This LTIT construction is an explicit example of a scenario in which late-time dynamics (rather than an early-time rescaling of $r_d$) can generate an effective CPL-like trend, while remaining testable through perturbation-level closure with growth and lensing. 

\section{Model comparison and robustness requirements}
\label{sec:stats}

Claims of evolving dark energy from combined low- and high-redshift data must be quantified with model-comparison metrics that are explicit about (i) the number of additional degrees of freedom, (ii) prior-volume dependence, and (iii) which probe(s) drive the improvement. This is particularly important for DESI-era combinations, where percent-level BAO distances/sub-percent calibration/selection effects in SNe can
project onto $(w_0,w_a)$ in a dataset-dependent way.

\subsection{Likelihood-ratio statistic and Gaussian-equivalent preference}

When quoting an ``$N_\sigma$ preference'' for an extension relative to $\Lambda$CDM, it is essential to
state the test statistic and the number of additional degrees of freedom.
For nested hypotheses and under standard regularity conditions (i.e.\ away from hard prior boundaries),
the likelihood-ratio statistic
\begin{equation}
\Delta\chi^2 \equiv -2\ln\!\left(\frac{\mathcal{L}_{\rm max}(\mathrm{restricted})}{\mathcal{L}_{\rm max}(\mathrm{extended})}\right)
\end{equation}
is asymptotically $\chi^2_k$-distributed with $k$ equal to the number of additional parameters.

We convert $\Delta\chi^2$ to a two-sided Gaussian-equivalent significance via
\begin{equation}
p = 1-F_{\chi^2_k}(\Delta\chi^2),\qquad
N_\sigma = \Phi^{-1}(1-p/2),
\label{eq:Nsigma_mapping}
\end{equation}
where $F_{\chi^2_k}$ is the $\chi^2$ CDF and $\Phi^{-1}$ is the inverse standard-normal CDF. Because evidence for evolving $w(z)$ is combination- and calibration-dependent \cite{DESIDR2Key,Efstathiou2025Systematics,Vincenzi2025CompareSN,Popovic2025DESDovekie},
claims of new physics should be supported by:
(i) $\Delta\chi^2$ or $\Delta\ln\cL_{\rm max}$, (ii) information criteria (AIC/BIC) and/or Bayesian evidence ratios, and
(iii) posterior predictive checks and explicit systematics marginalization. Perturbation-level closure tests using growth and lensing (e.g.\ DES Y6 $3\times2$\,pt) are essential to discriminate modified gravity or clustering dark energy from background-only modifications.

\begin{figure}[t]
  \centering
  \includegraphics[width=0.60\linewidth]{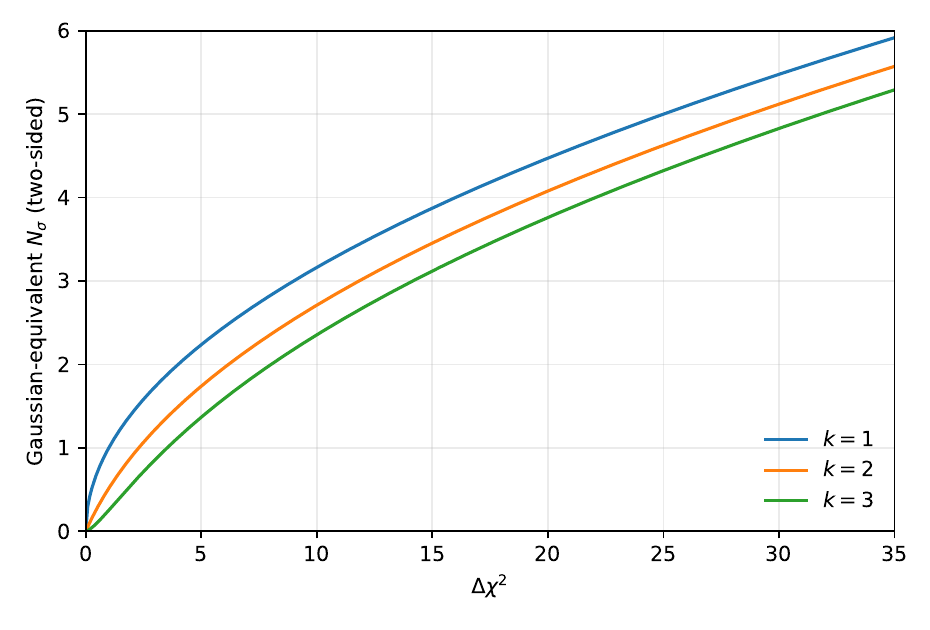}
  \caption{Conversion between likelihood-ratio improvements $\Delta\chi^2$ and
  Gaussian-equivalent significance $N_\sigma$ for nested models with $k=1$ and $k=2$
  additional parameters, using the two-sided mapping
  $p = 1-F_{\chi^2_k}(\Delta\chi^2)$ and $N_\sigma=\Phi^{-1}(1-p/2)$.
  This avoids the common (incorrect) shorthand $N_\sigma\simeq\sqrt{\Delta\chi^2}$
  when $k\neq 1$.}
  \label{fig:dchi2_to_sigma}
\end{figure}

For quick reference, Fig.~\ref{fig:dchi2_to_sigma} visualizes the mapping between $\Delta\chi^2$ and $N_\sigma$ for $k=1$ and $k=2$. In particular, for a two-parameter extension such as $(w_0,w_a)$, $N_\sigma=\{1,2,3,4\}$ corresponds to $\Delta\chi^2\simeq\{2.30,\,6.18,\,11.83,\,19.33\}$.

We note that, in addition to $\Delta\chi^2_{\min}$, it is useful to report information-criterion differences for transparency under parameter counting:
$\Delta{\rm AIC}\equiv{\rm AIC}_{\rm ext}-{\rm AIC}_{\Lambda{\rm CDM}}$ and
$\Delta{\rm BIC}\equiv{\rm BIC}_{\rm ext}-{\rm BIC}_{\Lambda{\rm CDM}}$.
Negative values indicate preference for the extension after complexity penalties. Because BIC penalizes additional parameters by $\ln N_{\rm data}$, it is typically more conservative than AIC for large combined data vectors.

\begin{table}[h]
\caption{\label{tab:desi_scorecard}
Model-comparison ``scorecard'' translating reported $N_\sigma$ preferences
for $w_0w_a$CDM (CPL dark-energy extension) over $\Lambda$CDM into implied $\Delta\chi^2$ (Wilks, $k=2$),
and the corresponding AIC gain $-\Delta{\rm AIC}=\Delta\chi^2-4$.
Larger positive values indicate stronger AIC preference for the extension.
These translations are valid only under the nested-model/Wilks assumptions and
should be complemented by probe-by-probe goodness-of-fit and systematics tests.}
\begin{tabular}{lccc}
\hline
Data combination & $N_\sigma$ (reported) & $\Delta\chi^2$ (implied) & $-\Delta{\rm AIC}$ (implied)\\
\hline\hline
DESI BAO + CMB & $3.1$ & $12.5$ & $8.5$ \\
DESI BAO + CMB + SNe & $2.8$--$4.2$ & $10.6$--$21.1$ & $6.6$--$17.1$ \\
\hline
\end{tabular}
\end{table}

\subsection{Information criteria and Bayesian evidence}

For Gaussian likelihoods one may define
\begin{equation}
\mathrm{AIC}=-2\ln\mathcal{L}_{\max}+2k_{\rm tot},\qquad
\mathrm{BIC}=-2\ln\mathcal{L}_{\max}+k_{\rm tot}\ln N_{\rm data},
\label{eq:AICBIC_defs}
\end{equation}
where $k_{\rm tot}$ is the total number of fitted parameters (including nuisance parameters)
and $N_{\rm data}$ is the dimension of the combined data vector.
AIC/BIC provide fast complexity penalties but do not replace Bayesian evidence when
priors are well-defined. The evidence is
\begin{equation}
Z \;=\; \int \! d\bm{\theta}\; \mathcal{L}(\bm{\theta})\,\pi(\bm{\theta}),
\qquad
\Delta\ln Z \equiv \ln\!\left(\frac{Z_{\rm ext}}{Z_{\Lambda{\rm CDM}}}\right),
\label{eq:evidence_def}
\end{equation}
which is sensitive to prior volumes and therefore must be quoted with the stated prior ranges.

Independent of the chosen metric (likelihood ratio, AIC/BIC, or evidence), a convincing
claim should be accompanied by:
(i) the best-fit $\chi^2_{\min}$ (or $-2\ln\mathcal{L}_{\max}$) in $\Lambda$CDM and the extension,
(ii) $\Delta\chi^2$ and the correct $k$ used in Eq.~(\ref{eq:Nsigma_mapping}),
(iii) a probe-by-probe decomposition $\Delta\chi^2=\sum_X\Delta\chi^2_X$ for
$X\in\{\mathrm{BAO,SN,CMB,growth/lensing}\}$ where available, and (iv) posterior predictive checks (PPCs) targeted at the specific residuals that motivate the extension (e.g.\ redshift-dependent distance residuals, or BAO anisotropy).  In SN-dominated combinations with $N_{\rm data}\sim\mathcal{O}(10^3)$--$\mathcal{O}(10^4)$,
BIC imposes a substantial penalty for additional dark-energy parameters: for $\Delta k=2$ (CPL), one needs $\Delta\chi^2 \gtrsim 2\ln N_{\rm data}\approx 14$--$18$ to be BIC-preferred,
whereas AIC requires only $\Delta\chi^2>4$. This clarifies why ``evolving-$w$'' claims that are significant under likelihood ratios may still be disfavored by BIC in large-$N_{\rm data}$ combinations.

In addition to AIC/BIC and, when feasible, Bayesian evidence, it is often informative to report the deviance information criterion (DIC), especially for posterior-sampled or hierarchical cosmological analyses where the effective number of parameters is more relevant than a fixed large-sample penalty \cite{SpiegelhalterEtAl2002DIC,RezaeiMalekjani2021ModelSelection}. We therefore recommend DIC as a useful complement to AIC/BIC rather than relying on any single information criterion in isolation.

\subsection{Goodness-of-fit, posterior predictive checks, and dataset consistency}
\label{sec:ppc_consistency}

A transparent robustness report should include (i) absolute goodness-of-fit for each probe block and (ii) internal consistency between probe combinations in the shared parameter subspace. For a data vector $\bm d$ partitioned into blocks $X\in\{\mathrm{BAO,SN,CMB,growth/lensing}\}$ with covariance blocks $\bm C_X$, we
define residuals $\bm r(\bm\theta)\equiv \bm d-\bm d_{\rm th}(\bm\theta)$ and
blockwise contributions
\begin{equation}
\chi_X^2(\bm\theta)\equiv \bm r_X^{T}\,\bm C_X^{-1}\,\bm r_X,
\qquad
\chi^2_{\rm tot}(\bm\theta)=\sum_X \chi_X^2(\bm\theta).
\label{eq:chi2_blocks}
\end{equation}
Reporting $\Delta\chi^2_X$ between nested models (alongside $\Delta\chi^2_{\rm tot}$) makes it explicit which probes drive any preference. For typical cosmological likelihoods with $N_{\rm data}\sim 10^2$--$10^3$, the BIC penalty per additional parameter is $\ln N_{\rm data}\simeq 4.6$--$6.9$
(i.e., substantially stronger than the AIC penalty of $2$ per parameter).

A Bayesian posterior predictive $p$-value for a chosen discrepancy statistic
$T(\bm d,\bm\theta)$ is
\begin{equation}
p_{\rm PPC}\equiv {\rm Pr}\!\left[T(\bm d_{\rm rep},\bm\theta)\ge
T(\bm d,\bm\theta)\mid \bm d\right]
=
\int {\rm d}\bm\theta\,p(\bm\theta\mid \bm d)
\int {\rm d}\bm d_{\rm rep}\,p(\bm d_{\rm rep}\mid \bm\theta)\,
\Theta\!\Big(T(\bm d_{\rm rep},\bm\theta)-T(\bm d,\bm\theta)\Big),
\label{eq:ppc}
\end{equation}
where $\bm d_{\rm rep}$ is a replicated dataset drawn from the likelihood $p(\bm d\mid \bm\theta)$ and $\Theta$ is the Heaviside step function. In practice, $p_{\rm PPC}$ is estimated by drawing $\bm\theta$ from the posterior and generating $\bm d_{\rm rep}$ from the corresponding likelihood, using the same covariance and systematics model.

To quantify dataset consistency in a common parameter subspace, consider two (approximately independent) probe combinations $A$ and $B$ with posterior means
$\bar{\bm\theta}_A$, $\bar{\bm\theta}_B$ and covariances $\bm\Sigma_A$, $\bm\Sigma_B$.
The parameter-shift statistic
\begin{equation}
T_{\rm AB}\equiv
(\bar{\bm\theta}_A-\bar{\bm\theta}_B)^{T}
(\bm\Sigma_A+\bm\Sigma_B)^{-1}
(\bar{\bm\theta}_A-\bar{\bm\theta}_B)
\label{eq:param_shift}
\end{equation}
is approximately $\chi^2_{N_p}$-distributed under the null (Gaussian posteriors, correct model, negligible cross-covariance), where $N_p$ is the number of shared
parameters. If the two inferences are correlated (e.g.\ shared calibration anchors), one should replace $\bm\Sigma_A+\bm\Sigma_B\rightarrow
\bm\Sigma_A+\bm\Sigma_B-\bm\Sigma_{AB}-\bm\Sigma_{BA}$. The corresponding $p$-value and Gaussian-equivalent $N_\sigma$ can be reported using Eq.~(\ref{eq:Nsigma_mapping}).

\subsection{Linear-response parameter shifts induced by systematic residuals}
\label{sec:paramshift}

In the Gaussian approximation around a best-fit point $\hat{\bm{\theta}}$, a (small) systematic residual
$\delta\bm{d}$ in the combined data vector induces a parameter shift that is well approximated by a linear response:
\begin{equation}
\delta\bm{\theta}
\simeq
\bm{F}^{-1}\,\bm{J}^{T}\bm{C}^{-1}\,\delta\bm{d},
\qquad
J_{i\alpha}\equiv
\left.\frac{\partial d_i}{\partial\theta_\alpha}\right|_{\hat{\bm{\theta}}},
\qquad
F_{\alpha\beta}=\bm{J}_{\alpha}^{T}\bm{C}^{-1}\bm{J}_{\beta},
\label{eq:theta_shift}
\end{equation}
where $\bm{C}$ is the total covariance of the data vector $\bm{d}$. For SN+BAO+CMB compressed combinations, $\delta\bm{d}$ can represent, for example, a redshift-dependent SN calibration drift $\delta\mu(z)$, a bias-correction residual mode, or a BAO systematic template mode. Eq.~\eqref{eq:theta_shift} provides a quantitative way to translate few$\times10^{-2}$ mag-level effects into expected shifts of $(w_0,w_a)$ (and derived parameters)
\emph{without rerunning} a full Markov chain Monte Carlo (MCMC), and it makes explicit which redshift ranges and probes dominate the projection.

Using the local response coefficients in Table~\ref{tab:sn_w0wa_sensitivity}, a redshift-dependent residual
$\Delta\mu_{\rm sys}(z\simeq 1)$ projects approximately as
$\Delta\mu_{\rm sys}\simeq (\partial\Delta\mu/\partial w_0)\,\delta w_0$ (holding $w_a$ fixed).
Thus, keeping $|\delta w_0|<0.05$ requires $|\Delta\mu_{\rm sys}(z\simeq 1)|\lesssim 0.05\times 0.31\simeq 0.016$ mag. A fully rigorous budget should use Eq.~(\ref{eq:sn_param_bias}) with the full covariance and the actual mode shape $\delta\mu_{\rm sys}(z)$.

In practice, $\bm{J}$ is obtained by finite differences of the pipeline prediction for the chosen data vector,
$\bm{C}$ is the published total covariance for the same observable definitions, and $\bm{F}=\bm{J}^T\bm{C}^{-1}\bm{J}$. Given any systematic template $\delta\bm{d}$ (e.g.\ a redshift-dependent SN $\delta\mu(z)$ mode or a BAO-template bias mode), Eq.~(\ref{eq:theta_shift}) yields the induced $\delta\bm{\theta}$, which can be compared directly to posterior uncertainties to define an explicit bias budget.

\section{Conclusions and outlook}
\label{sec:concl}

DESI DR2 moves late-time acceleration tests into a regime where the dominant limitation is no longer only the statistical precision of distance ratios, but the coherent propagation of a small set of \emph{calibration directions}:  the BAO ruler calibration (via $\rd$), redshift-dependent SN calibration/selection modes, and the CMB-side robustness budget that includes the treatment of lensing consistency. Three conclusions summarize the present status and justify the scope of this review.

\paragraph*{Observational status.}
DESI DR2 strengthens the case that a background-level anomaly exists in some joint low-$z$+CMB combinations, but the significance and physical interpretation remain dataset dependent. BAO+CMB combinations can already prefer evolving-$w$ parameterizations; SNe, especially the low-$z$ anchor and cross-calibration, materially affect the quoted significance and can localize the anomaly to very low redshift. Perturbation-sensitive closure tests (RSD, peculiar velocities, weak lensing, and CMB lensing) remain essential because any genuine late-time background deformation must also survive growth/lensing consistency.

\paragraph*{Reconstructions.}
Parametric and non-parametric reconstructions broadly agree that one should reconstruct at the level of $H(z)$, distances, or $\rho_{\rm DE}(z)$ rather than over-interpreting differentiated $w(z)$ curves. Apparent phantom crossing can emerge in CPL or other flexible reconstructions, but its statistical robustness depends on priors, parameterization, and data combination; it is therefore best treated as a phenomenological clue rather than as a microphysical result.

\paragraph*{Physical models.}
Current data do not yet single out a unique microphysical explanation. Canonical quintessence remains viable in some physical-model analyses, while a future robust detection of persistent phantom behavior would require non-canonical or multi-field physics, interacting dark sectors, or modified gravity. Equally, some of the apparent late-time preference may still be absorbed by an early-time ruler shift or by residual systematic modes. The decisive next step is therefore not simply a smaller $(w_0,w_a)$ contour, but an end-to-end sector-by-sector test combining ruler-independent BAO shape, calibrated SN systematics, full CMB likelihoods, and perturbation closure.

This paper contributes two lightweight, likelihood-faithful diagnostics intended to sharpen that interpretation:
\begin{itemize}
\item An explicitly $r_d$-independent BAO ``shape'' observable,
$F_{\rm AP}(z)\equiv D_{\rm M}(z)/D_{\rm H}(z)$, constructed directly from published anisotropic BAO products and their covariance.
This isolates late-time expansion-shape information from pure ruler-rescaling effects.
\item A linear-response mapping from redshift-dependent SN Hubble-diagram residuals $\delta\mu_{\rm sys}(z)$
to induced biases in $(w_0,w_a)$, including the covariance-projected form that marginalizes calibration-like
nuisance directions [Eqs.~(\ref{eq:SN_projection})--(\ref{eq:SN_bias_proj})].
This turns ``few$\times10^{-2}$\,mag matters'' into an explicit requirement on systematic mode amplitudes.
\end{itemize}

Three quantitative takeaways are worth emphasizing:
\begin{itemize}
\item In nested-model language, a quoted ``$N_\sigma$ preference'' for $w_0w_a$CDM over $\Lambda$CDM corresponds to a
\emph{specific} $\Delta\chi^2$ for $\Delta k=2$ degrees of freedom, not $\sqrt{\Delta\chi^2}$. Figure~\ref{fig:dchi2_to_sigma} provides a compact translation consistent with Eq.~(\ref{eq:Nsigma_mapping}).
\item SN redshift-dependent systematics at the level of $\Delta\mu_{\rm sys}\sim 0.02$ mag correspond to
$\Delta D_{\rm L}/D_{\rm L}\simeq 0.92\%$ and can shift $(w_0,w_a)$ at a level comparable to reported DESI-era ``evolving-$w$'' preferences (see Table~\ref{tab:sn_w0wa_sensitivity} and Sec.~\ref{sec:sn_bias_covproj}).
As a practical target, keeping $|\delta w_0|\lesssim 0.05$ typically requires controlling coherent
\emph{relative} modulus residuals to $\lesssim (1$--$2)\times 10^{-2}$ mag over $z\sim 0.5$--$1$,
with the exact bound determined by the covariance-projected mode shapes in Eq.~(\ref{eq:SN_bias_proj}).
\item $r_d$-independent BAO anisotropy provides a clean discriminator: if the anomaly is primarily a ruler shift,
$F_{\rm AP}(z)$ remains approximately unchanged, while statistically significant deviations in $F_{\rm AP}(z)$
point to a genuine late-time shape change (and/or curvature if allowed).
\end{itemize}

Looking forward, the most decisive near-term discriminants are those that break the ``geometry-only'' degeneracies:
(i) full-shape clustering and RSD tied to the same samples used for BAO (growth closure at fixed background),
(ii) end-to-end SN calibration/selection forward-modeling with explicit systematic-mode propagation into $(w_0,w_a)$,
(iii) joint growth+geometry lensing analyses ($3\times2$\,pt) testing perturbation-level consistency, and
(iv) standard sirens providing an absolute distance scale independent of both SN calibration and $r_d$,
while also testing modified-gravity signatures through $\DL^{\rm GW}/\DL^{\rm EM}$.

Finally, any scalar--tensor or long-range-force interpretation of late-time anomalies must also remain compatible with local tests of gravity (potentially via screening). This motivates a balanced discovery portfolio that combines cosmological distances, growth/lensing closure, and gravitational-wave propagation constraints with complementary laboratory, astrophysical, and Solar-System gravity tests (see Refs.~\cite{TuryshevChiowYu2024Tetrahedral,Turyshev2025SolarSystemDEDM}).

\section*{Acknowledgments} 
The work described here was carried out at the Jet Propulsion Laboratory, California Institute of Technology, Pasadena, California, under a contract with the National Aeronautics and Space Administration.
 \textcopyright 2026. California Institute of Technology. Government sponsorship acknowledged.

\appendix
\section{$r_d$-independent DESI DR2 $F_{\rm AP}(z)$ data vector and covariance}
\label{app:FAPvector}

For each anisotropic BAO redshift bin $z_i$ with published observables
$(D_{\rm M}(z_i)/r_d,\;D_{\rm H}(z_i)/r_d)$ and covariance $\bm{C}_d$ in the log-data vector
\begin{equation}
\label{eq:d_stat}
\bm{d}\equiv
\bigg(
\ln\!\frac{D_{\rm M}(z_1)}{r_d},\ldots,\ln\!\frac{D_{\rm M}(z_N)}{r_d},
\ln\!\frac{D_{\rm H}(z_1)}{r_d},\ldots,\ln\!\frac{D_{\rm H}(z_N)}{r_d}
\bigg)^T,
\end{equation}
we define the $r_d$-independent ``shape'' vector
\begin{equation}
\bm{y}\equiv \Big(\ln F_{\rm AP}(z_1),\ldots,\ln F_{\rm AP}(z_N)\Big)^T,
\qquad
\bm{y}=\bm{A}\bm{d},\quad \bm{A}=(\bm{I}_N,\,-\bm{I}_N),
\end{equation}
with covariance
\begin{equation}
\label{eq:d_cov}
\bm{C}_y=\bm{A}\bm{C}_d\bm{A}^T.
\end{equation}

The construction in Eqs.~(\ref{eq:d_stat})--(\ref{eq:d_cov}) is general and can be applied directly to any
published anisotropic BAO distance product that reports $(D_{\rm M}/r_d,\;D_{\rm H}/r_d)$ and the associated covariance. Table~\ref{tab:FAPvector} therefore serves as an example (Ly$\alpha$ at $z_{\rm eff}=2.33$); a full multi-bin DR2 $\{F_{\rm AP}(z_i)\}$ vector can be assembled straightforwardly from the public DESI DR2 anisotropic-BAO covariance matrices using Eq.~(\ref{eq:Cy_from_Cd}).

\begin{table*}[th]
\caption{\label{tab:FAPvector}
Example of $r_d$-independent anisotropic-BAO ``shape'' statistic for the DR2 Ly$\alpha$ bin,
$F_{\rm AP}(z)=D_{\rm M}(z)/D_{\rm H}(z)$, derived from published $(D_{\rm M}/r_d,\,D_{\rm H}/r_d)$ values and their covariance. Generated from the DESI DR2 distance products using (\ref{eq:d_stat})--(\ref{eq:d_cov}). 
We show the Ly$\alpha$ bin as an example.}
\setlength{\tabcolsep}{5pt}
\renewcommand{\arraystretch}{1.15}
\begin{tabular}{cccccc}
\hline
Tracer/bin & $z_{\rm eff}$ & $D_{\rm M}/r_d$ & $D_{\rm H}/r_d$ & $\rho_{\rm MH}$ & $F_{\rm AP}\pm\sigma_{\rm stat}\ (\pm\sigma_{\rm sys})$ \\
\hline\hline
Ly$\alpha$ & 2.33 & 38.99 & 8.632 & $-0.457$ & $4.518\pm0.095\ (\pm0.019)$ \\
\hline
\end{tabular}
\end{table*}


\begin{thebibliography}{91}%
\makeatletter
\providecommand \@ifxundefined [1]{%
 \@ifx{#1\undefined}
}%
\providecommand \@ifnum [1]{%
 \ifnum #1\expandafter \@firstoftwo
 \else \expandafter \@secondoftwo
 \fi
}%
\providecommand \@ifx [1]{%
 \ifx #1\expandafter \@firstoftwo
 \else \expandafter \@secondoftwo
 \fi
}%
\providecommand \natexlab [1]{#1}%
\providecommand \enquote  [1]{``#1''}%
\providecommand \bibnamefont  [1]{#1}%
\providecommand \bibfnamefont [1]{#1}%
\providecommand \citenamefont [1]{#1}%
\providecommand \href@noop [0]{\@secondoftwo}%
\providecommand \href [0]{\begingroup \@sanitize@url \@href}%
\providecommand \@href[1]{\@@startlink{#1}\@@href}%
\providecommand \@@href[1]{\endgroup#1\@@endlink}%
\providecommand \@sanitize@url [0]{\catcode `\\12\catcode `\$12\catcode
  `\&12\catcode `\#12\catcode `\^12\catcode `\_12\catcode `\%12\relax}%
\providecommand \@@startlink[1]{}%
\providecommand \@@endlink[0]{}%
\providecommand \url  [0]{\begingroup\@sanitize@url \@url }%
\providecommand \@url [1]{\endgroup\@href {#1}{\urlprefix }}%
\providecommand \urlprefix  [0]{URL }%
\providecommand \Eprint [0]{\href }%
\providecommand \doibase [0]{https://doi.org/}%
\providecommand \selectlanguage [0]{\@gobble}%
\providecommand \bibinfo  [0]{\@secondoftwo}%
\providecommand \bibfield  [0]{\@secondoftwo}%
\providecommand \translation [1]{[#1]}%
\providecommand \BibitemOpen [0]{}%
\providecommand \bibitemStop [0]{}%
\providecommand \bibitemNoStop [0]{.\EOS\space}%
\providecommand \EOS [0]{\spacefactor3000\relax}%
\providecommand \BibitemShut  [1]{\csname bibitem#1\endcsname}%
\let\auto@bib@innerbib\@empty
\bibitem [{\citenamefont {Riess}\ \emph {et~al.}(1998)\citenamefont {Riess}
  \emph {et~al.}}]{Riess1998}%
  \BibitemOpen
  \bibfield  {author} {\bibinfo {author} {\bibfnamefont {A.~G.}\ \bibnamefont
  {Riess}} \emph {et~al.},\ }\bibfield  {title} {\bibinfo {title}
  {{Observational Evidence from Supernovae for an Accelerating Universe and a
  Cosmological Constant}},\ }\href {https://doi.org/10.1086/300499} {\bibfield
  {journal} {\bibinfo  {journal} {Astron. J.}\ }\textbf {\bibinfo {volume}
  {116}},\ \bibinfo {pages} {1009} (\bibinfo {year} {1998})}\BibitemShut
  {NoStop}%
\bibitem [{\citenamefont {Perlmutter}\ \emph {et~al.}(1999)\citenamefont
  {Perlmutter} \emph {et~al.}}]{Perlmutter1999}%
  \BibitemOpen
  \bibfield  {author} {\bibinfo {author} {\bibfnamefont {S.}~\bibnamefont
  {Perlmutter}} \emph {et~al.},\ }\bibfield  {title} {\bibinfo {title}
  {{Measurements of $\Omega$ and $\Lambda$ from 42 High-Redshift Supernovae}},\
  }\href {https://doi.org/10.1086/307221} {\bibfield  {journal} {\bibinfo
  {journal} {Astrophys. J.}\ }\textbf {\bibinfo {volume} {517}},\ \bibinfo
  {pages} {565} (\bibinfo {year} {1999})}\BibitemShut {NoStop}%
\bibitem [{\citenamefont {Weinberg}(1989)}]{Weinberg1989CC}%
  \BibitemOpen
  \bibfield  {author} {\bibinfo {author} {\bibfnamefont {S.}~\bibnamefont
  {Weinberg}},\ }\bibfield  {title} {\bibinfo {title} {{The Cosmological
  Constant Problem}},\ }\href {https://doi.org/10.1103/RevModPhys.61.1}
  {\bibfield  {journal} {\bibinfo  {journal} {Rev. Mod. Phys.}\ }\textbf
  {\bibinfo {volume} {61}},\ \bibinfo {pages} {1} (\bibinfo {year}
  {1989})}\BibitemShut {NoStop}%
\bibitem [{\citenamefont {Carroll}(2001)}]{Carroll2001CCReview}%
  \BibitemOpen
  \bibfield  {author} {\bibinfo {author} {\bibfnamefont {S.~M.}\ \bibnamefont
  {Carroll}},\ }\bibfield  {title} {\bibinfo {title} {{The Cosmological
  Constant}},\ }\href {https://doi.org/10.12942/lrr-2001-1} {\bibfield
  {journal} {\bibinfo  {journal} {Living Rev. Relativ.}\ }\textbf {\bibinfo
  {volume} {4}},\ \bibinfo {pages} {1} (\bibinfo {year} {2001})}\BibitemShut
  {NoStop}%
\bibitem [{\citenamefont {Peebles}\ and\ \citenamefont
  {Ratra}(2003)}]{PeeblesRatra2003DE}%
  \BibitemOpen
  \bibfield  {author} {\bibinfo {author} {\bibfnamefont {P.~J.~E.}\
  \bibnamefont {Peebles}}\ and\ \bibinfo {author} {\bibfnamefont
  {B.}~\bibnamefont {Ratra}},\ }\bibfield  {title} {\bibinfo {title} {{The
  Cosmological Constant and Dark Energy}},\ }\href
  {https://doi.org/10.1103/RevModPhys.75.559} {\bibfield  {journal} {\bibinfo
  {journal} {Rev. Mod. Phys.}\ }\textbf {\bibinfo {volume} {75}},\ \bibinfo
  {pages} {559} (\bibinfo {year} {2003})}\BibitemShut {NoStop}%
\bibitem [{\citenamefont {Frieman}\ \emph {et~al.}(2008)\citenamefont
  {Frieman}, \citenamefont {Turner},\ and\ \citenamefont
  {Huterer}}]{Frieman2008DEReview}%
  \BibitemOpen
  \bibfield  {author} {\bibinfo {author} {\bibfnamefont {J.~A.}\ \bibnamefont
  {Frieman}}, \bibinfo {author} {\bibfnamefont {M.~S.}\ \bibnamefont
  {Turner}},\ and\ \bibinfo {author} {\bibfnamefont {D.}~\bibnamefont
  {Huterer}},\ }\bibfield  {title} {\bibinfo {title} {{Dark Energy and the
  Accelerating Universe}},\ }\href
  {https://doi.org/10.1146/annurev.astro.46.060407.145243} {\bibfield
  {journal} {\bibinfo  {journal} {Annu. Rev. Astron. Astrophys.}\ }\textbf
  {\bibinfo {volume} {46}},\ \bibinfo {pages} {385} (\bibinfo {year}
  {2008})}\BibitemShut {NoStop}%
\bibitem [{\citenamefont {{DESI
  Collaboration}}(2025{\natexlab{a}})}]{DESIDR2Key}%
  \BibitemOpen
  \bibfield  {author} {\bibinfo {author} {\bibnamefont {{DESI
  Collaboration}}},\ }\bibfield  {title} {\bibinfo {title} {{DESI DR2 Results.
  II. Measurements of Baryon Acoustic Oscillations and Cosmological
  Constraints}},\ }\href {https://doi.org/10.1103/tr6y-kpc6} {\bibfield
  {journal} {\bibinfo  {journal} {Phys. Rev. D}\ }\textbf {\bibinfo {volume}
  {112}},\ \bibinfo {pages} {083515} (\bibinfo {year}
  {2025}{\natexlab{a}})}\BibitemShut {NoStop}%
\bibitem [{\citenamefont {{DESI
  Collaboration}}(2025{\natexlab{b}})}]{DESIDR2LyA}%
  \BibitemOpen
  \bibfield  {author} {\bibinfo {author} {\bibnamefont {{DESI
  Collaboration}}},\ }\bibfield  {title} {\bibinfo {title} {{DESI DR2 Results.
  I. Baryon Acoustic Oscillations from the Lyman-$\alpha$ Forest}},\ }\href
  {https://doi.org/10.1103/2wwn-xjm5} {\bibfield  {journal} {\bibinfo
  {journal} {Phys. Rev. D}\ }\textbf {\bibinfo {volume} {112}},\ \bibinfo
  {pages} {083514} (\bibinfo {year} {2025}{\natexlab{b}})}\BibitemShut
  {NoStop}%
\bibitem [{\citenamefont {{DESI
  Collaboration}}(2025{\natexlab{c}})}]{DESIDR2BAOValidation}%
  \BibitemOpen
  \bibfield  {author} {\bibinfo {author} {\bibnamefont {{DESI
  Collaboration}}},\ }\bibfield  {title} {\bibinfo {title} {{Validation of the
  DESI DR2 Baryon Acoustic Oscillations Measurements from Galaxies and
  Quasars}},\ }\href {https://doi.org/10.1103/kdys-w8vl} {\bibfield  {journal}
  {\bibinfo  {journal} {Phys. Rev. D}\ }\textbf {\bibinfo {volume} {112}},\
  \bibinfo {pages} {083512} (\bibinfo {year} {2025}{\natexlab{c}})}\BibitemShut
  {NoStop}%
\bibitem [{\citenamefont {{DESI
  Collaboration}}(2025{\natexlab{d}})}]{Lodha2025ExtendedDE}%
  \BibitemOpen
  \bibfield  {author} {\bibinfo {author} {\bibnamefont {{DESI
  Collaboration}}},\ }\bibfield  {title} {\bibinfo {title} {{Extended Dark
  Energy Analysis Using the DESI DR2 BAO Measurements}},\ }\href
  {https://doi.org/10.1103/w4c6-1r5j} {\bibfield  {journal} {\bibinfo
  {journal} {Phys. Rev. D}\ }\textbf {\bibinfo {volume} {112}},\ \bibinfo
  {pages} {083511} (\bibinfo {year} {2025}{\natexlab{d}})}\BibitemShut
  {NoStop}%
\bibitem [{\citenamefont {Park}\ \emph {et~al.}(2024)\citenamefont {Park},
  \citenamefont {de~Cruz~Perez},\ and\ \citenamefont
  {Ratra}}]{ParkdeCruzPerezRatra2024NonDESI}%
  \BibitemOpen
  \bibfield  {author} {\bibinfo {author} {\bibfnamefont {C.-G.}\ \bibnamefont
  {Park}}, \bibinfo {author} {\bibfnamefont {J.}~\bibnamefont
  {de~Cruz~Perez}},\ and\ \bibinfo {author} {\bibfnamefont {B.}~\bibnamefont
  {Ratra}},\ }\bibfield  {title} {\bibinfo {title} {{Using non-DESI Data to
  Confirm and Strengthen the DESI 2024 Spatially-Flat {$w_0w_a$CDM}
  Cosmological Parameterization Result}},\ }\href
  {https://doi.org/10.1103/PhysRevD.110.123533} {\bibfield  {journal} {\bibinfo
   {journal} {Phys. Rev. D}\ }\textbf {\bibinfo {volume} {110}},\ \bibinfo
  {pages} {123533} (\bibinfo {year} {2024})}\BibitemShut {NoStop}%
\bibitem [{\citenamefont {Tada}\ and\ \citenamefont
  {Terada}(2024)}]{TadaTerada2024QuintessentialDESI}%
  \BibitemOpen
  \bibfield  {author} {\bibinfo {author} {\bibfnamefont {Y.}~\bibnamefont
  {Tada}}\ and\ \bibinfo {author} {\bibfnamefont {T.}~\bibnamefont {Terada}},\
  }\bibfield  {title} {\bibinfo {title} {{Quintessential Interpretation of the
  Evolving Dark Energy in Light of DESI}},\ }\href
  {https://doi.org/10.1103/PhysRevD.109.L121305} {\bibfield  {journal}
  {\bibinfo  {journal} {Phys. Rev. D}\ }\textbf {\bibinfo {volume} {109}},\
  \bibinfo {pages} {L121305} (\bibinfo {year} {2024})},\ \Eprint
  {https://arxiv.org/abs/2404.05722} {arXiv:2404.05722 [astro-ph.CO]}
  \BibitemShut {NoStop}%
\bibitem [{\citenamefont {Park}\ and\ \citenamefont
  {Ratra}(2025{\natexlab{a}})}]{ParkRatra2025PhiCDM}%
  \BibitemOpen
  \bibfield  {author} {\bibinfo {author} {\bibfnamefont {C.-G.}\ \bibnamefont
  {Park}}\ and\ \bibinfo {author} {\bibfnamefont {B.}~\bibnamefont {Ratra}},\
  }\href@noop {} {\bibinfo {title} {{Updated Observational Constraints on
  {$\phi$CDM} Dynamical Dark Energy Cosmological Models}}} (\bibinfo {year}
  {2025}{\natexlab{a}}),\ \Eprint {https://arxiv.org/abs/2509.25812}
  {arXiv:2509.25812 [astro-ph.CO]} \BibitemShut {NoStop}%
\bibitem [{\citenamefont {Wang}\ \emph {et~al.}(2026)\citenamefont {Wang},
  \citenamefont {Li}, \citenamefont {Liu},\ and\ \citenamefont
  {Du}}]{WangEtAl2026QuintessenceReconstruction}%
  \BibitemOpen
  \bibfield  {author} {\bibinfo {author} {\bibfnamefont {S.}~\bibnamefont
  {Wang}}, \bibinfo {author} {\bibfnamefont {T.-N.}\ \bibnamefont {Li}},
  \bibinfo {author} {\bibfnamefont {T.}~\bibnamefont {Liu}},\ and\ \bibinfo
  {author} {\bibfnamefont {G.-H.}\ \bibnamefont {Du}},\ }\href@noop {}
  {\bibinfo {title} {{Model-Independent Reconstruction of Quintessence
  Potential and Kinetic Energy from DESI DR2 and Pantheon+ Supernovae}}}
  (\bibinfo {year} {2026}),\ \Eprint {https://arxiv.org/abs/2603.21125}
  {arXiv:2603.21125 [astro-ph.CO]} \BibitemShut {NoStop}%
\bibitem [{\citenamefont {Shlivko}\ and\ \citenamefont
  {Poulin}(2026)}]{ShlivkoPoulin2026OmTugOfWar}%
  \BibitemOpen
  \bibfield  {author} {\bibinfo {author} {\bibfnamefont {D.}~\bibnamefont
  {Shlivko}}\ and\ \bibinfo {author} {\bibfnamefont {V.}~\bibnamefont
  {Poulin}},\ }\href@noop {} {\bibinfo {title} {{Phantom-Crossing Dark Energy
  and the {$\Omega_m$} Tug-of-War}}} (\bibinfo {year} {2026}),\ \Eprint
  {https://arxiv.org/abs/2603.22406} {arXiv:2603.22406 [astro-ph.CO]}
  \BibitemShut {NoStop}%
\bibitem [{\citenamefont {Efstathiou}(2025)}]{Efstathiou2025Systematics}%
  \BibitemOpen
  \bibfield  {author} {\bibinfo {author} {\bibfnamefont {G.}~\bibnamefont
  {Efstathiou}},\ }\bibfield  {title} {\bibinfo {title} {{Evolving dark energy
  or supernovae systematics?}},\ }\href {https://doi.org/10.1093/mnras/staf301}
  {\bibfield  {journal} {\bibinfo  {journal} {Mon. Not. R. Astron. Soc.}\
  }\textbf {\bibinfo {volume} {538}},\ \bibinfo {pages} {875} (\bibinfo {year}
  {2025})}\BibitemShut {NoStop}%
\bibitem [{\citenamefont {Vincenzi}\ \emph {et~al.}(2025)\citenamefont
  {Vincenzi} \emph {et~al.}}]{Vincenzi2025CompareSN}%
  \BibitemOpen
  \bibfield  {author} {\bibinfo {author} {\bibfnamefont {M.}~\bibnamefont
  {Vincenzi}} \emph {et~al.},\ }\bibfield  {title} {\bibinfo {title}
  {{Comparing the DES-SN5YR and Pantheon+ supernova cosmology analyses:
  investigation based on ``evolving dark energy or supernovae
  systematics''?}},\ }\href {https://doi.org/10.1093/mnras/staf943} {\bibfield
  {journal} {\bibinfo  {journal} {Mon. Not. R. Astron. Soc.}\ }\textbf
  {\bibinfo {volume} {541}},\ \bibinfo {pages} {2585} (\bibinfo {year}
  {2025})}\BibitemShut {NoStop}%
\bibitem [{\citenamefont {Popovic}\ \emph
  {et~al.}(2025{\natexlab{a}})\citenamefont {Popovic}, \citenamefont {Shah},
  \citenamefont {Kenworthy}, \citenamefont {Kessler}, \citenamefont {Davis},
  \citenamefont {Goobar}, \citenamefont {Scolnic}, \citenamefont {Vincenzi},
  \citenamefont {Wiseman}, \citenamefont {Chen} \emph
  {et~al.}}]{PopovicEtAl2025DESDovekie}%
  \BibitemOpen
  \bibfield  {author} {\bibinfo {author} {\bibfnamefont {B.}~\bibnamefont
  {Popovic}}, \bibinfo {author} {\bibfnamefont {P.}~\bibnamefont {Shah}},
  \bibinfo {author} {\bibfnamefont {W.~D.}\ \bibnamefont {Kenworthy}}, \bibinfo
  {author} {\bibfnamefont {R.}~\bibnamefont {Kessler}}, \bibinfo {author}
  {\bibfnamefont {T.~M.}\ \bibnamefont {Davis}}, \bibinfo {author}
  {\bibfnamefont {A.}~\bibnamefont {Goobar}}, \bibinfo {author} {\bibfnamefont
  {D.}~\bibnamefont {Scolnic}}, \bibinfo {author} {\bibfnamefont
  {M.}~\bibnamefont {Vincenzi}}, \bibinfo {author} {\bibfnamefont
  {P.}~\bibnamefont {Wiseman}}, \bibinfo {author} {\bibfnamefont
  {R.}~\bibnamefont {Chen}}, \emph {et~al.},\ }\href@noop {} {\bibinfo {title}
  {{The Dark Energy Survey Supernova Program: A Reanalysis Of Cosmology Results
  And Evidence For Evolving Dark Energy With An Updated Type Ia Supernova
  Calibration}}} (\bibinfo {year} {2025}{\natexlab{a}}),\ \Eprint
  {https://arxiv.org/abs/2511.07517} {arXiv:2511.07517 [astro-ph.CO]}
  \BibitemShut {NoStop}%
\bibitem [{\citenamefont {Dhawan}\ \emph {et~al.}(2025)\citenamefont {Dhawan},
  \citenamefont {Popovic},\ and\ \citenamefont {Goobar}}]{Dhawan2025Axis}%
  \BibitemOpen
  \bibfield  {author} {\bibinfo {author} {\bibfnamefont {S.}~\bibnamefont
  {Dhawan}}, \bibinfo {author} {\bibfnamefont {B.}~\bibnamefont {Popovic}},\
  and\ \bibinfo {author} {\bibfnamefont {A.}~\bibnamefont {Goobar}},\
  }\bibfield  {title} {\bibinfo {title} {{The axis of systematic bias in {SN}
  {Ia} cosmology and implications for {DESI} 2024 results}},\ }\href
  {https://doi.org/10.1093/mnras/staf779} {\bibfield  {journal} {\bibinfo
  {journal} {MNRAS}\ }\textbf {\bibinfo {volume} {540}},\ \bibinfo {pages}
  {1626} (\bibinfo {year} {2025})}\BibitemShut {NoStop}%
\bibitem [{\citenamefont {Gialamas}\ \emph {et~al.}(2025)\citenamefont
  {Gialamas}, \citenamefont {H{\"u}tsi}, \citenamefont {Kannike}, \citenamefont
  {Racioppi}, \citenamefont {Raidal}, \citenamefont {Vasar},\ and\
  \citenamefont {Veerm{\"a}e}}]{Gialamas2025DESI2024BAO_local}%
  \BibitemOpen
  \bibfield  {author} {\bibinfo {author} {\bibfnamefont {I.~D.}\ \bibnamefont
  {Gialamas}}, \bibinfo {author} {\bibfnamefont {G.}~\bibnamefont {H{\"u}tsi}},
  \bibinfo {author} {\bibfnamefont {K.}~\bibnamefont {Kannike}}, \bibinfo
  {author} {\bibfnamefont {A.}~\bibnamefont {Racioppi}}, \bibinfo {author}
  {\bibfnamefont {M.}~\bibnamefont {Raidal}}, \bibinfo {author} {\bibfnamefont
  {M.}~\bibnamefont {Vasar}},\ and\ \bibinfo {author} {\bibfnamefont
  {H.}~\bibnamefont {Veerm{\"a}e}},\ }\bibfield  {title} {\bibinfo {title}
  {{Interpreting DESI 2024 BAO: late-time dynamical dark energy or a local
  effect?}},\ }\href {https://doi.org/10.1103/PhysRevD.111.043540} {\bibfield
  {journal} {\bibinfo  {journal} {Phys. Rev. D}\ }\textbf {\bibinfo {volume}
  {111}},\ \bibinfo {pages} {043540} (\bibinfo {year} {2025})}\BibitemShut
  {NoStop}%
\bibitem [{\citenamefont {Abbott}\ \emph {et~al.}(2026)\citenamefont {Abbott}
  \emph {et~al.}}]{DESY6_3x2pt_2026}%
  \BibitemOpen
  \bibfield  {author} {\bibinfo {author} {\bibfnamefont {T.~M.~C.}\
  \bibnamefont {Abbott}} \emph {et~al.} (\bibinfo {collaboration} {DES}),\
  }\href {https://doi.org/10.48550/arXiv.2601.14559} {\bibinfo {title} {{Dark
  Energy Survey Year 6 Results: Cosmological Constraints from Galaxy Clustering
  and Weak Lensing}}} (\bibinfo {year} {2026}),\ \Eprint
  {https://arxiv.org/abs/2601.14559} {arXiv:2601.14559 [astro-ph.CO]}
  \BibitemShut {NoStop}%
\bibitem [{\citenamefont {Louis}\ \emph {et~al.}(2025)\citenamefont {Louis}
  \emph {et~al.}}]{ACTDR6Power2025}%
  \BibitemOpen
  \bibfield  {author} {\bibinfo {author} {\bibfnamefont {T.}~\bibnamefont
  {Louis}} \emph {et~al.},\ }\bibfield  {title} {\bibinfo {title} {{The Atacama
  Cosmology Telescope: DR6 Power Spectra and Cosmological Parameters}},\ }\href
  {https://doi.org/10.1088/1475-7516/2025/11/062} {\bibfield  {journal}
  {\bibinfo  {journal} {J. Cosmol. Astropart. Phys.}\ }\textbf {\bibinfo
  {volume} {2025}}\bibinfo  {number} { (11)},\ \bibinfo {pages}
  {062}}\BibitemShut {NoStop}%
\bibitem [{\citenamefont {Camphuis}\ \emph {et~al.}(2026)\citenamefont
  {Camphuis} \emph {et~al.}}]{SPT3GDR1_2025}%
  \BibitemOpen
\bibfield  {number} {  }\bibfield  {author} {\bibinfo {author} {\bibfnamefont
  {E.}~\bibnamefont {Camphuis}} \emph {et~al.},\ }\bibfield  {title} {\bibinfo
  {title} {{SPT-3G D1: CMB Temperature and Polarization Power Spectra and
  Cosmology from 2019 and 2020 Observations of the SPT-3G Main Field}},\ }\href
  {https://doi.org/10.1103/7wt3-9v2y} {\bibfield  {journal} {\bibinfo
  {journal} {Phys. Rev. D}\ }\textbf {\bibinfo {volume} {113}},\ \bibinfo
  {pages} {083504} (\bibinfo {year} {2026})}\BibitemShut {NoStop}%
\bibitem [{\citenamefont {Wright}\ \emph {et~al.}(2025)\citenamefont {Wright}
  \emph {et~al.}}]{KiDSLegacy2025}%
  \BibitemOpen
  \bibfield  {author} {\bibinfo {author} {\bibfnamefont {A.~H.}\ \bibnamefont
  {Wright}} \emph {et~al.},\ }\bibfield  {title} {\bibinfo {title}
  {{KiDS-Legacy: Cosmological constraints from cosmic shear with the complete
  Kilo-Degree Survey}},\ }\href {https://doi.org/10.1051/0004-6361/202554908}
  {\bibfield  {journal} {\bibinfo  {journal} {Astron. Astrophys.}\ }\textbf
  {\bibinfo {volume} {703}},\ \bibinfo {pages} {A158} (\bibinfo {year}
  {2025})}\BibitemShut {NoStop}%
\bibitem [{\citenamefont {Choppin~de Janvry}\ \emph {et~al.}(2025)\citenamefont
  {Choppin~de Janvry}, \citenamefont {Dai}, \citenamefont {Gontcho A~Gontcho},
  \citenamefont {Seljak},\ and\ \citenamefont {Zhang}}]{HSCY3ClusteringZ2025}%
  \BibitemOpen
  \bibfield  {author} {\bibinfo {author} {\bibfnamefont {J.}~\bibnamefont
  {Choppin~de Janvry}}, \bibinfo {author} {\bibfnamefont {B.}~\bibnamefont
  {Dai}}, \bibinfo {author} {\bibfnamefont {S.}~\bibnamefont {Gontcho
  A~Gontcho}}, \bibinfo {author} {\bibfnamefont {U.}~\bibnamefont {Seljak}},\
  and\ \bibinfo {author} {\bibfnamefont {T.}~\bibnamefont {Zhang}},\ }\href
  {https://doi.org/10.48550/arXiv.2511.18134} {\bibinfo {title} {{Cosmic Shear
  constraints from HSC Year 3 with clustering calibration of the tomographic
  redshift distributions from DESI}}} (\bibinfo {year} {2025}),\ \Eprint
  {https://arxiv.org/abs/2511.18134} {arXiv:2511.18134 [astro-ph.CO]}
  \BibitemShut {NoStop}%
\bibitem [{\citenamefont {{Euclid Collaboration}}(2025)}]{Euclid2025Overview}%
  \BibitemOpen
  \bibfield  {author} {\bibinfo {author} {\bibnamefont {{Euclid
  Collaboration}}},\ }\bibfield  {title} {\bibinfo {title} {{Euclid. I.
  Overview of the Euclid mission}},\ }\href
  {https://doi.org/10.1051/0004-6361/202450810} {\bibfield  {journal} {\bibinfo
   {journal} {Astron. Astrophys.}\ }\textbf {\bibinfo {volume} {697}},\
  \bibinfo {pages} {A1} (\bibinfo {year} {2025})}\BibitemShut {NoStop}%
\bibitem [{\citenamefont {Cao}\ and\ \citenamefont
  {Ratra}(2023)}]{CaoRatra2023LowerZExpansion}%
  \BibitemOpen
  \bibfield  {author} {\bibinfo {author} {\bibfnamefont {S.}~\bibnamefont
  {Cao}}\ and\ \bibinfo {author} {\bibfnamefont {B.}~\bibnamefont {Ratra}},\
  }\bibfield  {title} {\bibinfo {title} {{{$H_0=69.8\pm1.3$ km s$^{-1}$
  Mpc$^{-1}$, {$\Omega_{m0}=0.288\pm0.017$}, and Other Constraints from
  Lower-Redshift, Non-CMB, Expansion-Rate Data}}},\ }\href
  {https://doi.org/10.1103/PhysRevD.107.103521} {\bibfield  {journal} {\bibinfo
   {journal} {Phys. Rev. D}\ }\textbf {\bibinfo {volume} {107}},\ \bibinfo
  {pages} {103521} (\bibinfo {year} {2023})}\BibitemShut {NoStop}%
\bibitem [{\citenamefont {Dong}\ \emph {et~al.}(2023)\citenamefont {Dong},
  \citenamefont {Park}, \citenamefont {Hong}, \citenamefont {Kim},
  \citenamefont {Hwang}, \citenamefont {Park},\ and\ \citenamefont
  {Appleby}}]{DongEtAl2023TomographicAP}%
  \BibitemOpen
  \bibfield  {author} {\bibinfo {author} {\bibfnamefont {F.}~\bibnamefont
  {Dong}}, \bibinfo {author} {\bibfnamefont {C.}~\bibnamefont {Park}}, \bibinfo
  {author} {\bibfnamefont {S.~E.}\ \bibnamefont {Hong}}, \bibinfo {author}
  {\bibfnamefont {J.}~\bibnamefont {Kim}}, \bibinfo {author} {\bibfnamefont
  {H.~S.}\ \bibnamefont {Hwang}}, \bibinfo {author} {\bibfnamefont
  {H.}~\bibnamefont {Park}},\ and\ \bibinfo {author} {\bibfnamefont
  {S.}~\bibnamefont {Appleby}},\ }\bibfield  {title} {\bibinfo {title}
  {{Tomographic Alcock-Paczynski Test with Redshift-Space Correlation Function:
  Evidence for the Dark Energy Equation of State Parameter {$w>-1$}}},\ }\href
  {https://doi.org/10.3847/1538-4357/acd185} {\bibfield  {journal} {\bibinfo
  {journal} {Astrophys. J.}\ }\textbf {\bibinfo {volume} {953}},\ \bibinfo
  {pages} {98} (\bibinfo {year} {2023})}\BibitemShut {NoStop}%
\bibitem [{\citenamefont {Van~Raamsdonk}\ and\ \citenamefont
  {Waddell}(2024)}]{VanRaamsdonkWaddell2023DecreasingDE}%
  \BibitemOpen
  \bibfield  {author} {\bibinfo {author} {\bibfnamefont {M.}~\bibnamefont
  {Van~Raamsdonk}}\ and\ \bibinfo {author} {\bibfnamefont {C.}~\bibnamefont
  {Waddell}},\ }\bibfield  {title} {\bibinfo {title} {{Suggestions of
  Decreasing Dark Energy from Supernova and BAO Data}},\ }\href
  {https://doi.org/10.1088/1475-7516/2024/06/047} {\bibfield  {journal}
  {\bibinfo  {journal} {J. Cosmol. Astropart. Phys.}\ }\textbf {\bibinfo
  {volume} {2024}}\bibinfo  {number} { (06)},\ \bibinfo {pages}
  {047}}\BibitemShut {NoStop}%
\bibitem [{\citenamefont {Rubin}\ \emph
  {et~al.}(2025{\natexlab{a}})\citenamefont {Rubin}, \citenamefont {Aldering},
  \citenamefont {Betoule}, \citenamefont {Fruchter}, \citenamefont {Huang},
  \citenamefont {Kim}, \citenamefont {Lidman}, \citenamefont {Linder},
  \citenamefont {Perlmutter}, \citenamefont {Ruiz-Lapuente},\ and\
  \citenamefont {Suzuki}}]{RubinEtAl2023UnionThroughUNITY}%
  \BibitemOpen
\bibfield  {number} {  }\bibfield  {author} {\bibinfo {author} {\bibfnamefont
  {D.}~\bibnamefont {Rubin}}, \bibinfo {author} {\bibfnamefont
  {G.}~\bibnamefont {Aldering}}, \bibinfo {author} {\bibfnamefont
  {M.}~\bibnamefont {Betoule}}, \bibinfo {author} {\bibfnamefont
  {A.}~\bibnamefont {Fruchter}}, \bibinfo {author} {\bibfnamefont
  {X.}~\bibnamefont {Huang}}, \bibinfo {author} {\bibfnamefont {A.~G.}\
  \bibnamefont {Kim}}, \bibinfo {author} {\bibfnamefont {C.}~\bibnamefont
  {Lidman}}, \bibinfo {author} {\bibfnamefont {E.}~\bibnamefont {Linder}},
  \bibinfo {author} {\bibfnamefont {S.}~\bibnamefont {Perlmutter}}, \bibinfo
  {author} {\bibfnamefont {P.}~\bibnamefont {Ruiz-Lapuente}},\ and\ \bibinfo
  {author} {\bibfnamefont {N.}~\bibnamefont {Suzuki}},\ }\bibfield  {title}
  {\bibinfo {title} {{Union Through UNITY: Cosmology with 2,000 SNe Using a
  Unified Bayesian Framework}},\ }\href
  {https://doi.org/10.3847/1538-4357/adc0a5} {\bibfield  {journal} {\bibinfo
  {journal} {Astrophys. J.}\ }\textbf {\bibinfo {volume} {986}},\ \bibinfo
  {pages} {231} (\bibinfo {year} {2025}{\natexlab{a}})}\BibitemShut {NoStop}%
\bibitem [{\citenamefont {de~Cruz~Perez}\ \emph {et~al.}(2024)\citenamefont
  {de~Cruz~Perez}, \citenamefont {Park},\ and\ \citenamefont
  {Ratra}}]{deCruzPerezParkRatra2024XCDM}%
  \BibitemOpen
  \bibfield  {author} {\bibinfo {author} {\bibfnamefont {J.}~\bibnamefont
  {de~Cruz~Perez}}, \bibinfo {author} {\bibfnamefont {C.-G.}\ \bibnamefont
  {Park}},\ and\ \bibinfo {author} {\bibfnamefont {B.}~\bibnamefont {Ratra}},\
  }\bibfield  {title} {\bibinfo {title} {{Updated Observational Constraints on
  Spatially-Flat and Non-Flat {$\Lambda$CDM} and XCDM Cosmological Models}},\
  }\href {https://doi.org/10.1103/PhysRevD.110.023506} {\bibfield  {journal}
  {\bibinfo  {journal} {Phys. Rev. D}\ }\textbf {\bibinfo {volume} {110}},\
  \bibinfo {pages} {023506} (\bibinfo {year} {2024})}\BibitemShut {NoStop}%
\bibitem [{\citenamefont {{Turyshev}}(2025)}]{Turyshev_FunPhys_2025}%
  \BibitemOpen
  \bibfield  {author} {\bibinfo {author} {\bibfnamefont {S.~G.}\ \bibnamefont
  {{Turyshev}}},\ }\href {https://doi.org/10.48550/arXiv.2512.21445} {\bibinfo
  {title} {{{Fundamental Physics in 2025: Status, Decisive Targets, and Path
  Forward}}}} (\bibinfo {year} {2025}),\ \Eprint
  {https://arxiv.org/abs/2512.21445} {arXiv:2512.21445 [gr-qc]} \BibitemShut
  {NoStop}%
\bibitem [{\citenamefont {Alcock}\ and\ \citenamefont
  {Paczynski}(1979)}]{AlcockPaczynski1979}%
  \BibitemOpen
  \bibfield  {author} {\bibinfo {author} {\bibfnamefont {C.}~\bibnamefont
  {Alcock}}\ and\ \bibinfo {author} {\bibfnamefont {B.}~\bibnamefont
  {Paczynski}},\ }\bibfield  {title} {\bibinfo {title} {{An evolution free test
  for non-zero cosmological constant}},\ }\href
  {https://doi.org/10.1038/281358a0} {\bibfield  {journal} {\bibinfo  {journal}
  {Nature}\ }\textbf {\bibinfo {volume} {281}},\ \bibinfo {pages} {358}
  (\bibinfo {year} {1979})}\BibitemShut {NoStop}%
\bibitem [{\citenamefont {Aghanim}\ \emph {et~al.}(2020)\citenamefont {Aghanim}
  \emph {et~al.}}]{Planck2018VI}%
  \BibitemOpen
  \bibfield  {author} {\bibinfo {author} {\bibfnamefont {N.}~\bibnamefont
  {Aghanim}} \emph {et~al.} (\bibinfo {collaboration} {Planck}),\ }\bibfield
  {title} {\bibinfo {title} {{Planck 2018 results. {VI}. Cosmological
  parameters}},\ }\href {https://doi.org/10.1051/0004-6361/201833910}
  {\bibfield  {journal} {\bibinfo  {journal} {Astron. Astrophys.}\ }\textbf
  {\bibinfo {volume} {641}},\ \bibinfo {pages} {A6} (\bibinfo {year}
  {2020})}\BibitemShut {NoStop}%
\bibitem [{\citenamefont {Sahni}\ \emph {et~al.}(2008)\citenamefont {Sahni},
  \citenamefont {Shafieloo},\ and\ \citenamefont {Starobinsky}}]{Sahni2008Om}%
  \BibitemOpen
  \bibfield  {author} {\bibinfo {author} {\bibfnamefont {V.}~\bibnamefont
  {Sahni}}, \bibinfo {author} {\bibfnamefont {A.}~\bibnamefont {Shafieloo}},\
  and\ \bibinfo {author} {\bibfnamefont {A.~A.}\ \bibnamefont {Starobinsky}},\
  }\bibfield  {title} {\bibinfo {title} {{Two New Diagnostics of Dark
  Energy}},\ }\href {https://doi.org/10.1103/PhysRevD.78.103502} {\bibfield
  {journal} {\bibinfo  {journal} {Phys. Rev. D}\ }\textbf {\bibinfo {volume}
  {78}},\ \bibinfo {pages} {103502} (\bibinfo {year} {2008})}\BibitemShut
  {NoStop}%
\bibitem [{\citenamefont {Brout}\ \emph {et~al.}(2022)\citenamefont {Brout}
  \emph {et~al.}}]{Brout2022PantheonPlus}%
  \BibitemOpen
  \bibfield  {author} {\bibinfo {author} {\bibfnamefont {D.}~\bibnamefont
  {Brout}} \emph {et~al.},\ }\bibfield  {title} {\bibinfo {title} {{The
  Pantheon+ Analysis: Cosmological Constraints}},\ }\href
  {https://doi.org/10.3847/1538-4357/ac8e04} {\bibfield  {journal} {\bibinfo
  {journal} {Astrophys. J.}\ }\textbf {\bibinfo {volume} {938}},\ \bibinfo
  {pages} {110} (\bibinfo {year} {2022})}\BibitemShut {NoStop}%
\bibitem [{\citenamefont {Scolnic}\ \emph {et~al.}(2022)\citenamefont {Scolnic}
  \emph {et~al.}}]{Scolnic2022PantheonPlusData}%
  \BibitemOpen
  \bibfield  {author} {\bibinfo {author} {\bibfnamefont {D.}~\bibnamefont
  {Scolnic}} \emph {et~al.},\ }\bibfield  {title} {\bibinfo {title} {{The
  Pantheon+ Type Ia Supernova Sample: The Full Data Set and Light-curve
  Release}},\ }\href {https://doi.org/10.3847/1538-4357/ac8b7a} {\bibfield
  {journal} {\bibinfo  {journal} {Astrophys. J.}\ }\textbf {\bibinfo {volume}
  {938}},\ \bibinfo {pages} {113} (\bibinfo {year} {2022})}\BibitemShut
  {NoStop}%
\bibitem [{\citenamefont {Rubin}\ \emph
  {et~al.}(2025{\natexlab{b}})\citenamefont {Rubin} \emph
  {et~al.}}]{Rubin2025Union3}%
  \BibitemOpen
  \bibfield  {author} {\bibinfo {author} {\bibfnamefont {D.}~\bibnamefont
  {Rubin}} \emph {et~al.},\ }\bibfield  {title} {\bibinfo {title} {{Union
  Through {UNITY}: Cosmology with 2000 SNe Using a Unified Bayesian
  Framework}},\ }\href {https://doi.org/10.3847/1538-4357/adc0a5} {\bibfield
  {journal} {\bibinfo  {journal} {Astrophys. J.}\ }\textbf {\bibinfo {volume}
  {986}},\ \bibinfo {pages} {231} (\bibinfo {year}
  {2025}{\natexlab{b}})}\BibitemShut {NoStop}%
\bibitem [{\citenamefont {Abbott}\ \emph {et~al.}(2024)\citenamefont {Abbott}
  \emph {et~al.}}]{Abbott2024DESY5SN}%
  \BibitemOpen
  \bibfield  {author} {\bibinfo {author} {\bibfnamefont {T.~M.~C.}\
  \bibnamefont {Abbott}} \emph {et~al.} (\bibinfo {collaboration} {{DES}}),\
  }\bibfield  {title} {\bibinfo {title} {{The Dark Energy Survey Supernova
  Program: Cosmology Results with $\sim$1500 New High-redshift Type Ia
  Supernovae Using the Full 5-year Data Set}},\ }\href
  {https://doi.org/10.3847/2041-8213/ad6f9f} {\bibfield  {journal} {\bibinfo
  {journal} {Astrophys. J. Lett.}\ }\textbf {\bibinfo {volume} {973}},\
  \bibinfo {pages} {L14} (\bibinfo {year} {2024})}\BibitemShut {NoStop}%
\bibitem [{\citenamefont {Ballinger}\ \emph {et~al.}(1996)\citenamefont
  {Ballinger}, \citenamefont {Peacock},\ and\ \citenamefont
  {Heavens}}]{Ballinger1996}%
  \BibitemOpen
  \bibfield  {author} {\bibinfo {author} {\bibfnamefont {W.~E.}\ \bibnamefont
  {Ballinger}}, \bibinfo {author} {\bibfnamefont {J.~A.}\ \bibnamefont
  {Peacock}},\ and\ \bibinfo {author} {\bibfnamefont {A.~F.}\ \bibnamefont
  {Heavens}},\ }\bibfield  {title} {\bibinfo {title} {{Measuring the
  cosmological constant with redshift surveys}},\ }\href
  {https://doi.org/10.1093/mnras/282.3.877} {\bibfield  {journal} {\bibinfo
  {journal} {Mon. Not. R. Astron. Soc.}\ }\textbf {\bibinfo {volume} {282}},\
  \bibinfo {pages} {877} (\bibinfo {year} {1996})}\BibitemShut {NoStop}%
\bibitem [{\citenamefont {{{\'O} Colg{\'a}in}}\ \emph
  {et~al.}(2026)\citenamefont {{{\'O} Colg{\'a}in}}, \citenamefont
  {{Dainotti}}, \citenamefont {{Capozziello}}, \citenamefont {{Pourojaghi}},
  \citenamefont {{Sheikh-Jabbari}},\ and\ \citenamefont
  {{Stojkovic}}}]{OColgain:2024DESIconfirmLCDM}%
  \BibitemOpen
  \bibfield  {author} {\bibinfo {author} {\bibfnamefont {E.}~\bibnamefont
  {{{\'O} Colg{\'a}in}}}, \bibinfo {author} {\bibfnamefont {M.~G.}\
  \bibnamefont {{Dainotti}}}, \bibinfo {author} {\bibfnamefont
  {S.}~\bibnamefont {{Capozziello}}}, \bibinfo {author} {\bibfnamefont
  {S.}~\bibnamefont {{Pourojaghi}}}, \bibinfo {author} {\bibfnamefont {M.~M.}\
  \bibnamefont {{Sheikh-Jabbari}}},\ and\ \bibinfo {author} {\bibfnamefont
  {D.}~\bibnamefont {{Stojkovic}}},\ }\bibfield  {title} {\bibinfo {title}
  {{{Does DESI 2024 confirm {\ensuremath{\Lambda}}CDM?}}},\ }\href
  {https://doi.org/10.1016/j.jheap.2025.100428} {\bibfield  {journal} {\bibinfo
   {journal} {J. High Energy Astrophys.}\ }\textbf {\bibinfo {volume} {49}},\
  \bibinfo {eid} {100428} (\bibinfo {year} {2026})}\BibitemShut {NoStop}%
\bibitem [{\citenamefont {{\'O}~Colg{\'a}in}\ \emph {et~al.}(2025)\citenamefont
  {{\'O}~Colg{\'a}in}, \citenamefont {Pourojaghi}, \citenamefont
  {Sheikh-Jabbari},\ and\ \citenamefont
  {Yin}}]{OColgain:2025HowMuchEvolvedDR1}%
  \BibitemOpen
  \bibfield  {author} {\bibinfo {author} {\bibfnamefont {E.}~\bibnamefont
  {{\'O}~Colg{\'a}in}}, \bibinfo {author} {\bibfnamefont {N.}~\bibnamefont
  {Pourojaghi}}, \bibinfo {author} {\bibfnamefont {M.~M.}\ \bibnamefont
  {Sheikh-Jabbari}},\ and\ \bibinfo {author} {\bibfnamefont {G.}~\bibnamefont
  {Yin}},\ }\href@noop {} {\bibinfo {title} {{How much has DESI dark energy
  evolved since DR1?}}} (\bibinfo {year} {2025}),\ \Eprint
  {https://arxiv.org/abs/2504.04417} {arXiv:2504.04417 [astro-ph.CO]}
  \BibitemShut {NoStop}%
\bibitem [{\citenamefont {Park}\ \emph {et~al.}(2025)\citenamefont {Park},
  \citenamefont {de~Cruz~Perez},\ and\ \citenamefont
  {Ratra}}]{ParkdeCruzPerezRatra2024AL}%
  \BibitemOpen
  \bibfield  {author} {\bibinfo {author} {\bibfnamefont {C.-G.}\ \bibnamefont
  {Park}}, \bibinfo {author} {\bibfnamefont {J.}~\bibnamefont
  {de~Cruz~Perez}},\ and\ \bibinfo {author} {\bibfnamefont {B.}~\bibnamefont
  {Ratra}},\ }\bibfield  {title} {\bibinfo {title} {{Is the {$w_0w_a$CDM}
  Cosmological Parameterization Evidence for Dark Energy Dynamics Partially
  Caused by the Excess Smoothing of Planck CMB Anisotropy Data?}},\ }\href
  {https://doi.org/10.1142/S0218271825500580} {\bibfield  {journal} {\bibinfo
  {journal} {Int. J. Mod. Phys. D}\ }\textbf {\bibinfo {volume} {34}},\
  \bibinfo {pages} {2550058} (\bibinfo {year} {2025})},\ \Eprint
  {https://arxiv.org/abs/2410.13627} {arXiv:2410.13627 [astro-ph.CO]}
  \BibitemShut {NoStop}%
\bibitem [{\citenamefont {Roy~Choudhury}\ and\ \citenamefont
  {Okumura}(2024)}]{RoyChoudhuryOkumura2024ExtendedParams}%
  \BibitemOpen
  \bibfield  {author} {\bibinfo {author} {\bibfnamefont {S.}~\bibnamefont
  {Roy~Choudhury}}\ and\ \bibinfo {author} {\bibfnamefont {T.}~\bibnamefont
  {Okumura}},\ }\bibfield  {title} {\bibinfo {title} {{Updated Cosmological
  Constraints in Extended Parameter Space with Planck PR4, DESI Baryon Acoustic
  Oscillations, and Supernovae: Dynamical Dark Energy, Neutrino Masses, Lensing
  Anomaly, and the Hubble Tension}},\ }\href
  {https://doi.org/10.3847/2041-8213/ad8c26} {\bibfield  {journal} {\bibinfo
  {journal} {Astrophys. J. Lett.}\ }\textbf {\bibinfo {volume} {976}},\
  \bibinfo {pages} {L11} (\bibinfo {year} {2024})}\BibitemShut {NoStop}%
\bibitem [{\citenamefont
  {Roy~Choudhury}(2025)}]{RoyChoudhury2025DESIDR2Extended}%
  \BibitemOpen
  \bibfield  {author} {\bibinfo {author} {\bibfnamefont {S.}~\bibnamefont
  {Roy~Choudhury}},\ }\bibfield  {title} {\bibinfo {title} {{Cosmology in
  Extended Parameter Space with DESI Data Release 2 Baryon Acoustic
  Oscillations: A 2{$\sigma$}+ Detection of Nonzero Neutrino Masses with an
  Update on Dynamical Dark Energy and Lensing Anomaly}},\ }\href
  {https://doi.org/10.3847/2041-8213/ade1cc} {\bibfield  {journal} {\bibinfo
  {journal} {Astrophys. J. Lett.}\ }\textbf {\bibinfo {volume} {986}},\
  \bibinfo {pages} {L31} (\bibinfo {year} {2025})}\BibitemShut {NoStop}%
\bibitem [{\citenamefont {Roy~Choudhury}\ \emph {et~al.}(2025)\citenamefont
  {Roy~Choudhury}, \citenamefont {Okumura},\ and\ \citenamefont
  {Umetsu}}]{RoyChoudhuryOkumuraUmetsu2025NPDDE}%
  \BibitemOpen
  \bibfield  {author} {\bibinfo {author} {\bibfnamefont {S.}~\bibnamefont
  {Roy~Choudhury}}, \bibinfo {author} {\bibfnamefont {T.}~\bibnamefont
  {Okumura}},\ and\ \bibinfo {author} {\bibfnamefont {K.}~\bibnamefont
  {Umetsu}},\ }\bibfield  {title} {\bibinfo {title} {{Cosmological Constraints
  on Nonphantom Dynamical Dark Energy with DESI Data Release 2 Baryon Acoustic
  Oscillations: A 3{$\sigma$}+ Lensing Anomaly}},\ }\href
  {https://doi.org/10.3847/2041-8213/ae1a64} {\bibfield  {journal} {\bibinfo
  {journal} {Astrophys. J. Lett.}\ }\textbf {\bibinfo {volume} {994}},\
  \bibinfo {pages} {L26} (\bibinfo {year} {2025})},\ \Eprint
  {https://arxiv.org/abs/2509.26144} {arXiv:2509.26144 [astro-ph.CO]}
  \BibitemShut {NoStop}%
\bibitem [{\citenamefont {Park}\ and\ \citenamefont
  {Ratra}(2025{\natexlab{b}})}]{ParkRatra2025ExcessSmoothing}%
  \BibitemOpen
  \bibfield  {author} {\bibinfo {author} {\bibfnamefont {C.-G.}\ \bibnamefont
  {Park}}\ and\ \bibinfo {author} {\bibfnamefont {B.}~\bibnamefont {Ratra}},\
  }\href@noop {} {\bibinfo {title} {{Is Excess Smoothing of Planck CMB
  Anisotropy Data Partially Responsible for Evidence for Dark Energy Dynamics
  in Other {$w(z)$}CDM Parametrizations?}}} (\bibinfo {year}
  {2025}{\natexlab{b}}),\ \Eprint {https://arxiv.org/abs/2501.03480}
  {arXiv:2501.03480 [astro-ph.CO]} \BibitemShut {NoStop}%
\bibitem [{\citenamefont {Kaiser}(1987)}]{Kaiser1987}%
  \BibitemOpen
  \bibfield  {author} {\bibinfo {author} {\bibfnamefont {N.}~\bibnamefont
  {Kaiser}},\ }\bibfield  {title} {\bibinfo {title} {{Clustering in Real Space
  and in Redshift Space}},\ }\href {https://doi.org/10.1093/mnras/227.1.1}
  {\bibfield  {journal} {\bibinfo  {journal} {MNRAS}\ }\textbf {\bibinfo
  {volume} {227}},\ \bibinfo {pages} {1} (\bibinfo {year} {1987})}\BibitemShut
  {NoStop}%
\bibitem [{\citenamefont {Hamilton}(1998)}]{Hamilton1998}%
  \BibitemOpen
  \bibfield  {author} {\bibinfo {author} {\bibfnamefont {A.~J.~S.}\
  \bibnamefont {Hamilton}},\ }\bibfield  {title} {\bibinfo {title} {{Linear
  Redshift Distortions: A Review}},\ }in\ \href
  {https://doi.org/10.1007/978-94-011-4960-0_17} {\emph {\bibinfo {booktitle}
  {The Evolving Universe}}}\ (\bibinfo  {publisher} {Kluwer Academic},\
  \bibinfo {year} {1998})\ pp.\ \bibinfo {pages} {185--275}\BibitemShut
  {NoStop}%
\bibitem [{\citenamefont {Limber}(1953)}]{Limber1953}%
  \BibitemOpen
  \bibfield  {author} {\bibinfo {author} {\bibfnamefont {D.~N.}\ \bibnamefont
  {Limber}},\ }\bibfield  {title} {\bibinfo {title} {{The Analysis of Counts of
  the Extragalactic Nebulae in Terms of a Fluctuating Density Field}},\ }\href
  {https://doi.org/10.1086/145672} {\bibfield  {journal} {\bibinfo  {journal}
  {Astrophys. J.}\ }\textbf {\bibinfo {volume} {117}},\ \bibinfo {pages} {134}
  (\bibinfo {year} {1953})}\BibitemShut {NoStop}%
\bibitem [{\citenamefont {LoVerde}\ and\ \citenamefont
  {Afshordi}(2008)}]{LoVerdeAfshordi2008}%
  \BibitemOpen
  \bibfield  {author} {\bibinfo {author} {\bibfnamefont {M.}~\bibnamefont
  {LoVerde}}\ and\ \bibinfo {author} {\bibfnamefont {N.}~\bibnamefont
  {Afshordi}},\ }\bibfield  {title} {\bibinfo {title} {{Extended Limber
  Approximation}},\ }\href {https://doi.org/10.1103/PhysRevD.78.123506}
  {\bibfield  {journal} {\bibinfo  {journal} {Phys. Rev. D}\ }\textbf {\bibinfo
  {volume} {78}},\ \bibinfo {pages} {123506} (\bibinfo {year}
  {2008})}\BibitemShut {NoStop}%
\bibitem [{\citenamefont {Gubitosi}\ \emph {et~al.}(2013)\citenamefont
  {Gubitosi}, \citenamefont {Piazza},\ and\ \citenamefont
  {Vernizzi}}]{Gubitosi2013EFT}%
  \BibitemOpen
  \bibfield  {author} {\bibinfo {author} {\bibfnamefont {G.}~\bibnamefont
  {Gubitosi}}, \bibinfo {author} {\bibfnamefont {F.}~\bibnamefont {Piazza}},\
  and\ \bibinfo {author} {\bibfnamefont {F.}~\bibnamefont {Vernizzi}},\
  }\bibfield  {title} {\bibinfo {title} {{The Effective Field Theory of Dark
  Energy}},\ }\href {https://doi.org/10.1088/1475-7516/2013/02/032} {\bibfield
  {journal} {\bibinfo  {journal} {J. Cosmol. Astropart. Phys.}\ }\textbf
  {\bibinfo {volume} {2013}}\bibinfo  {number} { (02)},\ \bibinfo {pages}
  {032}}\BibitemShut {NoStop}%
\bibitem [{\citenamefont {Bellini}\ and\ \citenamefont
  {Sawicki}(2014)}]{BelliniSawicki2014Alpha}%
  \BibitemOpen
\bibfield  {number} {  }\bibfield  {author} {\bibinfo {author} {\bibfnamefont
  {E.}~\bibnamefont {Bellini}}\ and\ \bibinfo {author} {\bibfnamefont
  {I.}~\bibnamefont {Sawicki}},\ }\bibfield  {title} {\bibinfo {title}
  {{Maximal Freedom at Minimum Cost: Linear Large-Scale Structure in General
  Modifications of Gravity}},\ }\href
  {https://doi.org/10.1088/1475-7516/2014/07/050} {\bibfield  {journal}
  {\bibinfo  {journal} {J. Cosmol. Astropart. Phys.}\ }\textbf {\bibinfo
  {volume} {2014}}\bibinfo  {number} { (07)},\ \bibinfo {pages}
  {050}}\BibitemShut {NoStop}%
\bibitem [{\citenamefont {Clifton}\ \emph {et~al.}(2012)\citenamefont
  {Clifton}, \citenamefont {Ferreira}, \citenamefont {Padilla},\ and\
  \citenamefont {Skordis}}]{Clifton2012MG}%
  \BibitemOpen
\bibfield  {number} {  }\bibfield  {author} {\bibinfo {author} {\bibfnamefont
  {T.}~\bibnamefont {Clifton}}, \bibinfo {author} {\bibfnamefont {P.~G.}\
  \bibnamefont {Ferreira}}, \bibinfo {author} {\bibfnamefont {A.}~\bibnamefont
  {Padilla}},\ and\ \bibinfo {author} {\bibfnamefont {C.}~\bibnamefont
  {Skordis}},\ }\bibfield  {title} {\bibinfo {title} {{Modified Gravity and
  Cosmology}},\ }\href {https://doi.org/10.1016/j.physrep.2012.01.001}
  {\bibfield  {journal} {\bibinfo  {journal} {Phys. Rep.}\ }\textbf {\bibinfo
  {volume} {513}},\ \bibinfo {pages} {1} (\bibinfo {year} {2012})}\BibitemShut
  {NoStop}%
\bibitem [{\citenamefont {Joyce}\ \emph {et~al.}(2015)\citenamefont {Joyce},
  \citenamefont {Jain}, \citenamefont {Khoury},\ and\ \citenamefont
  {Trodden}}]{Joyce2015MGReview}%
  \BibitemOpen
  \bibfield  {author} {\bibinfo {author} {\bibfnamefont {A.}~\bibnamefont
  {Joyce}}, \bibinfo {author} {\bibfnamefont {B.}~\bibnamefont {Jain}},
  \bibinfo {author} {\bibfnamefont {J.}~\bibnamefont {Khoury}},\ and\ \bibinfo
  {author} {\bibfnamefont {M.}~\bibnamefont {Trodden}},\ }\bibfield  {title}
  {\bibinfo {title} {{Beyond the Cosmological Standard Model}},\ }\href
  {https://doi.org/10.1016/j.physrep.2014.12.002} {\bibfield  {journal}
  {\bibinfo  {journal} {Phys. Rep.}\ }\textbf {\bibinfo {volume} {568}},\
  \bibinfo {pages} {1} (\bibinfo {year} {2015})}\BibitemShut {NoStop}%
\bibitem [{\citenamefont {Abbott}\ \emph
  {et~al.}(2017{\natexlab{a}})\citenamefont {Abbott} \emph
  {et~al.}}]{Abbott2017GW170817}%
  \BibitemOpen
  \bibfield  {author} {\bibinfo {author} {\bibfnamefont {B.~P.}\ \bibnamefont
  {Abbott}} \emph {et~al.} (\bibinfo {collaboration} {LIGO Scientific
  Collaboration and Virgo Collaboration}),\ }\bibfield  {title} {\bibinfo
  {title} {{Multi-messenger Observations of a Binary Neutron Star Merger}},\
  }\href {https://doi.org/10.3847/2041-8213/aa91c9} {\bibfield  {journal}
  {\bibinfo  {journal} {Astrophys. J. Lett.}\ }\textbf {\bibinfo {volume}
  {848}},\ \bibinfo {pages} {L12} (\bibinfo {year}
  {2017}{\natexlab{a}})}\BibitemShut {NoStop}%
\bibitem [{\citenamefont {Baker}\ \emph {et~al.}(2017)\citenamefont {Baker}
  \emph {et~al.}}]{Baker2017GWConstraints}%
  \BibitemOpen
  \bibfield  {author} {\bibinfo {author} {\bibfnamefont {T.}~\bibnamefont
  {Baker}} \emph {et~al.},\ }\bibfield  {title} {\bibinfo {title} {{Strong
  Constraints on Cosmological Gravity from GW170817 and GRB 170817A}},\ }\href
  {https://doi.org/10.1103/PhysRevLett.119.251301} {\bibfield  {journal}
  {\bibinfo  {journal} {Phys. Rev. Lett.}\ }\textbf {\bibinfo {volume} {119}},\
  \bibinfo {pages} {251301} (\bibinfo {year} {2017})}\BibitemShut {NoStop}%
\bibitem [{\citenamefont {Wang}\ and\ \citenamefont
  {Steinhardt}(1998)}]{WangSteinhardt1998}%
  \BibitemOpen
  \bibfield  {author} {\bibinfo {author} {\bibfnamefont {L.}~\bibnamefont
  {Wang}}\ and\ \bibinfo {author} {\bibfnamefont {P.~J.}\ \bibnamefont
  {Steinhardt}},\ }\bibfield  {title} {\bibinfo {title} {{Cluster Abundance
  Constraints on Quintessence Models}},\ }\href
  {https://doi.org/10.1086/306436} {\bibfield  {journal} {\bibinfo  {journal}
  {Astrophys. J.}\ }\textbf {\bibinfo {volume} {508}},\ \bibinfo {pages} {483}
  (\bibinfo {year} {1998})}\BibitemShut {NoStop}%
\bibitem [{\citenamefont {Linder}(2005)}]{Linder2005Growth}%
  \BibitemOpen
  \bibfield  {author} {\bibinfo {author} {\bibfnamefont {E.~V.}\ \bibnamefont
  {Linder}},\ }\bibfield  {title} {\bibinfo {title} {{Cosmic Growth History and
  Expansion History}},\ }\href {https://doi.org/10.1103/PhysRevD.72.043529}
  {\bibfield  {journal} {\bibinfo  {journal} {Phys. Rev. D}\ }\textbf {\bibinfo
  {volume} {72}},\ \bibinfo {pages} {043529} (\bibinfo {year}
  {2005})}\BibitemShut {NoStop}%
\bibitem [{\citenamefont {Schutz}(1986)}]{Schutz1986}%
  \BibitemOpen
  \bibfield  {author} {\bibinfo {author} {\bibfnamefont {B.~F.}\ \bibnamefont
  {Schutz}},\ }\bibfield  {title} {\bibinfo {title} {{Determining the Hubble
  Constant from Gravitational Wave Observations}},\ }\href
  {https://doi.org/10.1038/323310a0} {\bibfield  {journal} {\bibinfo  {journal}
  {Nature}\ }\textbf {\bibinfo {volume} {323}},\ \bibinfo {pages} {310}
  (\bibinfo {year} {1986})}\BibitemShut {NoStop}%
\bibitem [{\citenamefont {Abbott}\ \emph
  {et~al.}(2017{\natexlab{b}})\citenamefont {Abbott} \emph
  {et~al.}}]{Abbott2017H0Siren}%
  \BibitemOpen
  \bibfield  {author} {\bibinfo {author} {\bibfnamefont {B.~P.}\ \bibnamefont
  {Abbott}} \emph {et~al.} (\bibinfo {collaboration} {{LIGO Scientific
  Collaboration and Virgo Collaboration}}),\ }\bibfield  {title} {\bibinfo
  {title} {{A gravitational-wave standard siren measurement of the {Hubble}
  constant}},\ }\href {https://doi.org/10.1038/nature24471} {\bibfield
  {journal} {\bibinfo  {journal} {Nature}\ }\textbf {\bibinfo {volume} {551}},\
  \bibinfo {pages} {85} (\bibinfo {year} {2017}{\natexlab{b}})}\BibitemShut
  {NoStop}%
\bibitem [{\citenamefont {Popovic}\ \emph
  {et~al.}(2025{\natexlab{b}})\citenamefont {Popovic} \emph
  {et~al.}}]{Popovic2025DESDovekie}%
  \BibitemOpen
  \bibfield  {author} {\bibinfo {author} {\bibfnamefont {B.}~\bibnamefont
  {Popovic}} \emph {et~al.},\ }\href
  {https://doi.org/10.48550/arXiv.2511.07517} {\bibinfo {title} {{DES-Dovekie:
  A Reanalysis of DES Supernova Cosmology with Updated Calibration and
  Systematics}}} (\bibinfo {year} {2025}{\natexlab{b}}),\ \bibinfo {note}
  {arXiv preprint},\ \Eprint {https://arxiv.org/abs/2511.07517}
  {arXiv:2511.07517 [astro-ph.CO]} \BibitemShut {NoStop}%
\bibitem [{\citenamefont {Turner}(2024)}]{Turner2024PeculiarVelocity}%
  \BibitemOpen
  \bibfield  {author} {\bibinfo {author} {\bibfnamefont {R.~J.}\ \bibnamefont
  {Turner}},\ }\href@noop {} {\bibinfo {title} {{Cosmology with Peculiar
  Velocity Surveys}}} (\bibinfo {year} {2024}),\ \Eprint
  {https://arxiv.org/abs/2411.19484} {arXiv:2411.19484 [astro-ph.CO]}
  \BibitemShut {NoStop}%
\bibitem [{\citenamefont {Said}\ \emph {et~al.}(2025)\citenamefont {Said},
  \citenamefont {Howlett}, \citenamefont {Davis}, \citenamefont {Lucey},
  \citenamefont {Saulder}, \citenamefont {Douglass}, \citenamefont {Kim},
  \citenamefont {Kremin}, \citenamefont {Ross}, \citenamefont {Aldering} \emph
  {et~al.}}]{SaidEtAl2024DESIPVFundamentalPlane}%
  \BibitemOpen
  \bibfield  {author} {\bibinfo {author} {\bibfnamefont {K.}~\bibnamefont
  {Said}}, \bibinfo {author} {\bibfnamefont {C.}~\bibnamefont {Howlett}},
  \bibinfo {author} {\bibfnamefont {T.}~\bibnamefont {Davis}}, \bibinfo
  {author} {\bibfnamefont {J.}~\bibnamefont {Lucey}}, \bibinfo {author}
  {\bibfnamefont {C.}~\bibnamefont {Saulder}}, \bibinfo {author} {\bibfnamefont
  {K.}~\bibnamefont {Douglass}}, \bibinfo {author} {\bibfnamefont {A.~G.}\
  \bibnamefont {Kim}}, \bibinfo {author} {\bibfnamefont {A.}~\bibnamefont
  {Kremin}}, \bibinfo {author} {\bibfnamefont {C.}~\bibnamefont {Ross}},
  \bibinfo {author} {\bibfnamefont {G.}~\bibnamefont {Aldering}}, \emph
  {et~al.},\ }\bibfield  {title} {\bibinfo {title} {{DESI Peculiar Velocity
  Survey -- Fundamental Plane}},\ }\href
  {https://doi.org/10.1093/mnras/staf700} {\bibfield  {journal} {\bibinfo
  {journal} {MNRAS}\ }\textbf {\bibinfo {volume} {539}},\ \bibinfo {pages}
  {3627} (\bibinfo {year} {2025})}\BibitemShut {NoStop}%
\bibitem [{\citenamefont {Moresco}(2024)}]{Moresco2024CosmicChronometers}%
  \BibitemOpen
  \bibfield  {author} {\bibinfo {author} {\bibfnamefont {M.}~\bibnamefont
  {Moresco}},\ }\href@noop {} {\bibinfo {title} {{Measuring the Expansion
  History of the Universe with Cosmic Chronometers}}} (\bibinfo {year}
  {2024}),\ \Eprint {https://arxiv.org/abs/2412.01994} {arXiv:2412.01994
  [astro-ph.CO]} \BibitemShut {NoStop}%
\bibitem [{\citenamefont {St{\"o}lzner}\ \emph {et~al.}(2025)\citenamefont
  {St{\"o}lzner}, \citenamefont {Wright}, \citenamefont {Asgari} \emph
  {et~al.}}]{Stolzner2025KiDSLegacyConsistency}%
  \BibitemOpen
  \bibfield  {author} {\bibinfo {author} {\bibfnamefont {B.}~\bibnamefont
  {St{\"o}lzner}}, \bibinfo {author} {\bibfnamefont {A.~H.}\ \bibnamefont
  {Wright}}, \bibinfo {author} {\bibfnamefont {M.}~\bibnamefont {Asgari}},
  \emph {et~al.},\ }\bibfield  {title} {\bibinfo {title} {{KiDS-Legacy:
  Consistency of Cosmic Shear Measurements and Joint Cosmological Constraints
  with External Probes}},\ }\href {https://doi.org/10.1051/0004-6361/202554893}
  {\bibfield  {journal} {\bibinfo  {journal} {Astron. Astrophys.}\ }\textbf
  {\bibinfo {volume} {702}},\ \bibinfo {pages} {A169} (\bibinfo {year}
  {2025})}\BibitemShut {NoStop}%
\bibitem [{\citenamefont {de~Cruz~Perez}\ \emph {et~al.}(2023)\citenamefont
  {de~Cruz~Perez}, \citenamefont {Park},\ and\ \citenamefont
  {Ratra}}]{deCruzPerezParkRatra2022Lensing}%
  \BibitemOpen
  \bibfield  {author} {\bibinfo {author} {\bibfnamefont {J.}~\bibnamefont
  {de~Cruz~Perez}}, \bibinfo {author} {\bibfnamefont {C.-G.}\ \bibnamefont
  {Park}},\ and\ \bibinfo {author} {\bibfnamefont {B.}~\bibnamefont {Ratra}},\
  }\bibfield  {title} {\bibinfo {title} {{Current Data Are Consistent with Flat
  Spatial Hypersurfaces in the {$\Lambda$CDM} Cosmological Model but Favor More
  Lensing than the Model Predicts}},\ }\href
  {https://doi.org/10.1103/PhysRevD.107.063522} {\bibfield  {journal} {\bibinfo
   {journal} {Phys. Rev. D}\ }\textbf {\bibinfo {volume} {107}},\ \bibinfo
  {pages} {063522} (\bibinfo {year} {2023})}\BibitemShut {NoStop}%
\bibitem [{\citenamefont {Chevallier}\ and\ \citenamefont
  {Polarski}(2001)}]{Chevallier2001}%
  \BibitemOpen
  \bibfield  {author} {\bibinfo {author} {\bibfnamefont {M.}~\bibnamefont
  {Chevallier}}\ and\ \bibinfo {author} {\bibfnamefont {D.}~\bibnamefont
  {Polarski}},\ }\bibfield  {title} {\bibinfo {title} {{Accelerating Universes
  with Scaling Dark Matter}},\ }\href
  {https://doi.org/10.1142/S0218271801000822} {\bibfield  {journal} {\bibinfo
  {journal} {Int. J. Mod. Phys. D}\ }\textbf {\bibinfo {volume} {10}},\
  \bibinfo {pages} {213} (\bibinfo {year} {2001})}\BibitemShut {NoStop}%
\bibitem [{\citenamefont {Linder}(2003)}]{Linder2003}%
  \BibitemOpen
  \bibfield  {author} {\bibinfo {author} {\bibfnamefont {E.~V.}\ \bibnamefont
  {Linder}},\ }\bibfield  {title} {\bibinfo {title} {{Exploring the Expansion
  History of the Universe}},\ }\href
  {https://doi.org/10.1103/PhysRevLett.90.091301} {\bibfield  {journal}
  {\bibinfo  {journal} {Phys. Rev. Lett.}\ }\textbf {\bibinfo {volume} {90}},\
  \bibinfo {pages} {091301} (\bibinfo {year} {2003})}\BibitemShut {NoStop}%
\bibitem [{\citenamefont {Albrecht}\ \emph {et~al.}(2006)\citenamefont
  {Albrecht} \emph {et~al.}}]{Albrecht2006DETF}%
  \BibitemOpen
  \bibfield  {author} {\bibinfo {author} {\bibfnamefont {A.}~\bibnamefont
  {Albrecht}} \emph {et~al.},\ }\href@noop {} {\emph {\bibinfo {title} {{Report
  of the Dark Energy Task Force}}}},\ \bibinfo {type} {Tech. Rep.}\ (\bibinfo
  {institution} {U.S. Department of Energy / NASA / NSF},\ \bibinfo {year}
  {2006})\ \bibinfo {note} {dETF Final Report},\ \Eprint
  {https://arxiv.org/abs/astro-ph/0609591} {arXiv:astro-ph/0609591}
  \BibitemShut {NoStop}%
\bibitem [{\citenamefont {Poulin}\ \emph {et~al.}(2019)\citenamefont {Poulin},
  \citenamefont {Smith}, \citenamefont {Karwal},\ and\ \citenamefont
  {Kamionkowski}}]{Poulin2019EDE}%
  \BibitemOpen
  \bibfield  {author} {\bibinfo {author} {\bibfnamefont {V.}~\bibnamefont
  {Poulin}}, \bibinfo {author} {\bibfnamefont {T.~L.}\ \bibnamefont {Smith}},
  \bibinfo {author} {\bibfnamefont {T.}~\bibnamefont {Karwal}},\ and\ \bibinfo
  {author} {\bibfnamefont {M.}~\bibnamefont {Kamionkowski}},\ }\bibfield
  {title} {\bibinfo {title} {Early dark energy can resolve the {Hubble}
  tension},\ }\href {https://doi.org/10.1103/PhysRevLett.122.221301} {\bibfield
   {journal} {\bibinfo  {journal} {Phys. Rev. Lett.}\ }\textbf {\bibinfo
  {volume} {122}},\ \bibinfo {pages} {221301} (\bibinfo {year}
  {2019})}\BibitemShut {NoStop}%
\bibitem [{\citenamefont {Hill}\ \emph {et~al.}(2020)\citenamefont {Hill},
  \citenamefont {McDonough}, \citenamefont {Toomey},\ and\ \citenamefont
  {Alexander}}]{Hill2020EDE}%
  \BibitemOpen
  \bibfield  {author} {\bibinfo {author} {\bibfnamefont {J.~C.}\ \bibnamefont
  {Hill}}, \bibinfo {author} {\bibfnamefont {E.}~\bibnamefont {McDonough}},
  \bibinfo {author} {\bibfnamefont {M.~W.}\ \bibnamefont {Toomey}},\ and\
  \bibinfo {author} {\bibfnamefont {S.}~\bibnamefont {Alexander}},\ }\bibfield
  {title} {\bibinfo {title} {{Early Dark Energy Does Not Restore Cosmological
  Concordance}},\ }\href {https://doi.org/10.1103/PhysRevD.102.043507}
  {\bibfield  {journal} {\bibinfo  {journal} {Phys. Rev. D}\ }\textbf {\bibinfo
  {volume} {102}},\ \bibinfo {pages} {043507} (\bibinfo {year}
  {2020})}\BibitemShut {NoStop}%
\bibitem [{\citenamefont {Kamionkowski}\ and\ \citenamefont
  {Riess}(2023)}]{Kamionkowski2023EDEReview}%
  \BibitemOpen
  \bibfield  {author} {\bibinfo {author} {\bibfnamefont {M.}~\bibnamefont
  {Kamionkowski}}\ and\ \bibinfo {author} {\bibfnamefont {A.~G.}\ \bibnamefont
  {Riess}},\ }\bibfield  {title} {\bibinfo {title} {{The Hubble Tension and
  Early Dark Energy}},\ }\href
  {https://doi.org/10.1146/annurev-nucl-111422-024107} {\bibfield  {journal}
  {\bibinfo  {journal} {Annu. Rev. Nucl. Part. Sci.}\ }\textbf {\bibinfo
  {volume} {73}},\ \bibinfo {pages} {153} (\bibinfo {year} {2023})}\BibitemShut
  {NoStop}%
\bibitem [{\citenamefont {Vagnozzi}(2021)}]{Vagnozzi2021eISW}%
  \BibitemOpen
  \bibfield  {author} {\bibinfo {author} {\bibfnamefont {S.}~\bibnamefont
  {Vagnozzi}},\ }\bibfield  {title} {\bibinfo {title} {{The early integrated
  Sachs--Wolfe effect as a probe of cosmological tensions}},\ }\href
  {https://doi.org/10.1103/PhysRevD.104.063524} {\bibfield  {journal} {\bibinfo
   {journal} {Phys. Rev. D}\ }\textbf {\bibinfo {volume} {104}},\ \bibinfo
  {pages} {063524} (\bibinfo {year} {2021})}\BibitemShut {NoStop}%
\bibitem [{\citenamefont {Pedrotti}\ and\ \citenamefont
  {Vagnozzi}(2025)}]{PedrottiVagnozzi2024HubbleMultiD}%
  \BibitemOpen
  \bibfield  {author} {\bibinfo {author} {\bibfnamefont {G.}~\bibnamefont
  {Pedrotti}}\ and\ \bibinfo {author} {\bibfnamefont {S.}~\bibnamefont
  {Vagnozzi}},\ }\bibfield  {title} {\bibinfo {title} {{Multidimensionality of
  the Hubble tension: the roles of $\Omega_m$ and $\omega_c$}},\ }\href
  {https://doi.org/10.1103/PhysRevD.111.023506} {\bibfield  {journal} {\bibinfo
   {journal} {Phys. Rev. D}\ }\textbf {\bibinfo {volume} {111}},\ \bibinfo
  {pages} {023506} (\bibinfo {year} {2025})}\BibitemShut {NoStop}%
\bibitem [{\citenamefont {Peebles}\ and\ \citenamefont
  {Ratra}(1988)}]{PeeblesRatra1988TimeVariableCC}%
  \BibitemOpen
  \bibfield  {author} {\bibinfo {author} {\bibfnamefont {P.~J.~E.}\
  \bibnamefont {Peebles}}\ and\ \bibinfo {author} {\bibfnamefont
  {B.}~\bibnamefont {Ratra}},\ }\bibfield  {title} {\bibinfo {title}
  {{Cosmology with a Time-Variable Cosmological ``Constant''}},\ }\href
  {https://doi.org/10.1086/185100} {\bibfield  {journal} {\bibinfo  {journal}
  {Astrophys. J. Lett.}\ }\textbf {\bibinfo {volume} {325}},\ \bibinfo {pages}
  {L17} (\bibinfo {year} {1988})}\BibitemShut {NoStop}%
\bibitem [{\citenamefont {Ratra}\ and\ \citenamefont
  {Peebles}(1988)}]{RatraPeebles1988RollingScalar}%
  \BibitemOpen
  \bibfield  {author} {\bibinfo {author} {\bibfnamefont {B.}~\bibnamefont
  {Ratra}}\ and\ \bibinfo {author} {\bibfnamefont {P.~J.~E.}\ \bibnamefont
  {Peebles}},\ }\bibfield  {title} {\bibinfo {title} {{Cosmological
  Consequences of a Rolling Homogeneous Scalar Field}},\ }\href
  {https://doi.org/10.1103/PhysRevD.37.3406} {\bibfield  {journal} {\bibinfo
  {journal} {Phys. Rev. D}\ }\textbf {\bibinfo {volume} {37}},\ \bibinfo
  {pages} {3406} (\bibinfo {year} {1988})}\BibitemShut {NoStop}%
\bibitem [{\citenamefont {Copeland}\ \emph {et~al.}(2006)\citenamefont
  {Copeland}, \citenamefont {Sami},\ and\ \citenamefont
  {Tsujikawa}}]{Copeland2006QuintessenceReview}%
  \BibitemOpen
  \bibfield  {author} {\bibinfo {author} {\bibfnamefont {E.~J.}\ \bibnamefont
  {Copeland}}, \bibinfo {author} {\bibfnamefont {M.}~\bibnamefont {Sami}},\
  and\ \bibinfo {author} {\bibfnamefont {S.}~\bibnamefont {Tsujikawa}},\
  }\bibfield  {title} {\bibinfo {title} {{Dynamics of Dark Energy}},\ }\href
  {https://doi.org/10.1142/S021827180600942X} {\bibfield  {journal} {\bibinfo
  {journal} {Int. J. Mod. Phys. D}\ }\textbf {\bibinfo {volume} {15}},\
  \bibinfo {pages} {1753} (\bibinfo {year} {2006})}\BibitemShut {NoStop}%
\bibitem [{\citenamefont {Zlatev}\ \emph {et~al.}(1999)\citenamefont {Zlatev},
  \citenamefont {Wang},\ and\ \citenamefont {Steinhardt}}]{Zlatev1999Tracker}%
  \BibitemOpen
  \bibfield  {author} {\bibinfo {author} {\bibfnamefont {I.}~\bibnamefont
  {Zlatev}}, \bibinfo {author} {\bibfnamefont {L.-M.}\ \bibnamefont {Wang}},\
  and\ \bibinfo {author} {\bibfnamefont {P.~J.}\ \bibnamefont {Steinhardt}},\
  }\bibfield  {title} {\bibinfo {title} {{Quintessence, Cosmic Coincidence, and
  the Cosmological Constant}},\ }\href
  {https://doi.org/10.1103/PhysRevLett.82.896} {\bibfield  {journal} {\bibinfo
  {journal} {Phys. Rev. Lett.}\ }\textbf {\bibinfo {volume} {82}},\ \bibinfo
  {pages} {896} (\bibinfo {year} {1999})}\BibitemShut {NoStop}%
\bibitem [{\citenamefont {Vikman}(2005)}]{Vikman2005NoGo}%
  \BibitemOpen
  \bibfield  {author} {\bibinfo {author} {\bibfnamefont {A.}~\bibnamefont
  {Vikman}},\ }\bibfield  {title} {\bibinfo {title} {{Can Dark Energy Evolve to
  the Phantom?}},\ }\href {https://doi.org/10.1103/PhysRevD.71.023515}
  {\bibfield  {journal} {\bibinfo  {journal} {Phys. Rev. D}\ }\textbf {\bibinfo
  {volume} {71}},\ \bibinfo {pages} {023515} (\bibinfo {year}
  {2005})}\BibitemShut {NoStop}%
\bibitem [{\citenamefont {Feng}\ \emph {et~al.}(2005)\citenamefont {Feng},
  \citenamefont {Wang},\ and\ \citenamefont {Zhang}}]{Feng2005Quintom}%
  \BibitemOpen
  \bibfield  {author} {\bibinfo {author} {\bibfnamefont {B.}~\bibnamefont
  {Feng}}, \bibinfo {author} {\bibfnamefont {X.}~\bibnamefont {Wang}},\ and\
  \bibinfo {author} {\bibfnamefont {X.}~\bibnamefont {Zhang}},\ }\bibfield
  {title} {\bibinfo {title} {{Dark Energy Constraints from the Cosmic Age and
  Supernova}},\ }\href {https://doi.org/10.1016/j.physletb.2004.12.071}
  {\bibfield  {journal} {\bibinfo  {journal} {Phys. Lett. B}\ }\textbf
  {\bibinfo {volume} {607}},\ \bibinfo {pages} {35} (\bibinfo {year}
  {2005})}\BibitemShut {NoStop}%
\bibitem [{\citenamefont {Amendola}(2000)}]{Amendola2000CoupledDE}%
  \BibitemOpen
  \bibfield  {author} {\bibinfo {author} {\bibfnamefont {L.}~\bibnamefont
  {Amendola}},\ }\bibfield  {title} {\bibinfo {title} {{Coupled
  Quintessence}},\ }\href {https://doi.org/10.1103/PhysRevD.62.043511}
  {\bibfield  {journal} {\bibinfo  {journal} {Phys. Rev. D}\ }\textbf {\bibinfo
  {volume} {62}},\ \bibinfo {pages} {043511} (\bibinfo {year}
  {2000})}\BibitemShut {NoStop}%
\bibitem [{\citenamefont {Dymnikova}\ and\ \citenamefont
  {Khlopov}(2000)}]{DymnikovaKhlopov2000VacuumEvap}%
  \BibitemOpen
  \bibfield  {author} {\bibinfo {author} {\bibfnamefont {I.~G.}\ \bibnamefont
  {Dymnikova}}\ and\ \bibinfo {author} {\bibfnamefont {M.~Y.}\ \bibnamefont
  {Khlopov}},\ }\bibfield  {title} {\bibinfo {title} {{Decay of cosmological
  constant as Bose condensate evaporation}},\ }\href
  {https://doi.org/10.1142/S0217732300002966} {\bibfield  {journal} {\bibinfo
  {journal} {Mod. Phys. Lett. A}\ }\textbf {\bibinfo {volume} {15}},\ \bibinfo
  {pages} {2305} (\bibinfo {year} {2000})}\BibitemShut {NoStop}%
\bibitem [{\citenamefont {Caldwell}(2002)}]{Caldwell2002PhantomMenace}%
  \BibitemOpen
  \bibfield  {author} {\bibinfo {author} {\bibfnamefont {R.~R.}\ \bibnamefont
  {Caldwell}},\ }\bibfield  {title} {\bibinfo {title} {{A Phantom Menace?
  Cosmological Consequences of a Dark Energy Component with Super-Negative
  Equation of State}},\ }\href {https://doi.org/10.1016/S0370-2693(02)02589-3}
  {\bibfield  {journal} {\bibinfo  {journal} {Phys. Lett. B}\ }\textbf
  {\bibinfo {volume} {545}},\ \bibinfo {pages} {23} (\bibinfo {year}
  {2002})}\BibitemShut {NoStop}%
\bibitem [{\citenamefont {Hu}\ and\ \citenamefont
  {Sawicki}(2007)}]{HuSawicki2007fR}%
  \BibitemOpen
  \bibfield  {author} {\bibinfo {author} {\bibfnamefont {W.}~\bibnamefont
  {Hu}}\ and\ \bibinfo {author} {\bibfnamefont {I.}~\bibnamefont {Sawicki}},\
  }\bibfield  {title} {\bibinfo {title} {{Models of $f(R)$ Cosmic Acceleration
  that Evade Solar-System Tests}},\ }\href
  {https://doi.org/10.1103/PhysRevD.76.064004} {\bibfield  {journal} {\bibinfo
  {journal} {Phys. Rev. D}\ }\textbf {\bibinfo {volume} {76}},\ \bibinfo
  {pages} {064004} (\bibinfo {year} {2007})}\BibitemShut {NoStop}%
\bibitem [{\citenamefont {Dvali}\ \emph {et~al.}(2000)\citenamefont {Dvali},
  \citenamefont {Gabadadze},\ and\ \citenamefont {Porrati}}]{Dvali2000DGP}%
  \BibitemOpen
  \bibfield  {author} {\bibinfo {author} {\bibfnamefont {G.~R.}\ \bibnamefont
  {Dvali}}, \bibinfo {author} {\bibfnamefont {G.}~\bibnamefont {Gabadadze}},\
  and\ \bibinfo {author} {\bibfnamefont {M.}~\bibnamefont {Porrati}},\
  }\bibfield  {title} {\bibinfo {title} {{4D Gravity on a Brane in 5D Minkowski
  Space}},\ }\href {https://doi.org/10.1016/S0370-2693(00)00669-9} {\bibfield
  {journal} {\bibinfo  {journal} {Phys. Lett. B}\ }\textbf {\bibinfo {volume}
  {485}},\ \bibinfo {pages} {208} (\bibinfo {year} {2000})}\BibitemShut
  {NoStop}%
\bibitem [{\citenamefont {Belgacem}\ \emph {et~al.}(2018)\citenamefont
  {Belgacem}, \citenamefont {Dirian}, \citenamefont {Foffa},\ and\
  \citenamefont {Maggiore}}]{Belgacem2018GWLumDist}%
  \BibitemOpen
  \bibfield  {author} {\bibinfo {author} {\bibfnamefont {E.}~\bibnamefont
  {Belgacem}}, \bibinfo {author} {\bibfnamefont {Y.}~\bibnamefont {Dirian}},
  \bibinfo {author} {\bibfnamefont {S.}~\bibnamefont {Foffa}},\ and\ \bibinfo
  {author} {\bibfnamefont {M.}~\bibnamefont {Maggiore}},\ }\bibfield  {title}
  {\bibinfo {title} {{The Gravitational-Wave Luminosity Distance in Modified
  Gravity Theories}},\ }\href {https://doi.org/10.1103/PhysRevD.97.104066}
  {\bibfield  {journal} {\bibinfo  {journal} {Phys. Rev. D}\ }\textbf {\bibinfo
  {volume} {97}},\ \bibinfo {pages} {104066} (\bibinfo {year}
  {2018})}\BibitemShut {NoStop}%
\bibitem [{\citenamefont {Spiegelhalter}\ \emph {et~al.}(2002)\citenamefont
  {Spiegelhalter}, \citenamefont {Best}, \citenamefont {Carlin},\ and\
  \citenamefont {Van Der~Linde}}]{SpiegelhalterEtAl2002DIC}%
  \BibitemOpen
  \bibfield  {author} {\bibinfo {author} {\bibfnamefont {D.~J.}\ \bibnamefont
  {Spiegelhalter}}, \bibinfo {author} {\bibfnamefont {N.~G.}\ \bibnamefont
  {Best}}, \bibinfo {author} {\bibfnamefont {B.~P.}\ \bibnamefont {Carlin}},\
  and\ \bibinfo {author} {\bibfnamefont {A.}~\bibnamefont {Van Der~Linde}},\
  }\bibfield  {title} {\bibinfo {title} {{Bayesian Measures of Model Complexity
  and Fit}},\ }\href {https://doi.org/10.1111/1467-9868.00353} {\bibfield
  {journal} {\bibinfo  {journal} {J. Roy. Stat. Soc. B}\ }\textbf {\bibinfo
  {volume} {64}},\ \bibinfo {pages} {583} (\bibinfo {year} {2002})}\BibitemShut
  {NoStop}%
\bibitem [{\citenamefont {Rezaei}\ and\ \citenamefont
  {Malekjani}(2021)}]{RezaeiMalekjani2021ModelSelection}%
  \BibitemOpen
  \bibfield  {author} {\bibinfo {author} {\bibfnamefont {M.}~\bibnamefont
  {Rezaei}}\ and\ \bibinfo {author} {\bibfnamefont {M.}~\bibnamefont
  {Malekjani}},\ }\bibfield  {title} {\bibinfo {title} {{Comparison between
  Different Methods of Model Selection in Cosmology}},\ }\href
  {https://doi.org/10.1140/epjp/s13360-021-01200-w} {\bibfield  {journal}
  {\bibinfo  {journal} {Eur. Phys. J. Plus}\ }\textbf {\bibinfo {volume}
  {136}},\ \bibinfo {pages} {219} (\bibinfo {year} {2021})},\ \Eprint
  {https://arxiv.org/abs/2102.10671} {arXiv:2102.10671 [astro-ph.CO]}
  \BibitemShut {NoStop}%
\bibitem [{\citenamefont {Turyshev}\ \emph {et~al.}(2024)\citenamefont
  {Turyshev}, \citenamefont {Chiow},\ and\ \citenamefont
  {Yu}}]{TuryshevChiowYu2024Tetrahedral}%
  \BibitemOpen
  \bibfield  {author} {\bibinfo {author} {\bibfnamefont {S.~G.}\ \bibnamefont
  {Turyshev}}, \bibinfo {author} {\bibfnamefont {S.-w.}\ \bibnamefont
  {Chiow}},\ and\ \bibinfo {author} {\bibfnamefont {N.}~\bibnamefont {Yu}},\
  }\bibfield  {title} {\bibinfo {title} {{Searching for new physics in the
  solar system with tetrahedral spacecraft formations}},\ }\href
  {https://doi.org/10.1103/PhysRevD.109.084059} {\bibfield  {journal} {\bibinfo
   {journal} {Phys. Rev. D}\ }\textbf {\bibinfo {volume} {109}},\ \bibinfo
  {pages} {084059} (\bibinfo {year} {2024})},\ \Eprint
  {https://arxiv.org/abs/2404.02096} {arXiv:2404.02096 [gr-qc]} \BibitemShut
  {NoStop}%
\bibitem [{\citenamefont {Turyshev}(2025)}]{Turyshev2025SolarSystemDEDM}%
  \BibitemOpen
  \bibfield  {author} {\bibinfo {author} {\bibfnamefont {S.~G.}\ \bibnamefont
  {Turyshev}},\ }\bibfield  {title} {\bibinfo {title} {{Solar System
  Experiments in the Search for Dark Energy and Dark Matter}},\ }\href
  {https://doi.org/10.1103/cmwl-xnhz} {\bibfield  {journal} {\bibinfo
  {journal} {Phys. Rev. D}\ }\textbf {\bibinfo {volume} {112}},\ \bibinfo
  {pages} {123003} (\bibinfo {year} {2025})},\ \Eprint
  {https://arxiv.org/abs/2509.05910} {arXiv:2509.05910 [astro-ph.CO]}
  \BibitemShut {NoStop}%
\end{thebibliography}

%

\end{document}